\documentclass[traditabstract]{aa}

\usepackage{natbib}
\usepackage{amsmath,amssymb,verbatim}
\usepackage{graphicx}
\usepackage{hyperref}
\usepackage{txfonts}
\usepackage{color}

\newcommand{\rom}[1]{\MakeUppercase{\romannumeral #1}}

\begin{document}
        
\title{The gravitational field of X-COP galaxy clusters}

\author{D. Eckert\inst{1} \and S. Ettori\inst{2,3} \and E. Pointecouteau\inst{4} \and R. F. J. van der Burg\inst{5} \and S. I. Loubser\inst{6}}
\institute{
        Department of Astronomy, University of Geneva, Ch. d'Ecogia 16, CH-1290 Versoix, Switzerland\\
        \email{Dominique.Eckert@unige.ch}
        \and
        INAF, Osservatorio di Astrofisica e Scienza dello Spazio, via Piero Gobetti 93/3, 40129 Bologna, Italy
        \and
        INFN, Sezione di Bologna, viale Berti Pichat 6/2, I-40127 Bologna, Italy
        \and
         IRAP, Université de Toulouse, CNRS, CNES, UPS, Toulouse, France
         \and
         European Southern Observatory, Karl-Schwarzschild-Str. 2, 85748, Garching, Germany
         \and
         Centre for Space Research, North-West University, Potchefstroom 2520, South Africa
}

\abstract{The mass profiles of massive dark matter halos are highly sensitive to the nature of dark matter and potential modifications of the theory of gravity on large scales. The $\Lambda$ cold dark matter (CDM) paradigm makes strong predictions on the shape of dark matter halos and on the dependence of the shape parameters on halo mass, such that any deviation from the predicted universal shape would have important implications for the fundamental properties of dark matter. Here we use a set of 12 galaxy clusters with available deep X-ray and Sunyaev-Zel'dovich data to constrain the shape of the gravitational field with an unprecedented level of precision over two decades in radius. We introduce a nonparametric framework to reconstruct the shape of the gravitational field under the assumption of hydrostatic equilibrium and compare the resulting mass profiles to the expectations of Navarro-Frenk-White (NFW) and Einasto parametric mass profiles. On average, we find that the NFW profile provides an excellent description of the recovered mass profiles, with deviations of less than 10\% over a wide radial range. However, there appears to be more diversity in the shape of individual profiles than can be captured by the NFW model. The average NFW concentration and its scatter agree very well with the prediction of the $\Lambda$CDM framework. For a subset of systems, we disentangle the gravitational field into the contribution of baryonic components (gas, brightest cluster galaxy, and satellite galaxies) and that of dark matter. The stellar content dominates the gravitational field inside $\sim0.02R_{500}$ but is responsible for only 1-2\% of the total gravitational field inside $R_{200}$. The total baryon fraction reaches the cosmic value at $R_{200}$ and slightly exceeds it beyond this point, possibly indicating a mild level of nonthermal pressure support ($10-20\%$) in cluster outskirts. Finally, the relation between observed and baryonic acceleration exhibits a complex shape that strongly departs from the radial acceleration relation in spiral galaxies, which shows that the aforementioned relation does not hold at the galaxy-cluster scale.}

\maketitle

\section{Introduction}a
Numerical simulations of cosmic structure formation have long predicted that the shape of collapsed halos is universal over a wide range of halo masses. In the cold dark matter (CDM) paradigm, dark matter (DM) halos are expected to follow the so-called Navarro-Frenk-White (NFW) profile \citep[][]{nfw96,nfw97}, in which the matter density forms a central ``cusp'' with $\rho(r) \propto r^{-1}$ and gradually steepens toward the outskirts as $\rho(r) \propto r^{-3}$, with a characteristic radius $r_s$ that defines the scale at which the logarithmic slope of the density profile has an isothermal value of $-$2 (i.e., $d \ln \rho/d \ln r | r_s = -2$). The characteristic shape is thought to arise from the universality of the structure formation process \citep{bullock01,maccio08}, and the distribution and redshift evolution of the halo structural parameters encode important cosmological information \citep[e.g.,][]{duffy08,dutton14}. As discussed in the original work by Navarro, Frenk, and White, the total mass at a given overdensity (typically 200 times the critical density of the universe enclosed in a sphere with radius $R_{200}$) correlates with the mass concentration of the halo, $c = R_{200} / r_s$. This correlation reflects the conditions in which a halo assembled, with halos formed earlier being more concentrated because the background density was larger back in cosmological time \citep{wechsler02}. More recently, several works \citep[see, e.g.,][]{dutton14,ludlow16,ragagnin20} have modeled and quantified the effects of halo accretion histories and of the parameters that describe the cosmological background model on the halo mass concentration. 
Recent N-body simulations have also shown that the radial profiles of massive halos deviate from the NFW prediction \citep{navarro04,diemer14,klypin16,ludlow16} and can be better fitted by functional forms that include a power-law dependence of the logarithmic radial slope, such as $d \ln \rho / d \ln r \propto r^{-\alpha}$. The \citet{einasto65} profile is an example of this revised functional form, where the Einasto index, $\alpha$, can be related to the power-law index of the primordial power spectrum, $n_s$ \citep{ludlow17,brown20}.

Additionally, deviations from the universal NFW shape can provide important clues to the nature of DM. For instance, in cases where the DM self-interaction (SIDM) cross section  is non-negligible, high-density regions are expected to be homogenized, leading to flatter cores compared to the cuspy profiles expected in the CDM scenario \citep{robertson20}. ``Warm'' dark matter (WDM) particle candidates with masses of a few keV would free-stream across the high-density regions of halos, thereby flattening the observed profiles \citep{maccio13,ludlow16}. Modified gravity theories such as MOdified Newtonian Dynamics \citep[MOND;][]{milgrom83} and its relativistic extensions \citep[e.g.,][]{bekenstein04,milgrom09,skordis21} can also be tested by comparing the observed gravitational field to predictions \citep[see][for a review]{famaey12}. 

As the most massive collapsed structures in the Universe, galaxy clusters are privileged sites for studying the properties of the gravitational field over a wide range of densities. Given their deep potential well, gravitational processes largely dominate over feedback processes, such that galaxy clusters are essentially closed boxes for baryons \citep{white93, evrard97, eke98, kravtsov05}. Recent studies demonstrate that the baryon fraction enclosed within the virial radius of massive clusters is very close to the universal baryon fraction \citep{gonzalez07,lagana13,mantz14,chiu17,eckert19}, with the majority of the baryonic mass ($\sim90\%$) in the hot intracluster medium (ICM) and the remaining 10\% in stars. The total baryonic content (gas and stars) of these structures is directly observable. Galaxy clusters thus represent ideal structures for studying the properties of the gravitational field in DM halos and its interplay with the various baryonic components.

The mass profiles of galaxy clusters have been the subject of numerous recent studies \citep[see][for a review]{pratt19}. The shape of the mass distribution can be estimated through weak and strong gravitational lensing \citep{okabe13,umetsu16,umetsu17,tam20}, the dynamics of cluster galaxies \citep{biviano13,munari14,geller14}, and the hydrostatic equilibrium (HSE) equation \citep{pointecouteau05,vikhlinin06,ettori10,bartalucci19,ettori19}. The NFW profile usually provides an adequate description of the data at hand, irrespective of the adopted method \citep{umetsu16,ettori19}. Models that include a large central core, such as the Burkert \citep{salucci00} or isothermal sphere models \citep{king62}, provide a significantly worse representation of the data \citep{umetsu17,ettori19}. However, the data quality is usually insufficient to distinguish between the various models that include a central cusp, in particular between the NFW and Einasto profiles. Finally, galaxy cluster mass profiles were also extensively used to test modified gravity theories \citep[e.g.,][]{sanders99,sanders03,pointecouteau05b,wilcox15,ettori17,mitchell18}.

In this paper we investigate in detail the shape of the gravitational field in local galaxy clusters under the HSE assumption. We make use of the data acquired for the \emph{XMM Cluster Outskirts Project} \citep[X-COP;][]{xcop}, a very large program on \emph{XMM-Newton} that provides complete X-ray mapping out to the virial radius for a sample of 12 nearby systems selected from the \emph{Planck} Sunyaev-Zel'dovich (SZ) survey. We present a novel nonparametric technique for recovering hydrostatic mass profiles from joint X-ray and SZ data, which we include in a global framework for the estimation of hydrostatic mass profiles. In Sect. \ref{sec:mock} we validate our reconstruction technique using mock \emph{XMM-Newton} observations of a synthetic NFW cluster. We apply our technique to the 12 X-COP systems with the aim of comparing NFW and Einasto profiles. For a subset of systems, we use measurements of the gas mass as well as the stellar mass within the brightest cluster galaxy (BCG) and satellite galaxies to break down the gravitational field into its baryonic and DM components. Finally, we study the relation between the baryonic and total gravitational acceleration and compare it with the radial acceleration relation (RAR) of rotationally supported galaxies \citep{mcgaugh16}. In a companion paper (hereafter Paper \rom{2}), we discuss our measurements of the Einasto shape parameter and perform a detailed comparison to cosmological simulations that include self-interacting DM.

Throughout the paper we assume a Planck 2015 $\Lambda$CDM cosmology \citep{planck15_13} with $H_0=67.8$ km/s/Mpc and $\Omega_m=0.308$. The codes developed in the framework of this project are made publicly available. In particular, we release our mass reconstruction code in the form of an easy-to-use Python package named \texttt{hydromass}. Our \emph{XMM-Newton} mock data generator can be downloaded from GitHub.

\section{Sample and data reduction}
\label{sec:sample}

\subsection{The X-COP sample}

X-COP \citep{xcop} is a very large program on \emph{XMM-Newton,} totaling 1.5 Ms of observing time. The original data set was awarded during \emph{XMM-Newton} AO-13 (proposal ID 074441, PI: Eckert) and completed in 2015. The primary goal of the project is to use the combination of X-ray data from \emph{XMM-Newton} with SZ data from \emph{Planck} to probe the state of the ICM out to the virial radius. The sample was selected from the first \emph{Planck} SZ catalog \citep[PSZ1;][]{psz1} according to the following criteria: (i) \emph{Planck} PSZ1 signal-to-noise greater than 12, (ii) apparent size $\theta_{500}>10^{\prime}$ to ensure that all systems are spatially resolved by \emph{Planck}, (iii) redshift in the range $0.04<z<0.1$, and Galactic absorption column density $N_H<10^{21}$ cm$^{-2}$. 

For more details about the sample we refer the reader to \citet{xcop}. The full observation log was presented in Appendix F of \citet{ghirardini19}. For the present work, we add a set of 6 additional \emph{XMM-Newton} pointings that were obtained during AO-17 (observation ID 082365, PI: Eckert) on two clusters (A3266 and A2029) to homogenize the data set. 

\subsection{XMM-Newton data analysis}
\label{sec:xmm}

With the exception of the two systems with new observations (A3266 and A2029), we employ the publicly available data products presented in \citet{ghirardini19}\footnote{\href{https://dominiqueeckert.wixsite.com/xcop}{https://dominiqueeckert.wixsite.com/xcop}} and already used to study the gravitational field of galaxy clusters in several previous works \citep{ettori17,ettori19,pradyumna21,haridasu21,harikumar22}. Here we briefly recall the data reduction chain developed for X-COP. More details are available in Sect. 2 of \citet{ghirardini19}. For the remaining two systems, we applied the same analysis procedure to the newly available data and jointly analyzed them together with the previously existing data set.

The data were processed using the \texttt{XMMSAS} v13.5 package and the corresponding calibration database. The standard event screening chains were run to extract lists of valid events for the three detectors of the European Photon Imaging Camera (EPIC). We used the \texttt{mos-filter} and \texttt{pn-filter} tasks to filter out time periods affected by strong soft proton flares. We extracted count images from the clean event files in the [0.7-1.2] keV band, which maximizes the source-to-background ratio and allows us to minimize the systematics associated with background subtraction. To model the high-energy particle background, we used a large collection of filter-wheel-closed observations available in the calibration database, and rescaled the normalization of the closed data such that the high-energy count rate matches the count rate measured in the corners of the detectors, which are located outside of the instrument's field of view (FOV) and thus do not record any incoming photon. We also attempted to model the residual soft proton background by measuring the ratio of the high-energy count rates inside and outside the FOV, which can be related to the expected soft proton rate \citep{lm08}. This relation was calibrated using a large set of blank-sky pointings and presented in Appendix B of \citet{ghirardini18}.

For each X-COP object, the count images from the various individual pointings were co-added to create a mosaic combining all the available data. The same procedure was repeated for the individual exposure maps and non X-ray background maps. We masked the detected point sources and used the azimuthal median technique \citep{eckert15} to remove the contribution of unresolved gas clumps that might bias the measured X-ray density \citep{nagai}. To compute the median surface brightness profile, we created adaptively binned surface brightness maps using Voronoi tessellation \citep{cappellari03}, and measured the median surface brightness value in concentric annuli centered on the surface brightness peak.

To determine the plasma temperature, we accumulated X-ray spectra in concentric annuli and extracted the appropriate response files and non X-ray background spectra using the Extended Source Analysis Software (ESAS) package \citep[][]{snowden08} available within \texttt{XMMSAS}. We followed the XSPEC spectral fitting procedure outlined in \citet{eckert14b}, in which the particle background is fitted with a phenomenological model and adjusted to the corners of the detectors, and the sky background is described by a three-component model \citep[cosmic X-ray background, Galactic halo, and local hot bubble;][]{mccammon02}. The free parameters of the sky background model are fitted to the spectrum of a background region located far outside of the cluster ($>2R_{500}$) and then rescaled appropriately to the source region. The source is described by a single-temperature thin-plasma emission model using the APEC code \citep{apec} with the temperature, normalization and metal abundance left free while fitting. Finally, the source model is modified by the Galactic absorption along the line of sight, with the Galactic column density fixed to the HI column density estimated by the LAB survey \citep{kalberla}.

\subsection{Planck data analysis}

The pressure profiles are recovered from the \emph{Planck} survey data. We made use of a custom all-sky SZ map constructed in a very similar way as the public map of the full \emph{Planck} survey delivered by the \citet{planckymap2015}. Both maps were obtained from an internal linear combination of the \emph{Planck} frequency maps, using the modified internal linear combination algorithm (MILCA) method \citep[][]{hur13}. 
Our SZ map has a spatial resolution of 7~arcmin full width half maximum and by construction bears intrinsic correlated noise. 

The SZ signal is provided in unit-less Comptonization parameter, integrating the gas pressure along the line of sight, 
\begin{equation}
y= \frac{\sigma_{\rm T}}{m_{\rm e} c^2} \int_{\ell} P(\ell)\, d\ell.
\end{equation}

The $y$-profiles and 3D pressure profiles are obtained following the method presented in \citet{planck5}, and subsequently used in studies by the X-COP collaboration  \citep[e.g.,][]{tchernin16,ghirardini19, eckert19}. We recall here the main steps of the $y$ and pressure profile computation.
 
$y$-profiles are computed over local maps of size $20\times \theta_\textrm{500}$ on a side, centered on the X-COP cluster position and extracted from the all-sky SZ map.  The patches are over-sampled with respect to the \emph{Planck} SZ map pixel. The induced correlation between the points of the $y$-profile is accounted for and propagated through the covariance matrix of the profile.  We performed an azimuthal mean of the pixel fluxes falling in each radial bin, and subtracted a background residual offset estimated from radii larger than $5\times R_\textrm{500}$ safe from any cluster signal.
Obvious positive or negative (arising from the linear combination of frequencies) point sources were manually masked. We then  performed a pixel clipping of the local map using a $2.5\sigma$ criterion with respect to the mean of the flux beyond $5\times R_\textrm{500}$.
The noise of the \emph{Planck} SZ map can be considered as Gaussian and correlated. For each cluster we characterized the  statistical properties of the local noise over the individual patches in regions excluding the cluster emission, that is, $\theta>5\times R_\textrm{500}$. From the derived noise power spectrum, we draw a thousand realizations of the noise over each patch. Each realization is folded in the procedure to extract the $y$-profile and then used to compute the profile covariance matrix. The latter hence propagates the intrinsic properties of the noise in the SZ \emph{Planck} map as well as the correlation introduced by our choice of pixel size.
  
To recover the 3D pressure profiles, we optimally binned the $y$-profiles. We assumed spherical symmetry and applied a deconvolution from the Gaussian \emph{Planck} point spread function (PSF) and a geometrical deprojection following the method detailed by \citet{cro06}. Uncertainties on the $y$-profile are propagated through the covariance matrix by generating $10,000$ realizations of the noise profile. Each are co-added to the $y$-profile, individually deconvolved and deprojected, and used to compute the covariance matrix of the 3D pressure profile. 

\subsection{Stellar mass}

\subsubsection{BCG}

For the BCGs of A644, A1795, A2029, A2142, and A2319, we used stellar mass distributions from \citet{loubser20}. The stellar masses were derived from $r$-band imaging obtained on the Canada-France-Hawaii Telescope (CFHT; see \citealt{sand2012}). The observations and corrections (e.g.,\ extinction) are described in \citet{loubser20}, and we only briefly summarize the BCG stellar mass modeling here.  

We used the multi-Gaussian expansion (MGE) method, as implemented by \citet{cappellari2002}, to obtain the stellar mass distributions from the $r$-band imaging. The PSF-convolved MGE models extend beyond the effective radii ($R_{e}$) for the BCGs. In the case of A1795, A2142 and A2319, we masked some foreground or background features on the outer edges of the images. 

The surface brightness obtained from the MGE method was converted to surface brightness density, and the deprojection from surface density to intrinsic density was performed for the axisymmetric case using the Jeans anisotropic method (\citealt{cappellari2008}) assuming a constant stellar mass-to-light ratio, $\Upsilon_{* \rm DYN}$. We assumed an inclination of $i$ = 90$\degr$ (see \citealt{smith2017,Fasano2010,VanDerMarel1991}). In addition to the stellar mass component, the BCG mass models include a central mass component for a supermassive black hole, and a DM halo that follows a spherically symmetric NFW profile. The DM mass ($M_{\rm DM}$) within $R_{200}$ was approximated from weak lensing results \citep{herbonnet2020}, resulting in only two free parameters: (i) the stellar velocity anisotropy ($\beta_{z}$) and (ii) the stellar mass-to-light ratio ($\Upsilon_{* \rm DYN}$). The BCG mass model was compared to the observed kinematic profiles and the two parameters $\beta_{z}$ and $\Upsilon_{* \rm DYN}$ were adjusted until the predicted stellar kinematics best matched the spectroscopic observations (as described in \citealt{loubser20}), placing constraints on those parameters. We use this best fitting stellar mass-to-light ratio to derive the enclosed stellar mass of the BCG within spheres of different radii. 

For A644 and A2319, we did not have weak lensing masses available from \citet{herbonnet2020} but used the average DM distribution (of the 23 other clusters in \citealt{loubser20}) to calculate the effect of a DM mass component on the best fitting stellar mass-to-light ratio ($\Upsilon_{* \rm DYN}$). When a DM mass component is included in the dynamical mass models for the BCGs, the best fitting $\Upsilon_{* \rm DYN}$ decrease by 8.3 $\pm$ 2.9 per cent on average.

Our surface brightness measurement are accurate at the level of $< 0.25$ mag arcsec$^{-2}$, and our stellar kinematic measurement errors are $\pm4.7$ per cent on average (see \citealt{Loubser2018}). We also find that our dynamical modeling is robust against the DM distribution profile that we assume, as well as against assumptions regarding the black hole mass component. However, we find that the observational errors on $M_{\rm DM}$ and $R_{200}$ (obtained from the weak lensing results by \citealt{herbonnet2020}) are by far the dominant uncertainty. We derived the errors on the best fitting parameters by incorporating the 1$\sigma$ errors on the weak lensing masses as well as a $\pm$ 10 per cent uncertainty on the calculated value for the concentration parameter. The uncertainties on the stellar mass profiles are therefore constant with radius. 

\subsubsection{Satellite galaxies}
The stellar mass content contained in the satellite population of the same five clusters (A644, A1795, A2029, A2142, and A2319), is probed using $g$- and $r$-band imaging data obtained with MegaCam at the CFHT, and additional $u$- and $i$-band photometry taken with the Wide Field Camera at the Isaac Newton Telescope. Aperture magnitude limits (5-$\sigma$)are typically 24.3, 24.8, 24.2, and 23.3 in the $ugri$-filters, respectively, when measured on PSF-homogenized images using Gaussian weight functions. The stellar mass-to-light ratios that form the basis of the stellar mass measurements are obtained using spectral energy distribution fitting of these aperture fluxes. For this we assume the stellar population libraries from \citet{bruzual03}, an exponentially declining star-formation history, and a \citet{chabrier03} initial mass function. Further details are provided in \citet{vdb15}. 

While initially all galaxies are assumed to be members of the cluster, we clean the galaxy distribution from interlopers (i.e.,~foreground and background galaxies) by performing a statistical subtraction of sources observed in the COSMOS field \citep{muzzin13}. Details are provided in \citet{vdb15} and \citet{ghizzardi21}. In summary, to make a fair accounting, we redo the photometry in the COSMOS field, only utilizing the $ugri$-filters. We also accounted for field-to-field (often called “cosmic”) variance, given that the COSMOS field is relatively small. We estimate this uncertainty using \citet{moster11}, as detailed in \citet{vdb15}.

\section{Mass modeling}
\label{sec:modeling}

Here we present the scheme that we developed to recover HSE masses and the various methods used to reconstruct the mass profiles. The framework presented here implements several methods to estimate the underlying mass profile: parametric mass models (Sect. \ref{sec:massmod}), forward fitting with parametric functions (Sect. \ref{sec:forward}), and nonparametric log-normal mixture reconstruction (Sect. \ref{sec:GP}). We distribute our code jointly with this paper in the form of the publicly available Python package \texttt{hydromass} \footnote{\href{https://github.com/domeckert/hydromass}{https://github.com/domeckert/hydromass}}.

\subsection{Framework}

The gravitational field of galaxy clusters can be estimated from the measured thermodynamic profiles under the assumption that the gas is in HSE within the potential well of the underlying halo. We assume that the ICM is fully thermalized and that the gas pressure locally balances the gravitational force,

\begin{equation}
        \frac{dP_{\rm gas}}{dr} = -\rho_{\rm gas}\frac{GM_{\rm tot}(<r)}{r^2}
\label{eq:hse}
,\end{equation}

\noindent with $\rho_{\rm gas}, P_{\rm gas}$ the gas density and pressure, respectively, and $M_{\rm tot}(<r)$ the total (Newtonian) mass enclosed within a radius $r$. To take advantage of the sum of information provided by the combination of X-ray and SZ data, we performed joint fits to the two data sets by constructing 3D model profiles for the gas density, temperature, and pressure, which are related to one another through the ideal gas equation of state,

\begin{equation}
        P_{\rm gas}=\frac{k}{\mu m_p}\rho_{\rm gas}T
        \label{eq:idealgas}
.\end{equation}

Solving the HSE equation requires a good control over the gradient of the thermodynamic quantities, which can be highly sensitive to statistical fluctuations. We use three complementary approaches to model the radial temperature and pressure profiles and reconstruct $M_{HSE}$, which we describe in Sects. \ref{sec:massmod} through \ref{sec:GP}. We integrate the three methods within a common Bayesian fitting framework. The X-ray (surface brightness and spectroscopic temperature) and SZ (electron pressure) observables are fitted jointly to optimize the model parameters. We adopt a forward modeling approach in which the model 3D quantities are projected along the line of sight and compared to the projected data.

The gas density profile is modeled using the nonparametric multi-scale approach implemented in \citet{eckert20}. Namely, the 3D X-ray emissivity profile is described as the linear combination of a large number $N_K$ of King profiles,

\begin{equation}
\epsilon(r)=\sum_{k=1}^{N_K}\alpha_k\Phi_k(r),
\label{eq:multiscale}
\end{equation}

\noindent with $\{\Phi_k\}_{k=1}^{N_K}$ the dictionary of input King functions \citep{king62},

\begin{equation}
\Phi_k(r) = \left(1 + \frac{r}{r_{c,k}}^2 \right)^{-3\beta_k},
\end{equation}

\noindent and $\{\alpha_k\}_{k=1}^{N_K}$ their multiplicative coefficients. Following \citet{eckert20} the King function parameters are set adaptively on a grid of values for the parameters $r_c$ and $\beta$, with one value of $r_c$ for every set of 4 surface brightness points and 10 values of $\beta$ sampling the range 0.6-3. Our final dictionary contains $N_K = 5N/2$ functions, with $N$ the number of surface brightness points. The 3D model is projected onto the line of sight and convolved with the instrumental PSF to create a model for the observed surface brightness. The coefficients $\alpha_k$ are then fitted to the data. More details on the method and an extensive validation with numerical simulations are provided in \citet{eckert20}. 

The observed spectroscopic temperature is the line-of-sight average of the 3D temperature profile. The average temperature is weighted by the local emissivity $\epsilon\propto n_e^2$. On top of that, the temperatures obtained through single-temperature fits to multiphase spectra tend to be lower than mass-weighted temperatures because of the response of X-ray instruments.  \citet{mazzotta04} showed that for temperatures of 3 keV and above, a condition that is satisfied by all X-COP clusters, the average spectroscopic temperature can be written as

\begin{equation}
T_{spec}(r) = \frac{\int T_{3D}\,n_e^2T_{3D}^{-3/4}d\ell}{\int n_e^2 T_{3D}^{-3/4}d\ell}
\label{eq:mazzotta}
.\end{equation}

For a given model 3D temperature profile, the profile is projected onto the line of sight using Eq. \ref{eq:mazzotta}, convolved with the instrumental PSF and adjusted onto the measured temperatures. The model 3D pressure profile is fitted jointly to the SZ pressure profile, taking into account the correlated noise of the \emph{Planck} SZ signal \citep{planck5}. Therefore, our joint fitting procedure takes advantage of both the X-ray and SZ information by constructing a joint likelihood comparing the 3D model pressure to the SZ pressure and the spectroscopic-like temperature profile to the X-ray temperatures \citep[see Eq. 2 of ][]{ettori19}.

\subsection{Mass model}
\label{sec:massmod}

If a parametric form for the mass model can be defined, the HSE equation (Eq. \ref{eq:hse}) can be integrated to predict the expected pressure and temperature profiles,

\begin{equation}
        P_{3D}(r) = P_0 + \int_{r}^{r_0} \frac{\rho_{\rm gas}GM_{\rm mod }(r^{\prime})}{r^{\prime 2}}\,dr^{\prime}\label{eq:p0}
,\end{equation}

\noindent with $M_{\rm mod}(r)$ the model density profile integrated over the volume out to radius $r$, $r_0$ the outermost radius out to which the pressure can be measured, and $P_0=P(r_0)$ the integration constant at the outer boundary. This method is based on the ``backward'' approach introduced by \citet{ettori10}, with the notable difference that the integration constant $P_0$ is left free while fitting. We set a Gaussian prior on $P_0$ with mean and standard deviation set to the outermost pressure value and its 1-sigma uncertainty. 

In \citet{ettori19} (hereafter E19) we applied a similar technique to the X-COP sample and compared five different model profiles. We found that models with a central cusp are usually preferred by the data over cored models. For this reason, here we focus on the NFW model,
\begin{equation}
        \rho_{NFW}(r) = \frac{\rho_0}{\left(\frac{r}{r_s}\right)\left(1+\frac{r}{r_s}\right)^2}
        \label{eq:nfw}
,\end{equation}

\noindent and the Einasto model,
\begin{equation}
        \rho_{Einasto}(r) = \rho_{s}\exp\left[-\frac{2}{\alpha}\left(\left(\frac{r}{r_{s}}\right)^\alpha-1\right)\right].
        \label{eq:einasto}
\end{equation}

Instead of the scale radius $r_s$ and density normalization $\rho_s$ we optimize for the overdensity radius $R_{200}$ and the concentration $c_{200}=R_{200}/r_{s}$. We expand on the work of E19 in several ways. First, by correcting for the telescope's PSF (see Sect. \ref{sec:optim}), which allows us to accurately trace the DM profile in the inner regions. Second, in the Einasto case we leave the Einasto index $\alpha$ free, whereas previously its value was fixed to $1/\alpha\equiv\mu=5$ \citep{navarro04,mamon05}. We set weakly informative independent Gaussian priors on the model parameters of the NFW and Einasto models to make sure the fitting procedure quickly converges toward a reasonable solution whilst not biasing the resulting posteriors. In Table \ref{tab:priors} we give the details of the priors on the NFW parameters; a similar description of the priors on the Einasto model is provided in Paper \rom{2}. In case the available data are of lower quality, our framework also allows the user to set informative priors on the model parameters, for example by including prior information on the mass-concentration relation.

\begin{table}
\caption{\label{tab:priors}Normal priors on the NFW fit parameters. Here $P_{m}$ and $dP_{m}$ denote the outermost SZ pressure value and its error.}
\begin{tabular}{ccccc}
\hline
Parameter & Mean & $\sigma$ & Min & Max \\
\hline
\hline
$R_{200}$ [kpc] & 2000 & 1000 & 300 & 4000 \\
$c_{200}$ & 4 & 4 & 0 & 15\\
$P_0$ & $P_{m}$ & $dP_{m}$ & $P_{m}-2dP_{m}$ & $P_{m}+2dP_{m}$\\
\hline
\end{tabular}
\end{table}

Whenever information on the baryonic mass profile (stellar and gas) is available, we perform an additional reconstruction in which we attempt to model the gravitational field as the sum of baryonic and DM profiles,
\begin{equation}
        \rho_{\rm tot}(r) = \rho_{DM} + \rho_{\rm gas} + \rho_{\star, BCG} + \rho_{\star, sat}
.\end{equation}

In this case, the DM profile $\rho_{DM}$ is described by a mass model (NFW or Einasto) and the contribution of the baryonic components is provided as input.

\subsection{Parametric forward model}
\label{sec:forward}

In this case, we introduce a parametric form for the pressure profile and attempt to reconstruct the shape of the mass profile without prior assumption on its shape. This method is similar to the approach introduced by \citet{vikhlinin06}. We model the pressure profile as a generalized NFW (gNFW) profile \citep{nagai07},

\begin{equation}
        P_{forw}(r) = \frac{P_0}{(r/r_s)^{\gamma}\left(1+(r/r_s)^{\alpha}\right)^{\frac{\beta-\gamma}{\alpha}}}
        \label{eq:gnfw}
,\end{equation}

\noindent with $P_0$, $r_s$, $\alpha$, $\beta,$ and $\gamma$ as free parameters. This functional form was found to provide a good representation of gas pressure profiles determined from both X-ray and SZ observations \citep{arnaud10,planck5,sayers13,bourdin17,ghirardini19,pointecouteau21}. The pressure gradient can be computed analytically from the reconstructed functional form to recover the mass profile from Eq. \ref{eq:hse}.

\subsection{Nonparametric log-normal mixture reconstruction}
\label{sec:GP}

Here we introduce a nonparametric method for temperature deprojection and HSE mass reconstruction. Our method attempts at the same time to make minimal assumptions on the shape of the mass profile and limit the fluctuations introduced by the deprojection process and the computation of the pressure gradient. To this aim, we describe the 3D temperature profile as a linear combination of a large number $N_{g}$ of log-normal functions,

\begin{equation}
        T_{NP}(r)=\sum_{i=1}^{N_g} G_i \frac{1}{\sqrt{2\pi\sigma_i^2}}\exp\left(-\frac{(\ln(r)-\ln(\mu_i))^2}{2\sigma_i^2}\right)\label{eq:gp}
.\end{equation}

The mean values $\{\mu_i\}_{i=1}^{N_g}$ and standard deviations $\{\sigma_i\}_{i=1}^{N_g}$ of the Gaussians are set a priori to allow as much freedom as possible to the fitted function, whilst at the same time preventing small, unphysical fluctuations induced by the statistical nature of the problem. For large values of $N_g$ the model is essentially independent of the choice of $N_g$ and $\mu_i$, whereas the standard deviations $\sigma_i$ act as effective smoothing scales. We implement the model with $N_g=200$ and the values of $\mu_i$ logarithmically spaced from the innermost to the outermost data points. We set the values of $\sigma_i$ to the width of the nearest spectroscopic annulus or SZ radial bin, such that fluctuations on scales smaller than the bin size are suppressed and the model follows closely the data on larger scales. 

Once the number of log-normal functions $N_g$, the mean values $\mu_i$ and the standard deviations $\sigma_i$ are adaptively set, the model can be projected onto the line of sight including spectroscopic-like weights (Eq. \ref{eq:mazzotta}) and the normalizations $\{G_i\}_{i=1}^{N_g}$ can be fit to the spectroscopic X-ray temperatures and SZ pressure profile. The mass profile is then reconstructed by combining the 3D temperature and density profiles,

\begin{equation}
        M(<r) =  -\frac{rkT_{3D}(r)}{G\mu m_p}\left(\frac{\partial \log T_{3D}}{\partial \log r} + \frac{\partial\log n_{\rm gas}}{\partial\log r}\right).
\end{equation}

\noindent The temperature gradient can be computed analytically from Eq. \ref{eq:gp},

\begin{equation}
        \frac{\partial T_{3D}}{\partial r} = \sum_{i=1}^{N_g} G_i \frac{1}{\sqrt{2\pi}\sigma_i^3} \exp\left(-\frac{(\ln(r)-\ln(\mu_i))^2}{2\sigma_i^2}\right) \frac{\ln(\mu_i)-\ln(r)}{r}.
\end{equation}

\noindent Similarly, since the King functions $\{\Phi_k\}_{k=1}^{N_K}$ are analytical functions, the gas density gradient can be computed analytically from Eq. \ref{eq:multiscale},

\begin{equation}
\frac{\partial n_{\rm gas}}{\partial r} = \sum_{k=1}^{N_K}\alpha_k \frac{\partial \Phi_k}{\partial r}.
\end{equation}

\subsection{Optimization}
\label{sec:optim}

Since our models can contain several hundred parameters, we require the use of a statistical sampler that is suitable for high-dimensional optimization problems. We use the No U-Turn Sampler \citep[NUTS;][]{nuts}, a Hamiltonian Monte Carlo (HMC) algorithm implemented in the \texttt{PyMC3} package \citep{pymc3}. The convergence of HMC algorithms is typically much faster than that of traditional Markov chain Monte Carlo samplers thanks to the use of gradient information. At the same time, NUTS includes advanced self-tuning features, such that the sampler requires minimal input from the user. 

For each cluster and model, we start by searching for the maximum likelihood using a gradient descent method, and then run four parallel HMC chains with 2,000 tuning steps and 2,000 samples each. The four chains are then combined to measure the posterior distributions of the parameters. Running multiple chains in parallel improves the computational efficiency of the code and reduces the risk that the posterior be drawn from a single chain that has not fully converged.

\begin{figure*}
        \centerline{\resizebox{\hsize}{!}{\vbox{
                                \includegraphics[width=0.45\textwidth]{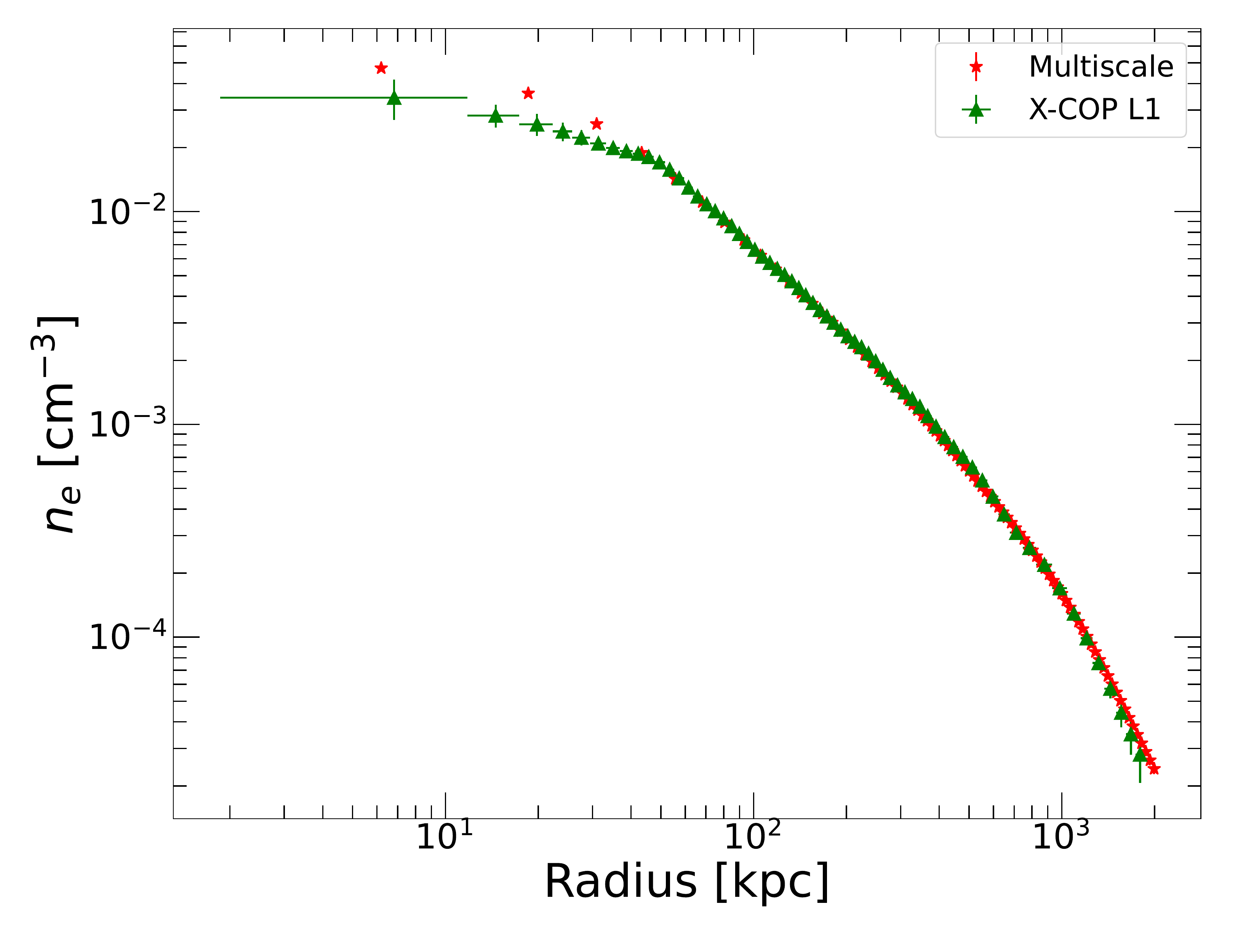}
                                \includegraphics[width=0.45\textwidth]{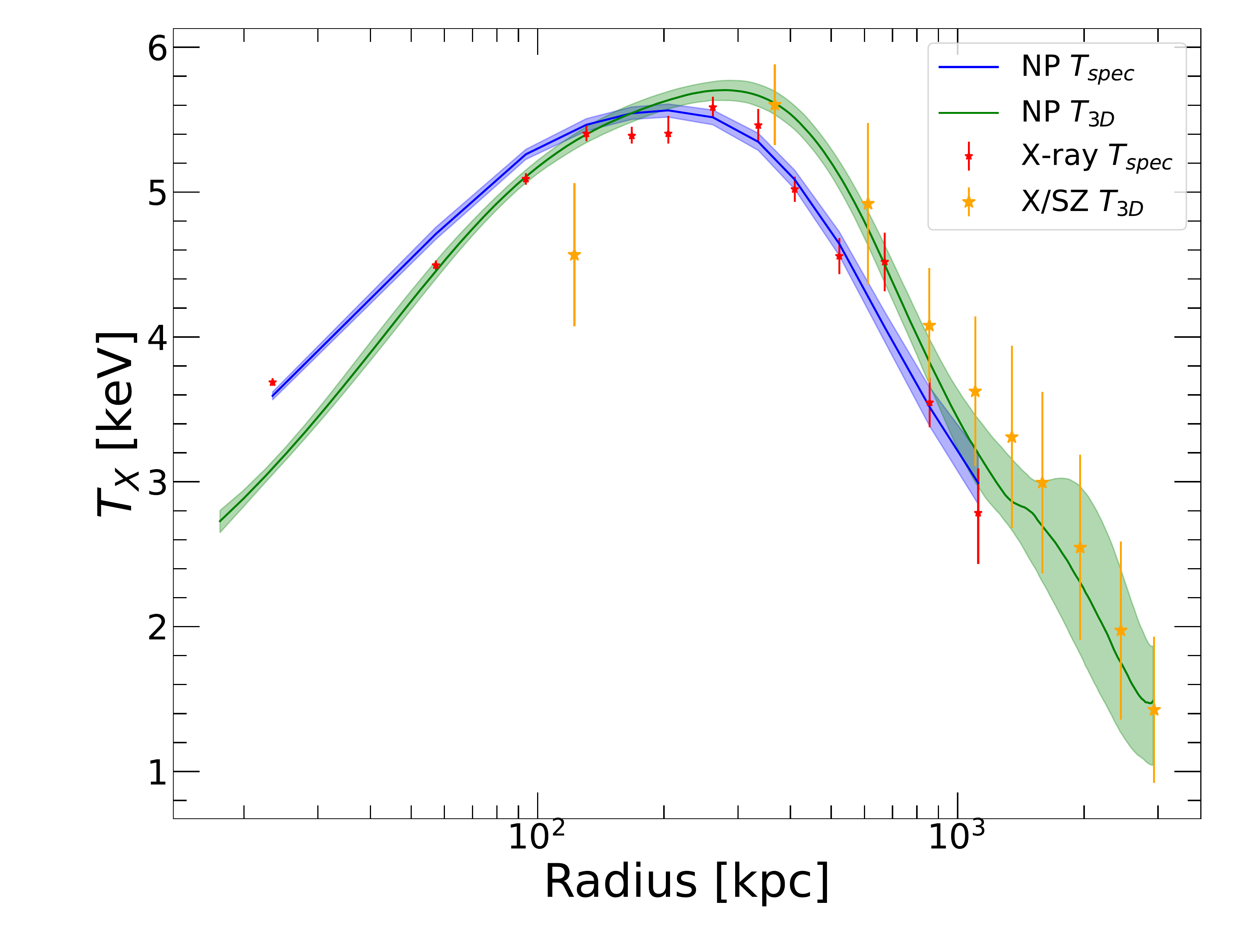}\\
                                
                                \includegraphics[width=0.45\textwidth]{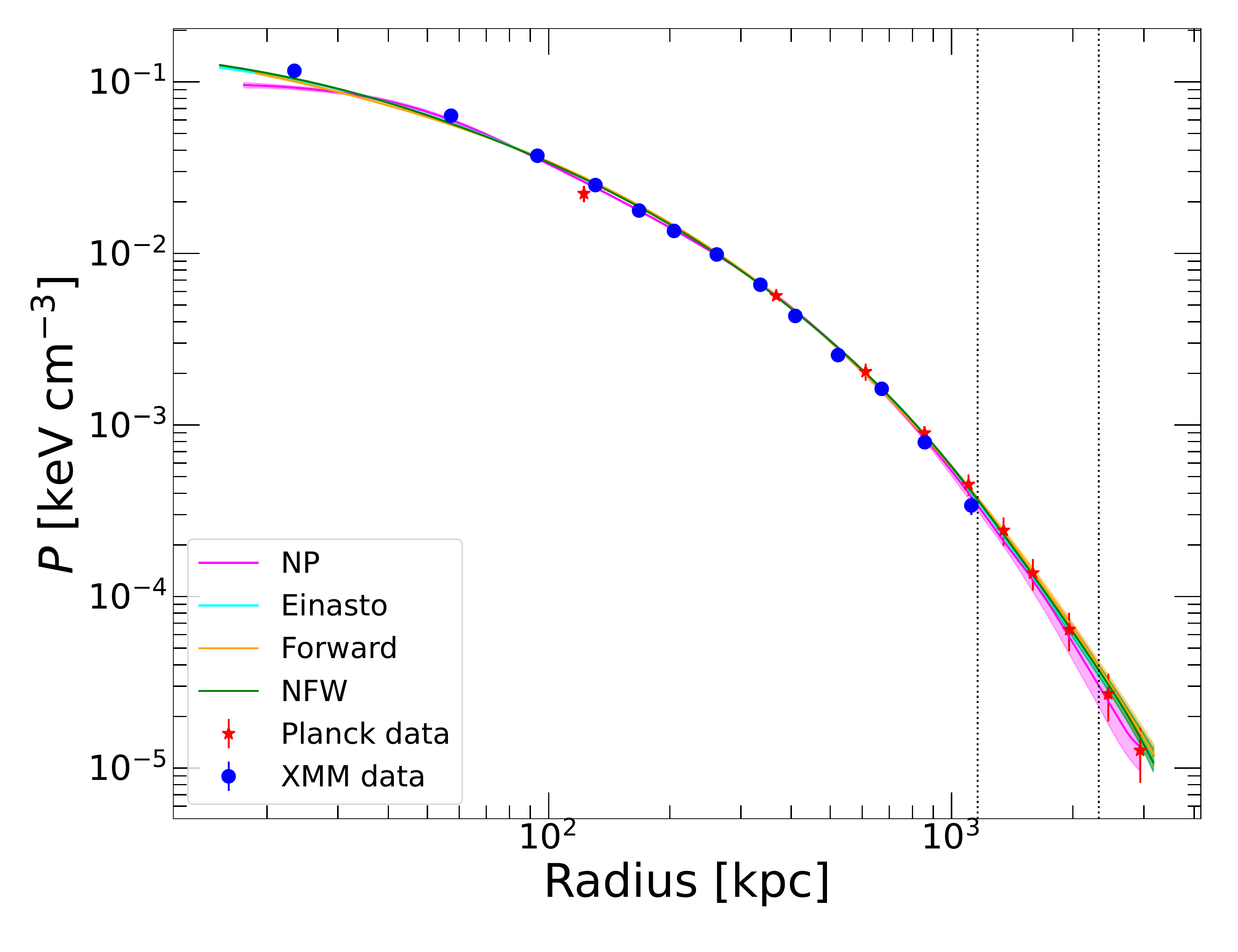}
                                \includegraphics[width=0.45\textwidth]{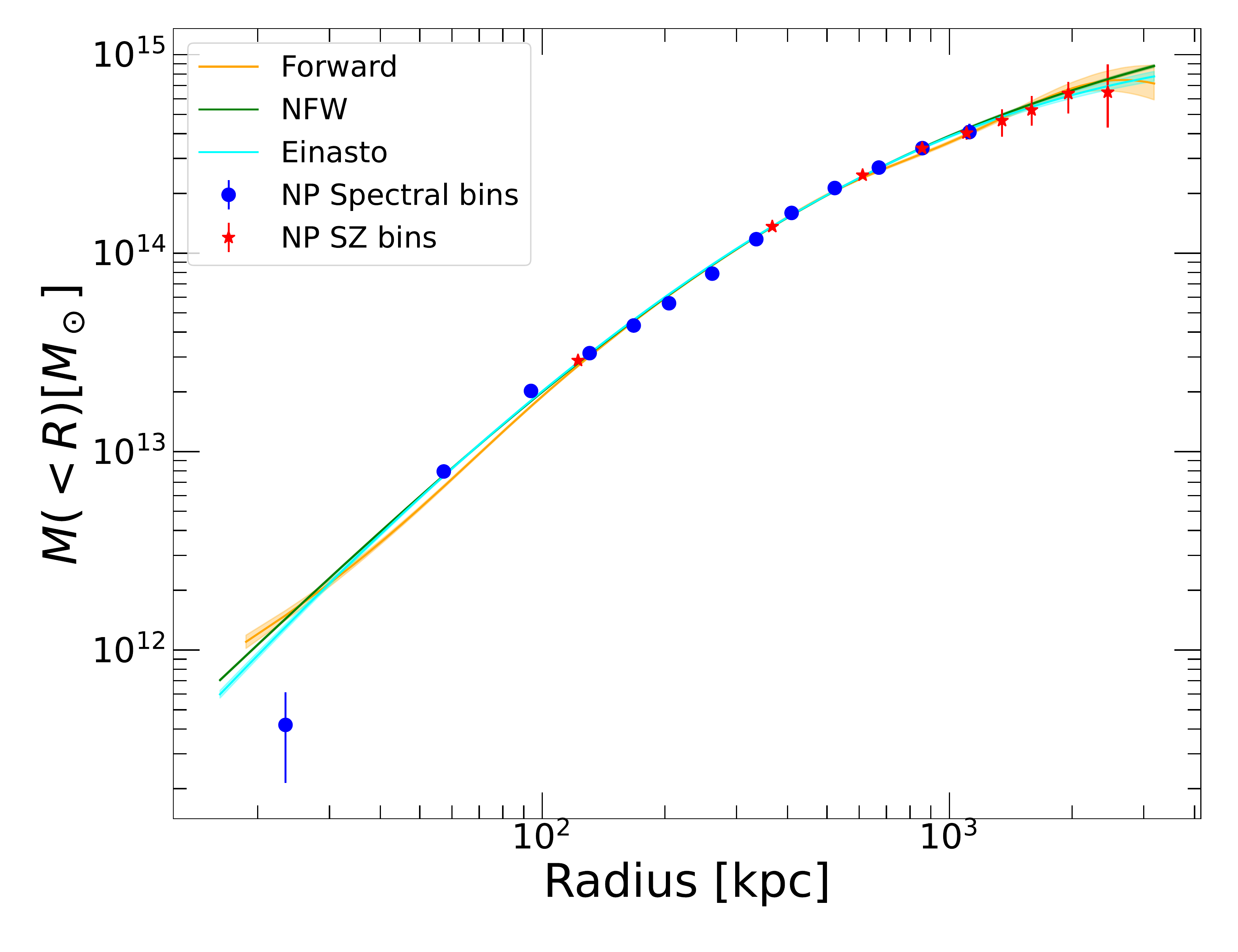}\\
                                
                                \includegraphics[width=0.45\textwidth]{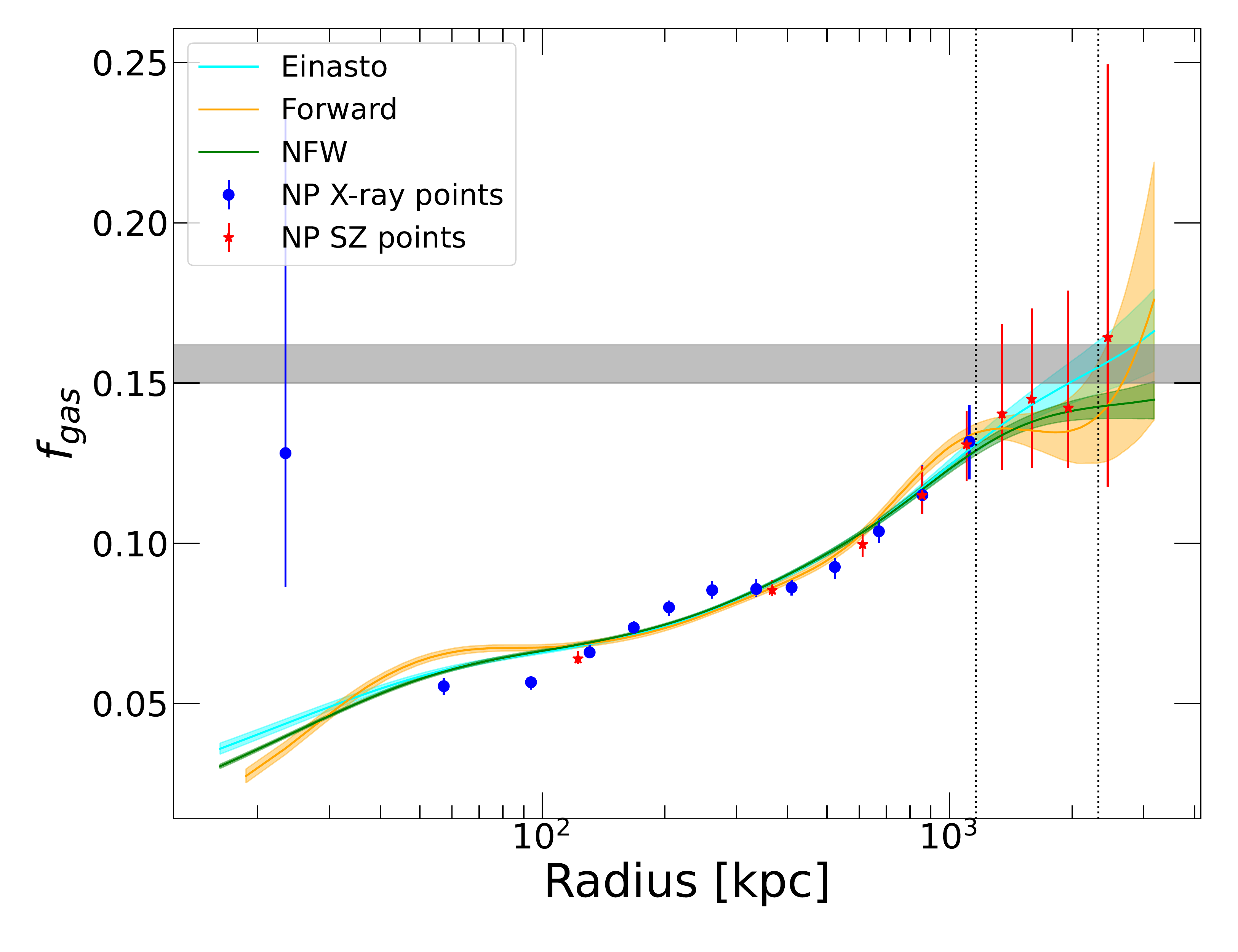}
                                \hspace{0.5cm}
                                \includegraphics[width=0.37\textwidth]{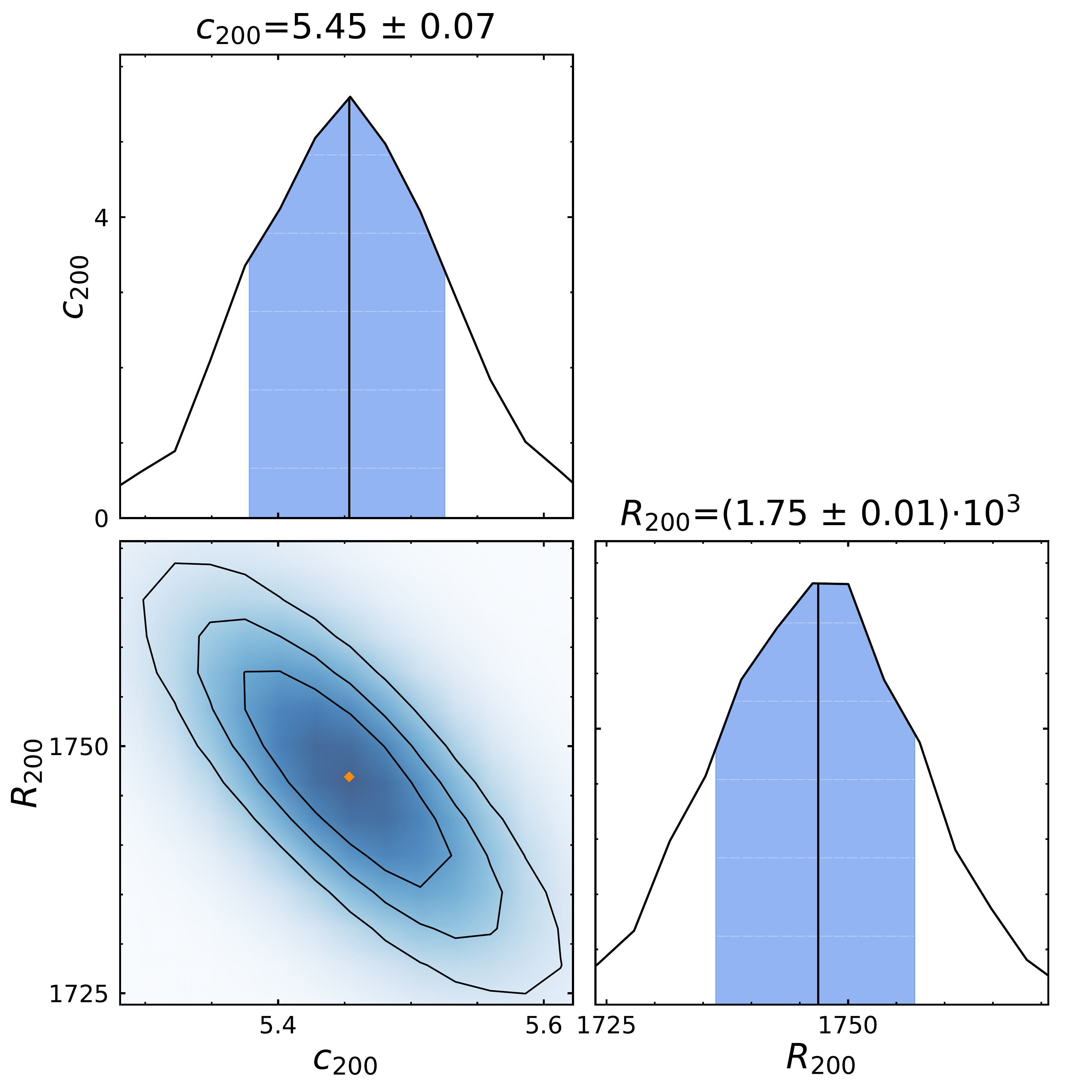}
        }}}
        \caption{\label{fig:a1795} Mass reconstruction results for A1795. \emph{Top left:} Electron density profile reconstructed with the multi-scale method used here \citep{eckert20} compared to the L1 regularization results presented in \citet{ghirardini19} and publicly released by the X-COP team. \emph{Top right:} Nonparametric reconstruction of the 3D temperature profile (green) compared to the spectroscopic X-ray measurements (red) and the 3D temperature profile obtained by dividing the SZ pressure by the X-ray density (orange). See Sect. \ref{sec:GP} for details. The blue curve shows the projected, spectroscopically weighted, and PSF convolved temperature profile, to be compared to the red data points. \emph{Middle left:} Electron pressure profile measured by \emph{Planck} and \emph{XMM-Newton} compared to the reconstructions obtained from the NFW and Einasto mass models (Sect. \ref{sec:massmod}), the parametric forward approach (labeled ``Forward''; see Sect. \ref{sec:forward}), and the NP reconstruction (Sect. \ref{sec:GP}). The dotted vertical lines show the location of $R_{500}$ and $R_{200}$ from the NFW fit. \emph{Middle right:} Output mass profiles obtained from the four individual reconstructions (NFW, Einasto, Forward, and NP). The data points show the NP results evaluated at the radius of the X-ray spectroscopic measurements (labeled ``NP Spectral bins," blue) and the SZ pressure data (``NP SZ bins," red). \emph{Bottom left:} Same as previous but for the gas mass fraction. The gray shaded area shows the cosmic baryon fraction $\Omega_b/\Omega_m$ in \citet{planck15_13} cosmology. \emph{Bottom right:} Posterior distributions and correlations for the parameters of the NFW profile.} 
\end{figure*}

In Fig. \ref{fig:a1795} we show an example of reconstruction for A1795 ($z=0.0622$). The results obtained from the NFW and Einasto mass models are compared to the results of the parametric forward approach (hereafter labeled as ``Forward'') and the nonparametric log-normal mixture method (hereafter NP). All the methods provide similar results across most of the radial range, although differences arise primarily in the innermost 100 kpc. Since the NP method is designed to trace the observed values as closely as possible, it cannot be extrapolated outside of the fitted range. Thus, for the remainder of the paper we evaluate it at the radii of the observed data and present them in the form of data points rather than continuous models. In the bottom-right panel we show the posterior distributions for the parameters of the NFW profile and the correlations between the parameters. Given the excellent statistical quality of the X-COP data, the posterior distributions are much narrower than the prior and clearly pull away from it, indicating that our data are able to break the degeneracy between NFW mass and concentration. To help convergence of the Einasto model, we optimize for the inverse of the Einasto index, $\mu=1/\alpha$ (see Paper \rom{2}). The details of the Einasto fitting procedure, the adopted priors, and the resulting best-fit parameters are provided in Paper \rom{2}.

\section{Validation using mock data}
\label{sec:mock}

To validate our numerical framework, we created synthetic observations of an idealized cluster in HSE and ran our entire pipeline on the mock data. In the following, we describe the various steps of this validation process and present the accuracy of our reconstruction technique.

\begin{figure*}
        \centerline{\resizebox{\hsize}{!}{\includegraphics{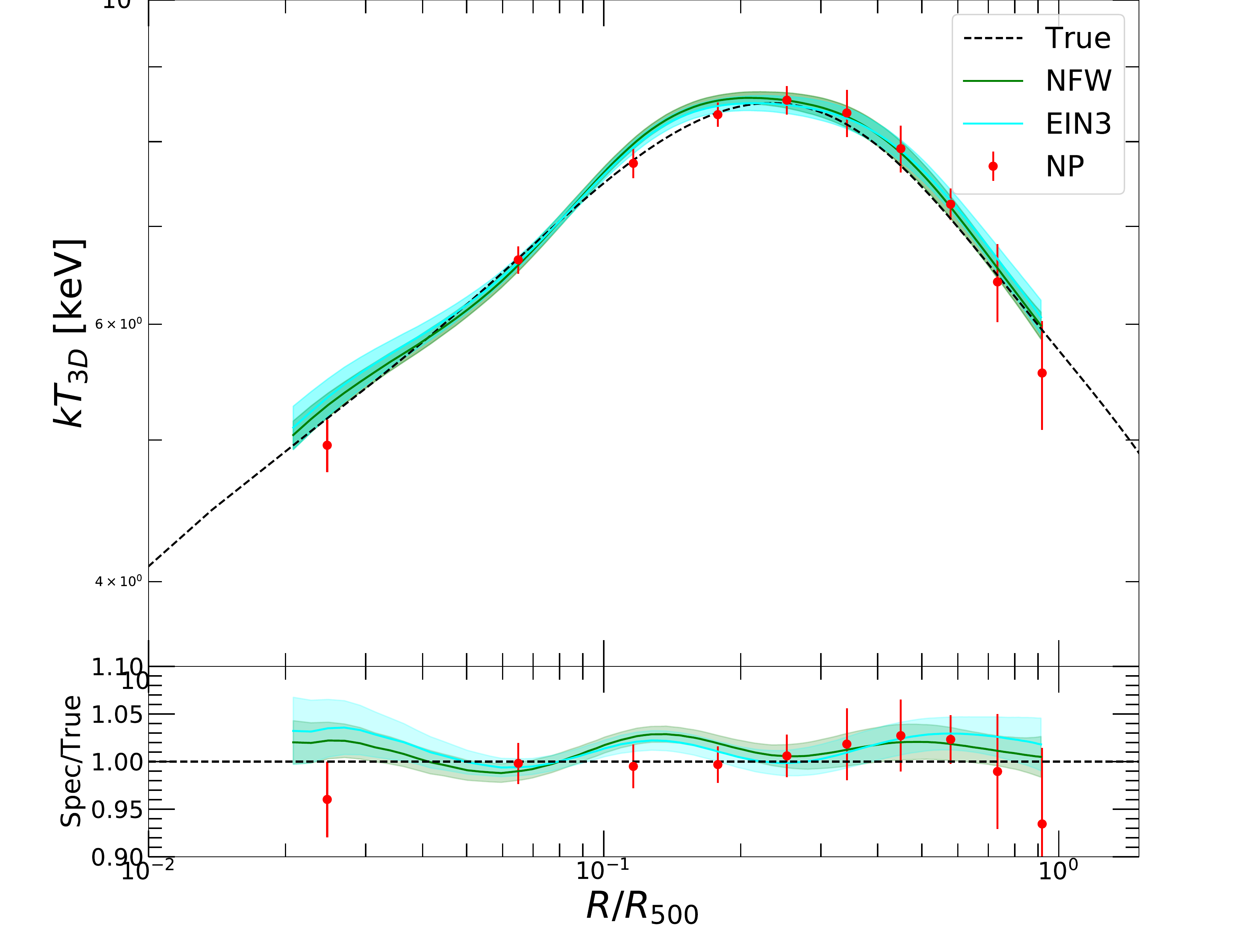}\includegraphics{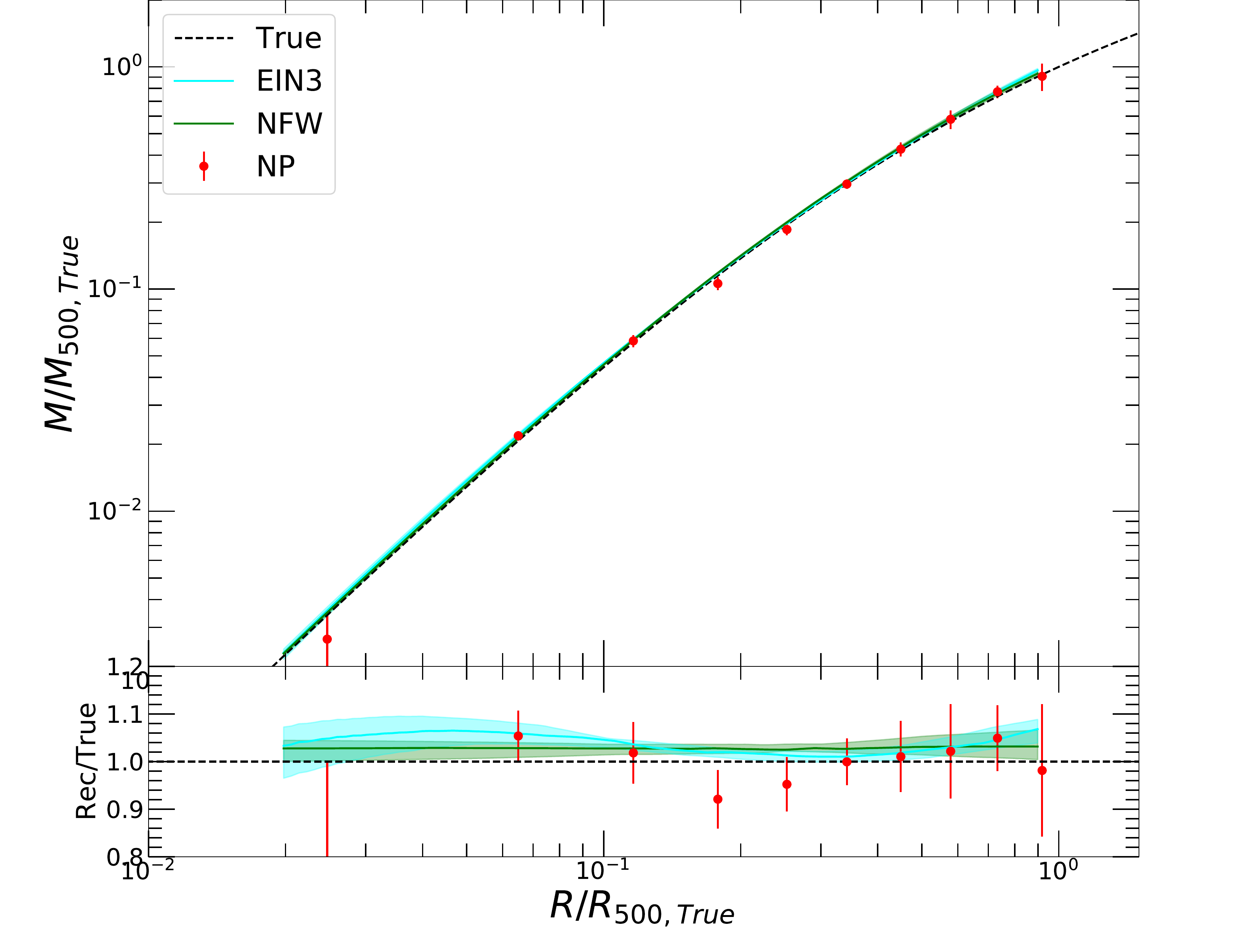}}}
        \caption{\label{fig:mock}Result of temperature deprojection (left) and HSE mass reconstruction (right) reconstructed from a set of ten mock \emph{XMM-Newton} observations of a synthetic NFW cluster (see Sect. \ref{sec:mock}). The curves show the mean and standard deviation of the ten realizations for the various reconstruction techniques (NFW, green; Einasto, cyan; and NP, red). The bottom panels show the ratio of reconstructed profiles to true input profiles. }
\end{figure*}

\subsection{Synthetic galaxy cluster simulations}

We consider a fiducial DM halo with a mass profile described by a generic NFW profile (Eq. \ref{eq:nfw}) with a concentration $c_{NFW}=4$ and an overdensity radius $R_{200}=2000$ kpc at a redshift $z=0.2$, such that the simulated cluster fits well within the \emph{XMM-Newton} FOV. The halo is filled with an ICM in HSE within the input NFW potential well. The input gas density profile follows the ``universal'' average X-COP gas density profile \citep[see Fig. 3 and Table 3 of][]{ghirardini19}, which was fitted to a collection of individual nonparametric gas density profiles using a single common functional form adapted from \citet{vikhlinin06} and jointly accounting for the intrinsic scatter among the cluster population. 

Given the input NFW mass profile and the universal gas density profile, we used Eq. \ref{eq:p0} to compute the corresponding thermal pressure profile. The profiles were computed out to $5\times R_{500}$, which is well beyond the range covered by our data to avoid uncertainties linked with the integration constant $P_0$. We then generated a 3D box inside which we set the gas density and temperature of the ICM inside each cell following spherical symmetry. The cell size was set to 2 kpc, which is substantially smaller than the \emph{XMM-Newton} resolution. Inside each box cell, we used the APEC model \citep{apec} as implemented in 3ML \citep{3ml,3ml2}\footnote{\href{https://threeml.readthedocs.io/en/latest/notebooks/APEC\_doc.html}{https://threeml.readthedocs.io/en/latest/notebooks/APEC\_doc.html}} to generate model spectra in each cell. A constant metal abundance of $0.3Z_\odot$ relative to the \citet{ag89} Solar abundance table was used in all cells. The spectra were absorbed by Galactic absorption with $N_H=5\times10^{20}$ cm$^{-2}$. Finally, the 3D model spectra were projected onto the line of sight. As a result, a spectrum made of the linear combination of individual APEC spectra is associated with each individual pixel of a projected 2D map.

\subsection{Mock data creation}

From the projected model spectra and images, we generated mock \emph{XMM-Newton} observations of our synthetic cluster including a wide variety of instrumental effects: effective area, energy redistribution, vignetting, PSF smearing, and a realistic background model. Our mock data generation tool is in the form of a Python package, which is made publicly available together with this paper\footnote{\href{https://github.com/domeckert/xmm\_simulator}{https://github.com/domeckert/xmm\_simulator}}. The tool reads the on-axis effective area of each of the three \emph{XMM-Newton} instruments from the calibration database, and combines the on-axis area with the telescope's energy-dependent vignetting curve to generate a local effective area curve at each position in the FOV. The projected model spectra described above are multiplied by the local effective area curve and convolved with the instrument's redistribution matrix. The spectra are then convolved with the instrumental PSF. Finally, we generate a mock source event list from Poisson realizations of the model spectra.

On top of the source spectrum, we add a realistic background spectrum made of a vignetted component for the sky background and an un-vignetted component for the particle-induced background. The un-vignetted component is drawn out of the collection of filter-wheel-closed observations available in the calibration database. The vignetted component is made of a three-component model (see Sect. \ref{sec:xmm}) consisting of an absorbed power law for the cosmic X-ray background and two APEC models describing the foregrounds. The normalization of these components is set to be representative of a typical extragalactic field. A Poisson realization of the sky background is then generated and merged with the source event list to create a total event file including our source, the astrophysical background and the cosmic-ray induced background. Images and spectra can then be extracted from the final event list and analyzed in the same way as real observations.

\subsection{Code performance}

We performed a set of ten individual 50 ks mock observations of our synthetic cluster using the procedure described above and analyzed the simulated data using the exact same analysis pipeline as for the actual data. Namely, for each realization, an image of the synthetic cluster in the [0.7-1.2] keV band is extracted from the simulated events. From the resulting image, we extract a background subtracted surface brightness profile in the same way as for real observations. We extract mock X-ray spectra in ten logarithmically spaced annuli and a background region located at $R>2R_{500}$ from the cluster center, and we fit the spectra using a single-temperature APEC model and the sum of vignetted and un-vignetted background components. We then run our mass reconstruction code on the surface brightness and spectroscopic temperature profiles. In each case, we fit the data using the NFW and Einasto mass models as well as the nonparametric log-normal mixture model. 

The performance of our code on the mock data is shown in Fig. \ref{fig:mock}. In the left-hand panel we show the mean and dispersion of the ten deprojected 3D temperature profiles compared with the input profile. The nonparametric profile was evaluated at the average radii of the spectroscopic bins. We can see that all three methods are able to trace the true profile with minimal bias (i.e., less than 3\% at each radius). In particular, the nonparametric method provides a good description of the profile with a scatter of $\sim5\%$, even though it makes no assumption on the shape of the mass profile. The NFW model traces the true profile very closely, which is expected since the true input mass profile was assumed to be an NFW. The Einasto profile also provides an excellent description of the data, although it overestimates the temperature by 3\% on average. 

The comparison between the mean of the fitted mass profiles and the true profile can be seen in the right-hand panel of Fig. \ref{fig:mock}, with the residuals shown in the bottom panel. Again, as expected the NFW reconstruction matches very well the input profile, with a bias of at most 3\% at all radii but consistent within the uncertainties. The Einasto and nonparametric profiles closely trace the true profile, with the Einasto exceeding the true mass by about 6\% on average, which can be explained by the application of a different parametric model compared to the true one. The nonparametric technique follows the true profile with minimal bias (<5\%) but somewhat higher uncertainties given the freedom allowed to the model. The uncertainties are higher at the inner and outer edges of the profile, since the temperature gradient cannot be directly constrained there.

Overall, the tests presented here show that our analysis pipeline and HSE reconstruction framework are able to reconstruct the properties of the gravitational field in our synthetic cluster with a bias of at most 3\% when adopting the NFW method. The uncertainties associated with the reconstruction technique are therefore subdominant with respect to other sources of systematic uncertainty such as hydrostatic bias \citep[10-20\%; e.g.,][]{rasia06,lau09}, temperature inhomogeneities \citep[$\sim10\%$,][]{rasia14} and effective area calibration \citep[$15\%$,][]{schellenberger15}.

\section{Results}

\begin{table*}
        \caption{\label{tab:master} Sample properties and results of NFW fits. The columns labeled ``DM'' refer to the NFW fits to the DM only for the five systems with complete information on the baryonic components (see Sect. \ref{sec:fbar}), whereas the ``TOT'' subscripts refer to the fits of the total gravitational field with an NFW model (see Fig. \ref{fig:mod_vs_gp}). In this case, the quoted total mass is computed from the sum of DM and baryonic density profiles.}
        \centering
\begin{tabular}{lcccccccc}
\hline
Cluster & z & $N_{H}$ & $R_{500, TOT}$ & $M_{500, TOT}$ & $M_{200, TOT}$ & $c_{200, TOT}$ & $M_{200, DM}$ & $c_{200, DM}$ \\ 
  &   & $10^{20}$ cm$^{-2}$ & kpc  & $10^{14}M_{\odot}$ & $10^{14}M_{\odot}$ &  & $10^{14}M_{\odot}$ &  \\ 
\hline
\hline
A85 & 0.0555 & 2.8 & $1214_{-4}^{+4}$ & $5.03_{-0.05}^{+0.05}$ & $7.04_{-0.07}^{+0.08}$ & $4.708_{-0.044}^{+0.042}$ & - & - \\ 
A644 & 0.0704 & 7.5 & $1398_{-16}^{+16}$ & $7.80_{-0.27}^{+0.26}$ & $10.97_{-0.43}^{+0.45}$ & $4.597_{-0.156}^{+0.163}$ & $10.98_{-0.41}^{+0.45}$ & $4.458_{-0.169}^{+0.171}$ \\ 
A1644 & 0.0473 & 4.1 & $1031_{-10}^{+11}$ & $3.06_{-0.09}^{+0.09}$ & $4.93_{-0.20}^{+0.21}$ & $2.548_{-0.135}^{+0.140}$ & - & - \\ 
A1795 & 0.0622 & 1.2 & $1160_{-6}^{+6}$ & $4.42_{-0.07}^{+0.07}$ & $6.04_{-0.11}^{+0.11}$ & $5.454_{-0.076}^{+0.072}$ & $6.82_{-0.13}^{+0.13}$ & $4.510_{-0.072}^{+0.074}$ \\ 
A2029 & 0.0766 & 3.2 & $1340_{-9}^{+9}$ & $6.91_{-0.14}^{+0.14}$ & $9.29_{-0.21}^{+0.22}$ & $5.929_{-0.101}^{+0.102}$ & $11.67_{-0.28}^{+0.29}$ & $4.185_{-0.091}^{+0.093}$ \\ 
A2142 & 0.09 & 3.8 & $1453_{-9}^{+9}$ & $8.93_{-0.16}^{+0.17}$ & $13.33_{-0.30}^{+0.32}$ & $3.440_{-0.070}^{+0.076}$ & $13.94_{-0.31}^{+0.30}$ & $3.046_{-0.074}^{+0.076}$ \\ 
A2255 & 0.0809 & 2.5 & $1202_{-16}^{+17}$ & $5.01_{-0.20}^{+0.22}$ & $8.94_{-0.56}^{+0.66}$ & $1.800_{-0.180}^{+0.193}$ & - & - \\ 
A2319 & 0.0557 & 3.2 & $1424_{-4}^{+5}$ & $8.12_{-0.08}^{+0.08}$ & $11.61_{-0.14}^{+0.14}$ & $4.233_{-0.070}^{+0.068}$ & $12.16_{-0.14}^{+0.15}$ & $4.098_{-0.079}^{+0.079}$ \\ 
A3158 & 0.059 & 1.4 & $1146_{-11}^{+12}$ & $4.25_{-0.13}^{+0.14}$ & $6.46_{-0.25}^{+0.28}$ & $3.178_{-0.146}^{+0.145}$ & - & - \\ 
A3266 & 0.0589 & 1.6 & $1381_{-10}^{+10}$ & $7.43_{-0.15}^{+0.16}$ & $12.22_{-0.36}^{+0.41}$ & $2.352_{-0.098}^{+0.097}$ & - & - \\ 
RXC1825 & 0.065 & 9.4 & $1109_{-8}^{+9}$ & $3.88_{-0.08}^{+0.09}$ & $5.66_{-0.17}^{+0.18}$ & $3.807_{-0.159}^{+0.165}$ & - & - \\ 
Zw1215 & 0.0766 & 1.7 & $1346_{-21}^{+21}$ & $7.00_{-0.32}^{+0.35}$ & $11.44_{-0.70}^{+0.76}$ & $2.413_{-0.148}^{+0.155}$ & - & - \\ 
\hline
\end{tabular}\end{table*}

\subsection{Total density profiles}

We used our \texttt{hydromass} Python package introduced in Sect. \ref{sec:modeling} to analyze the gravitational field of the 12 X-COP clusters. In each case, we ran the reconstruction with the two mass models (NFW and Einasto), the parametric forward method, and the nonparametric log-normal mixture technique, similar to the example shown in Fig. \ref{fig:a1795} for A1795. In the case of the NP reconstruction, the flexibility of the model makes it highly uncertain away from the fitted data points. Therefore, we restrict specifically to the radii corresponding to the existing spectral and SZ data points to avoid extrapolating. The masses estimated from the NFW fits as well as the basic properties of the sample are shown in Table \ref{tab:master}. The results obtained with the Einasto model are presented and discussed in Paper \rom{2}.

We started by comparing the output of our new framework to the masses previously published in E19 in the case of the ``standard'' NFW reconstruction. As explained in Sect. \ref{sec:massmod}, the main difference with respect to our previous analysis is the implementation of a modeling scheme for the PSF, both for the surface brightness and the temperature profile. Correcting for PSF smearing increases the concentration parameter of the NFW profile, which results in slightly higher masses in the core and lower masses in the outskirts. The average ratio between the E19 masses and the results presented here is given in Table \ref{tab:mass_ratio} for several overdensities. Our new masses are slightly higher inside $R_{2500}$ and slightly lower inside $R_{200}$; the difference, however, is always within 10\%. The validation on mock data presented in Sect. \ref{sec:mock} shows that our framework is able to recover the NFW shape very accurately, and thus we conclude that the NFW concentrations given in E19 were slightly underestimated.

\begin{table}
\caption{\label{tab:mass_ratio}Ratio between the masses published in E19 ($M_{\rm E19}$) and our new results (labeled $M_{\rm hydromass}$) for several overdensity factors. For consistency, the masses estimated here were converted into the same cosmology as that used in E19.}
\begin{tabular}{lcccc}
\hline
Overdensity & 2500 & 1000 & 500 & 200\\
\hline
\hline
mean($M_{\rm hydromass}/M_{\rm E19}$) & 1.08 & 1.01 & 0.96 & 0.93\\
median($M_{\rm hydromass}/M_{\rm E19}$) & 1.05 & 1.02 & 0.95 & 0.90\\
\hline
\end{tabular}
\end{table}

\begin{figure*}
        \centerline{\resizebox{\hsize}{!}{\includegraphics[width=0.5\textwidth]{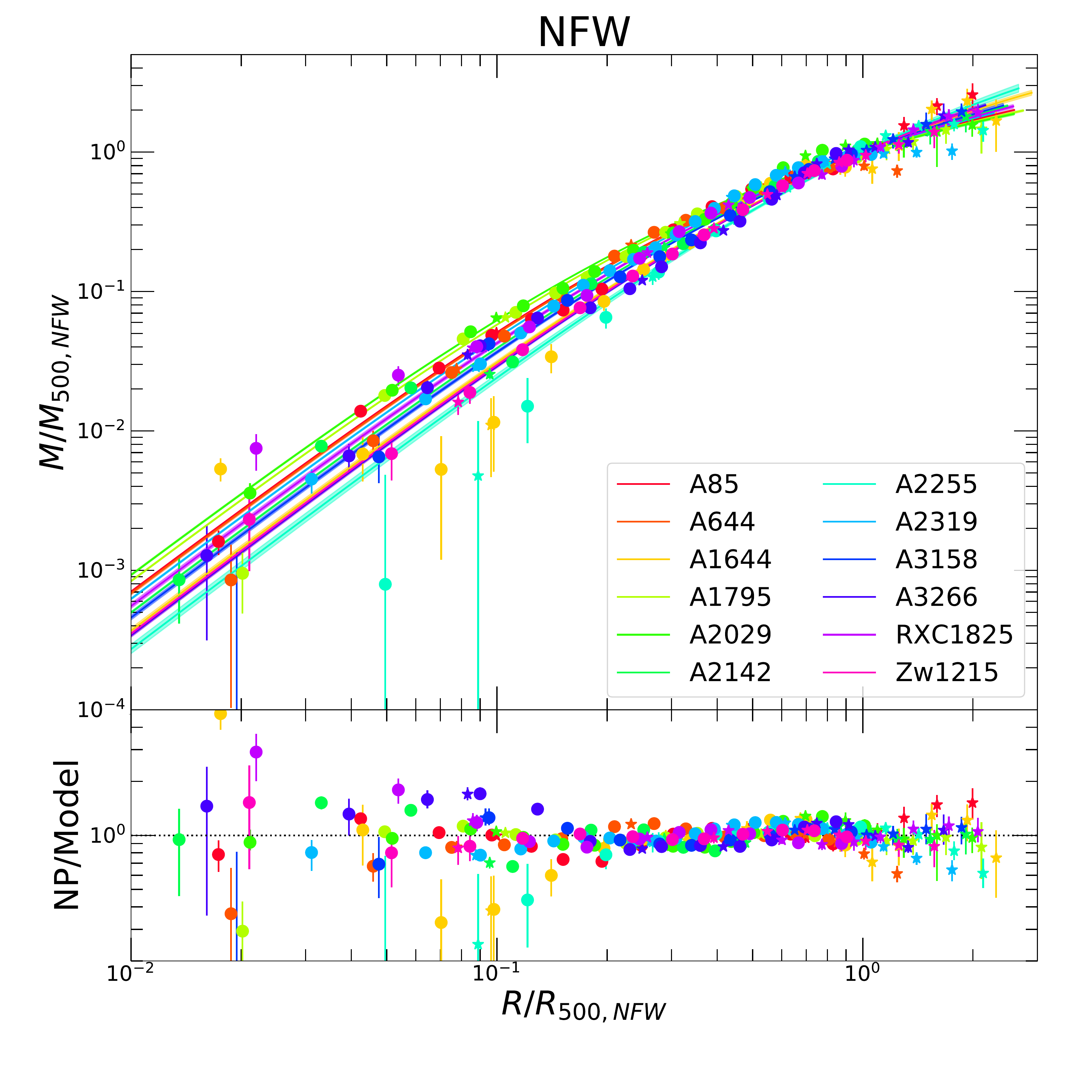}
                        \includegraphics[width=0.5\textwidth]{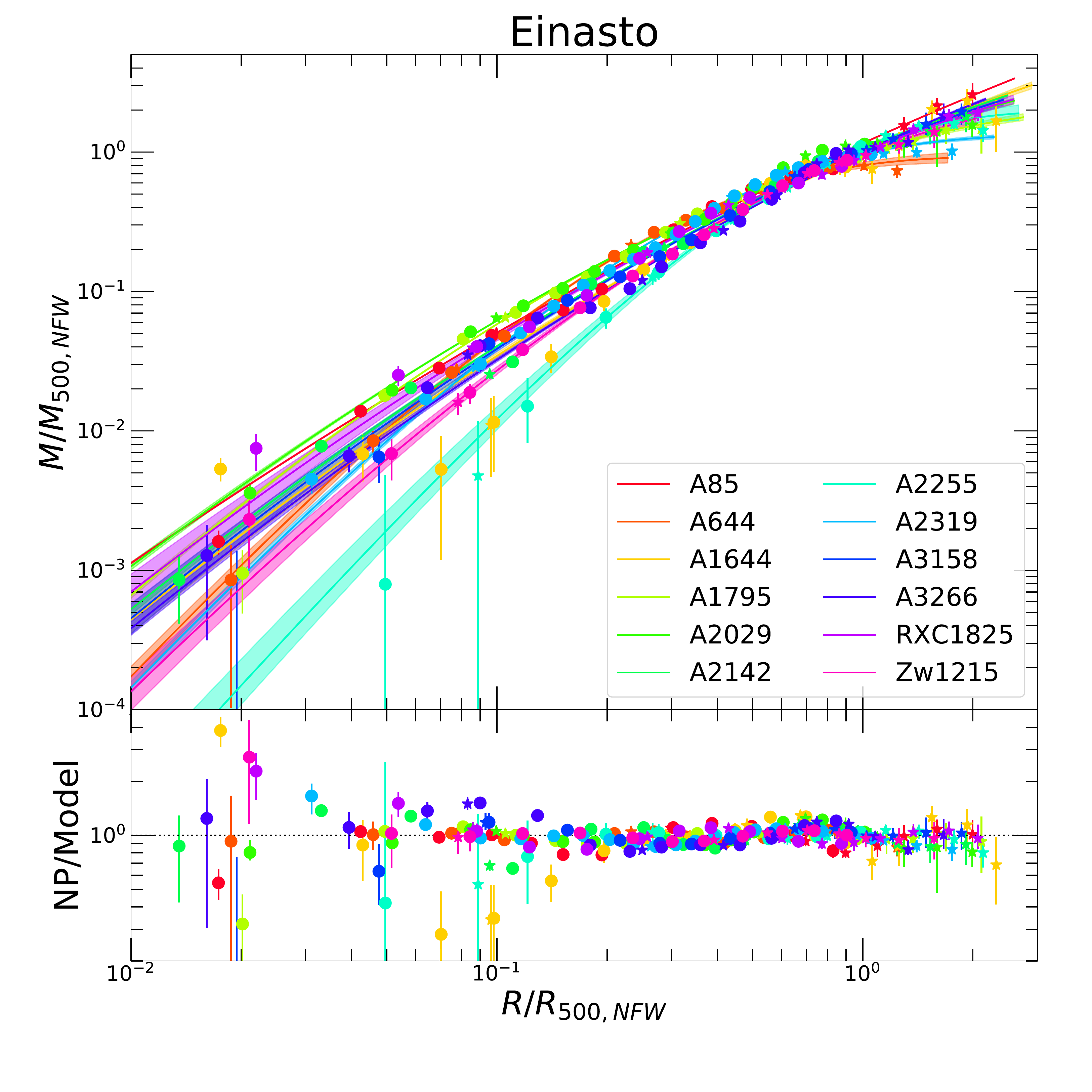}}}
        \caption{Comparison between the mass profiles reconstructed using mass models (solid curves and shaded areas) and the nonparametric log-normal mixture method (data points). The results obtained for the NFW and Einasto profile are shown in the left and right panels, respectively. The data points show the posterior NP mass estimates at the radii of the X-ray spectral data (circles) and the SZ pressure data (asterisks). For clarity, all the profiles are scaled by the NFW values for $R_{500}$ and $M_{500}$. The bottom panels show the local ratio of NP to model. }
        \label{fig:mod_vs_gp}
\end{figure*}

In Fig. \ref{fig:mod_vs_gp} we present a detailed comparison between the NFW and Einasto mass models and the nonparametric reconstruction. All the masses and radii were scaled by the NFW $M_{500}$ and $R_{500}$ values to visualize the differences in the shape of the profiles. The choice of the NFW results as scaling factors is motivated by the greater stability of the NFW masses compared to the other methods. Conversely, to study the radial shape of the profiles, the results obtained with the NP method are used as a benchmark of the local gravitational field because the method is fully data driven. In the left- and right-hand panels of the figure we show the reconstructed NFW and Einasto profiles, respectively, for the entire X-COP sample. We immediately notice that both parametric mass models trace the NP points closely, albeit with substantial scatter in the core and the outskirts. The additional degree of freedom afforded by the Einasto model allows it to trace the NP points more closely than the NFW, as there appears to be more diversity in the profile shapes than can be captured by the NFW model. 

\begin{figure*}
        \resizebox{\hsize}{!}{\includegraphics[width=0.5\textwidth]{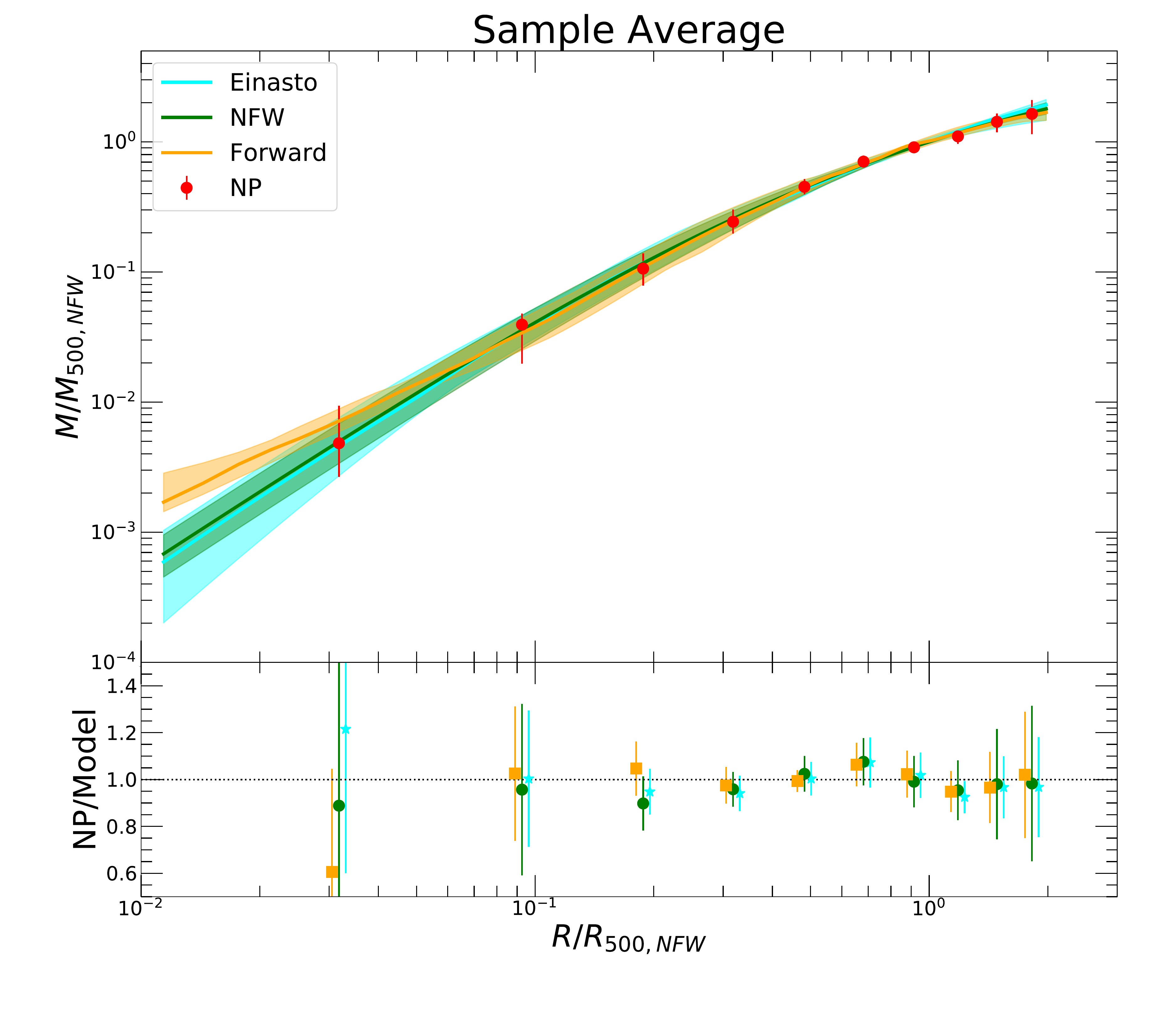}\includegraphics[width=0.5\textwidth]{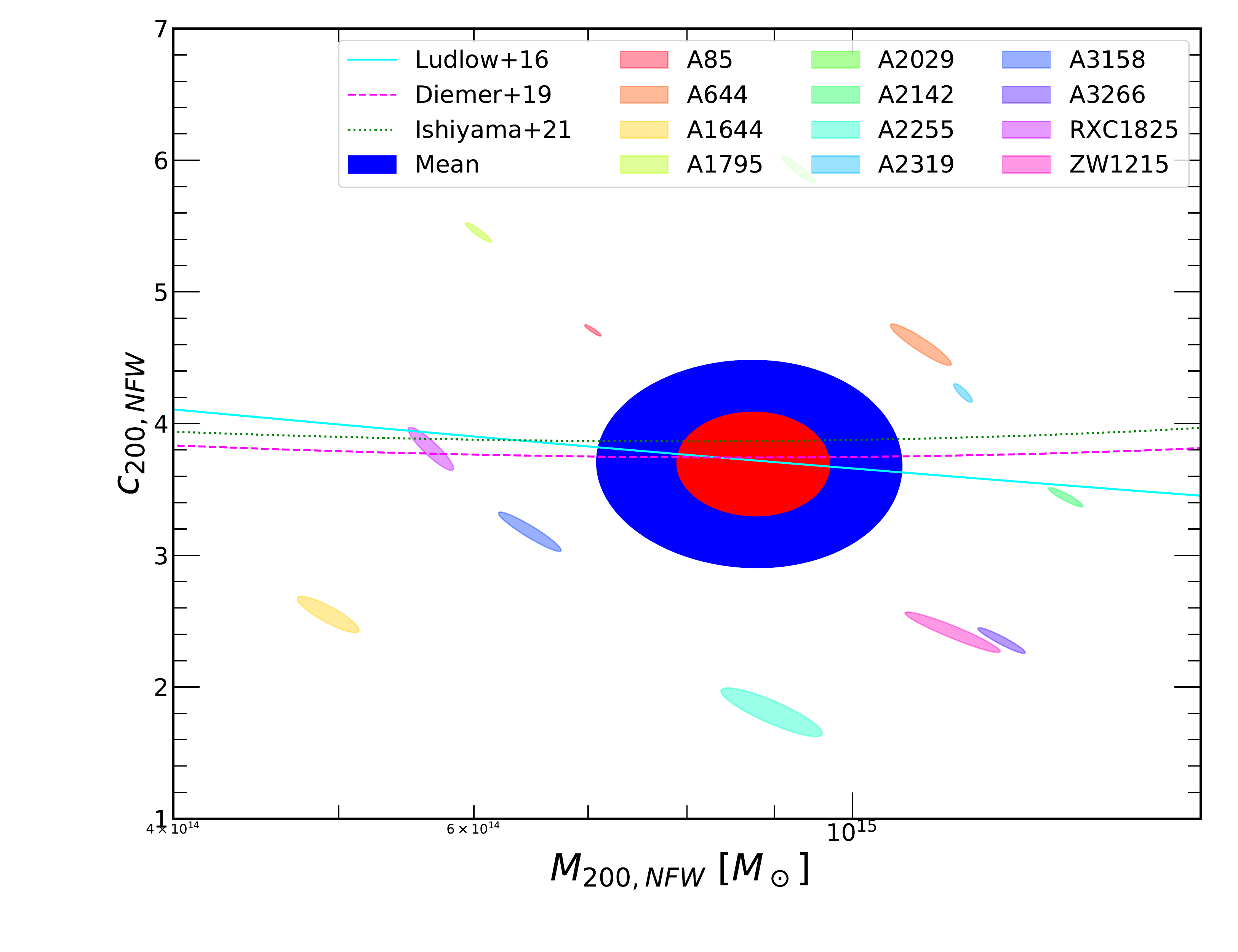}}
        \caption{\label{fig:c-m}Average profile shapes. \emph{Left:} Median scaled mass profiles for the four methods investigated here. The bottom panel shows the median ratio of NP to model, with the error bars indicating the 68\% interval. The ratios are evaluated at the same radius but are slightly shifted for better readability. \emph{Right:} NFW mass-concentration relation for our sample. The small shaded ellipses indicate the $1\sigma$ contours of the fitted individual values for the 12 X-COP clusters individually, whereas the solid red and blue ellipses show the fitted sample mean and its $1\sigma$ and $2\sigma$ confidence regions, respectively. The curves show the predictions of N-body simulations \citep{ludlow16,diemer19,ishiyama21}. }
\end{figure*}

This visual impression is confirmed in a quantitative way in the left-hand panel of Fig. \ref{fig:c-m}, where we show the sample median profiles and 68\% percentiles for the various methods. The range of values in the cluster core is clearly broader for Einasto than NFW. The forward-fitting method also closely traces the NP profile, with the exception of the innermost regions, where the power-law shape of the gNFW pressure profile leads to excess pressure, and hence excess mass, compared to the NP data. In the left-hand panel of Fig. \ref{fig:c-m} we show the median NP-to-model ratio across the sample, as well as the dispersion of the model with respect to the reference NP value. The mass profiles estimated with all the methods are in remarkable agreement, with the median masses in the range $[0.05-2]R_{500}$ being within 3\% of one another, even though the various methods make vastly different underlying assumptions. On average, we do not observe systematic deviations from an NFW shape, with the relative difference between NFW and NP points being less than 10\% at all radii and consistent with the level of deviations found in the mock observations (see Sect. \ref{sec:mock}). Similarly, the Einasto profile accurately reproduces the average shape of the gravitational field in our systems. This study is therefore in agreement with that of E19, which concluded that the mass profiles of X-COP clusters are better represented by models that exhibit a rising density profile in their inner regions, whereas models that include large central cores (Burkert, isothermal sphere) are statistically disfavored. In all cases, the scatter is minimal in the radial range [0.2-1]$R_{500}$, where all the methods converge within less than 10\% of one another. The difference between the models is largest in the outer regions, where the scatter of the NFW values relative to the NP points is nearly twice as large as for the other two methods. This result shows that the properties of the gas in cluster outskirts require more diversity in the DM profile shape than can be described by the NFW model, although the validity of the HSE assumption at large radii is unclear (see the discussion in Sect. \ref{sec:syst}). Our results agree with the prediction of \citet{neto07}, who showed that while the NFW profile provides a good description of $\Lambda$CDM density profiles on average, deviations from an NFW shape can be substantial in individual systems, with average deviations on the order of $0.1$ dex. Conversely, the Einasto profile traces the measured ICM properties closely even beyond $R_{500}$.

\subsection{Concentration-mass relation}

The relation between NFW concentration and mass is a clear prediction of the $\Lambda$CDM paradigm \citep[e.g.,][]{duffy08,bhattacharya13,dutton14,diemer14}. It arises from the universality of the structure formation process through mergers and accretion. Our precise measurements of halo concentrations therefore allow us to test the $\Lambda$CDM framework.

Given that our sample spans a narrow mass range (only a factor of $\sim2$), it is not sensitive to the shallow slope of the NFW mass-concentration relation. However, it is well suited to measure the mean and scatter of the relation around $M_{200}=10^{15}M_\odot$. To this end, we performed a Bayesian multivariate analysis of our sample, with the fractional log-normal scatter in mass and concentration as free parameters. Namely, we describe our sample as a set of values drawn from a distribution centered on the mean values for $R_{200}$ and $c_{200}$ with free intrinsic log-normal scatter on both axes. We set Gaussian priors on the two mean values, with the mean and standard deviation of the priors set to the mean and scatter of the sample values. A positive half-Cauchy prior with $
\beta=0.5$ is set on the fractional intrinsic scatter along both axes. Adopting instead a half-normal prior on the scatter has little impact on the results. To take the strong intrinsic correlation between mass and concentration in the NFW model into account, for each cluster we compute the covariance matrix between the two parameters from the output chains, and we set a global likelihood as the product of the Gaussian multivariate likelihoods of each object including the covariance matrix. We then sample the total likelihood using \texttt{PyMC3}. 

\begin{figure}
        \resizebox{\hsize}{!}{\includegraphics{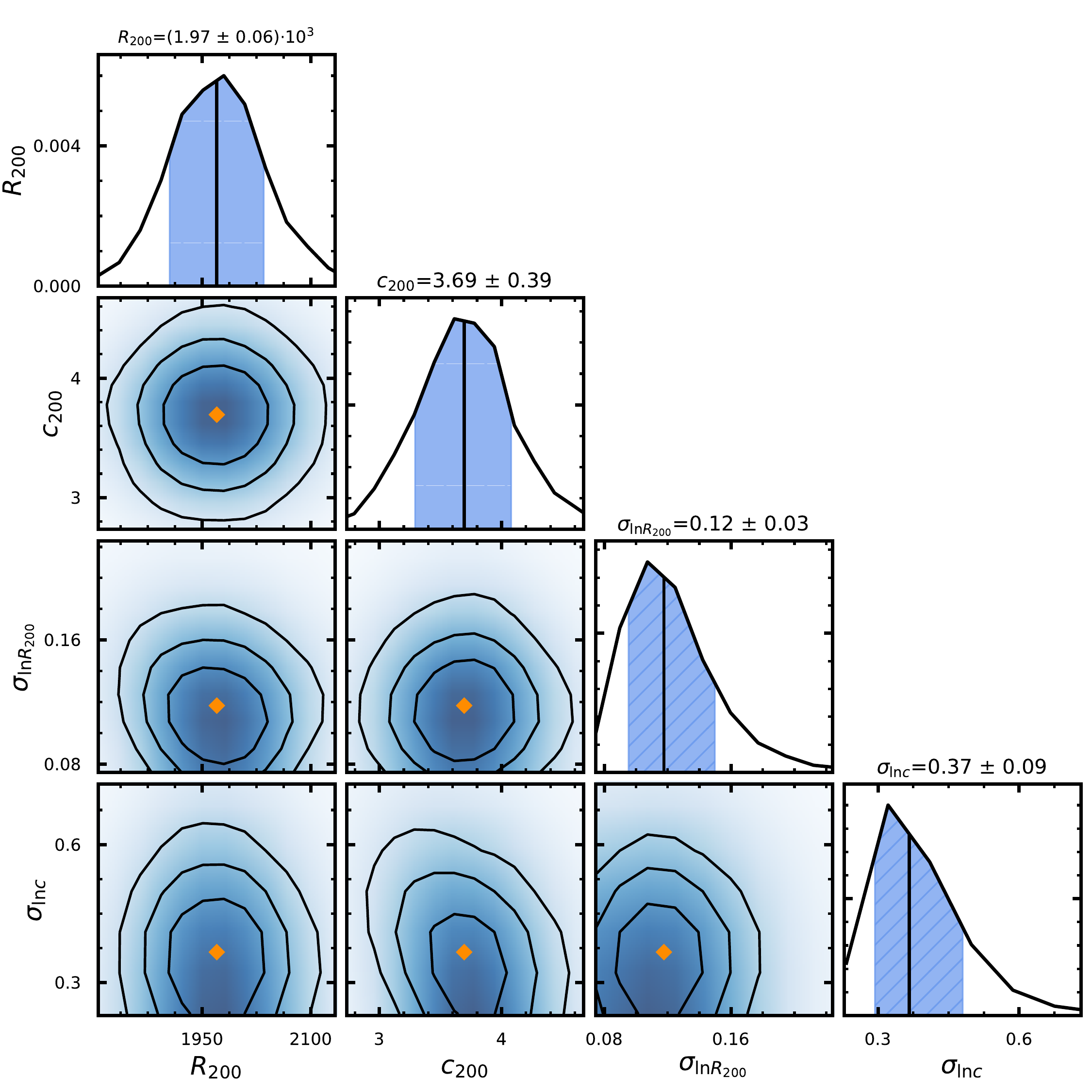}}
        \caption{\label{fig:cm_corner_plot}Posterior distributions for the fitted concentration-mass relation, with $R_{200}$ and $c_{200}$ the mean values of $R_{200}$ and NFW concentration in our sample and $\sigma_{R_{200}}$ and $\sigma_{c}$ the fractional log-normal intrinsic scatter of the points around the mean value.}
\end{figure}

The measured $1\sigma$ contours for all X-COP clusters in the $M_{200}-c_{200}$ plane are shown in Fig. \ref{fig:c-m} together with the fitted mean values. The strong intrinsic anticorrelation between mass and concentration can be clearly seen on the plot. The posterior distributions for the parameters of interest are shown in Fig. \ref{fig:cm_corner_plot}. The mass-concentration relation for our sample is centered on $M_{200}=8.64_{-0.81}^{+0.91}\times10^{14}M_\odot$ and $c_{200}=3.69_{-0.36}^{+0.39}$ with an intrinsic scatter $\sigma_{\ln c_{200}}=0.37_{-0.07}^{+0.11}$. 

We used the Python package \texttt{Colossus} \citep{diemer18} to compare our results with the predictions of various N-body simulation suites. The curves in the right-hand panel of Fig. \ref{fig:c-m} show the predictions of three different sets of simulations \citep{ludlow16,diemer19,ishiyama21}. The curves were calculated at the median redshift of the X-COP sample ($z=0.065$) for a proper comparison with the data. We find a remarkably good agreement between our average values and the predictions of numerical simulations, with all three simulations considered here agreeing with our measurements within $1\sigma$. On the other hand, numerical simulations also predict a large scatter in concentration at given mass, lower in more relaxed systems. \citet{neto07} find that a log-normal distribution represents the estimated concentrations in a given mass bin well, with a mean and a dispersion that decrease at higher masses, with the dispersion on the order of $\sigma_{\ln c} \sim 0.21$ for more relaxed systems. For the most massive clusters, \citet{bhattacharya13} estimate $\sigma_{\ln c}=0.33$, in excellent agreement with our measurements. \citet{diemer15} measure a scatter of about 0.16 dex ($\sigma_{\ln c}=0.37$) at all redshifts, masses, and for all mass definition, when the entire ensemble of halos is considered, again in agreement with the X-COP data. Different halo collapse times, with higher concentration associated with a halo assembled earlier, can account for most of the measured scatter \citep[e.g.,][]{nfw97,ludlow16}.

All together, the shape parameters retrieved here agree well with the predictions of the $\Lambda$CDM framework, both for the average NFW concentration and the scatter of the c-M relation. A discussion of profile shape in the central regions of X-COP clusters as traced by the Einasto shape parameter is provided in Paper \rom{2,} where we present our measurements of $\alpha$ and discuss in detail the implications of our measurements. 

\subsection{Breaking down into baryonic and DM components}

\begin{figure*}
\resizebox{\hsize}{!}{\includegraphics{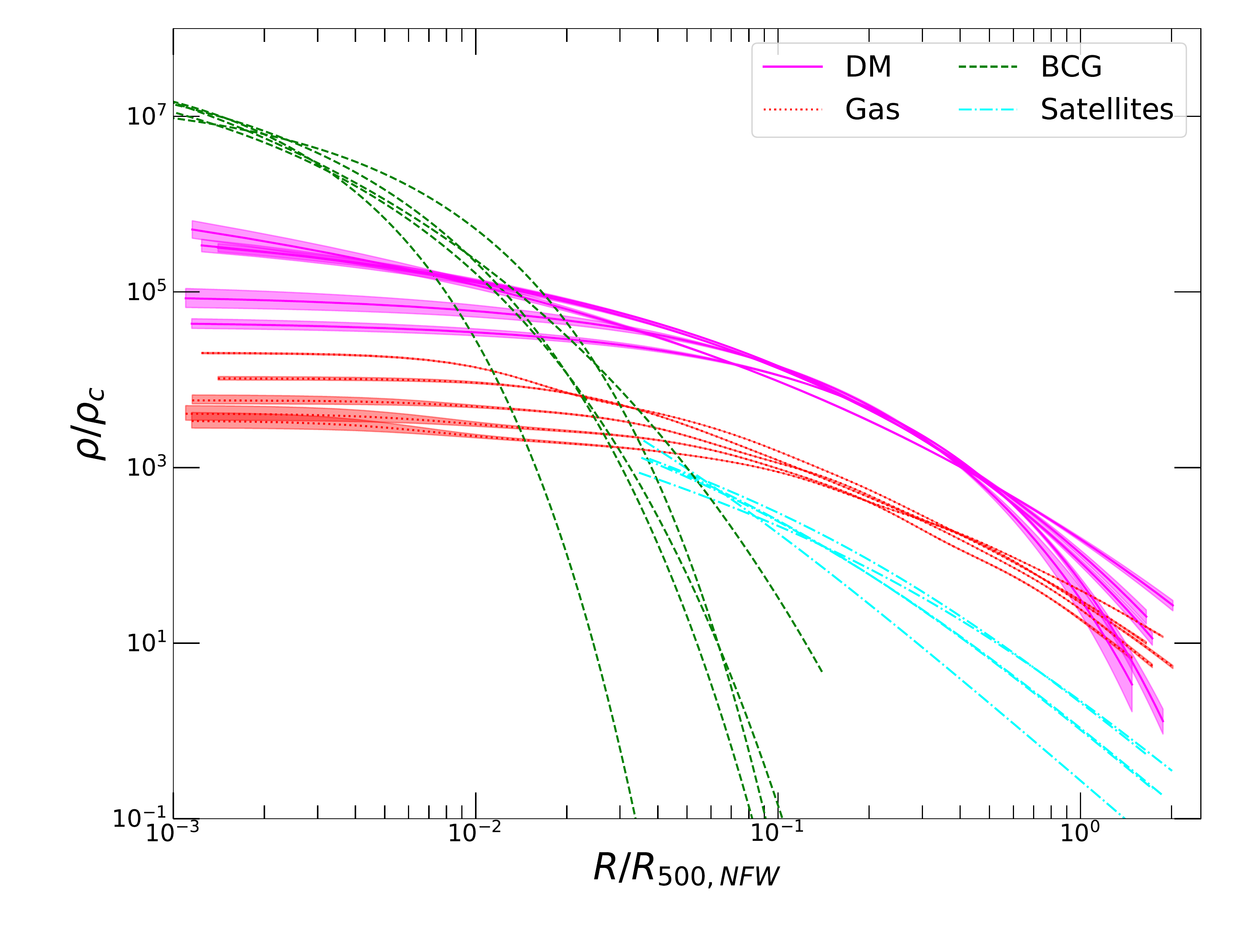}\includegraphics{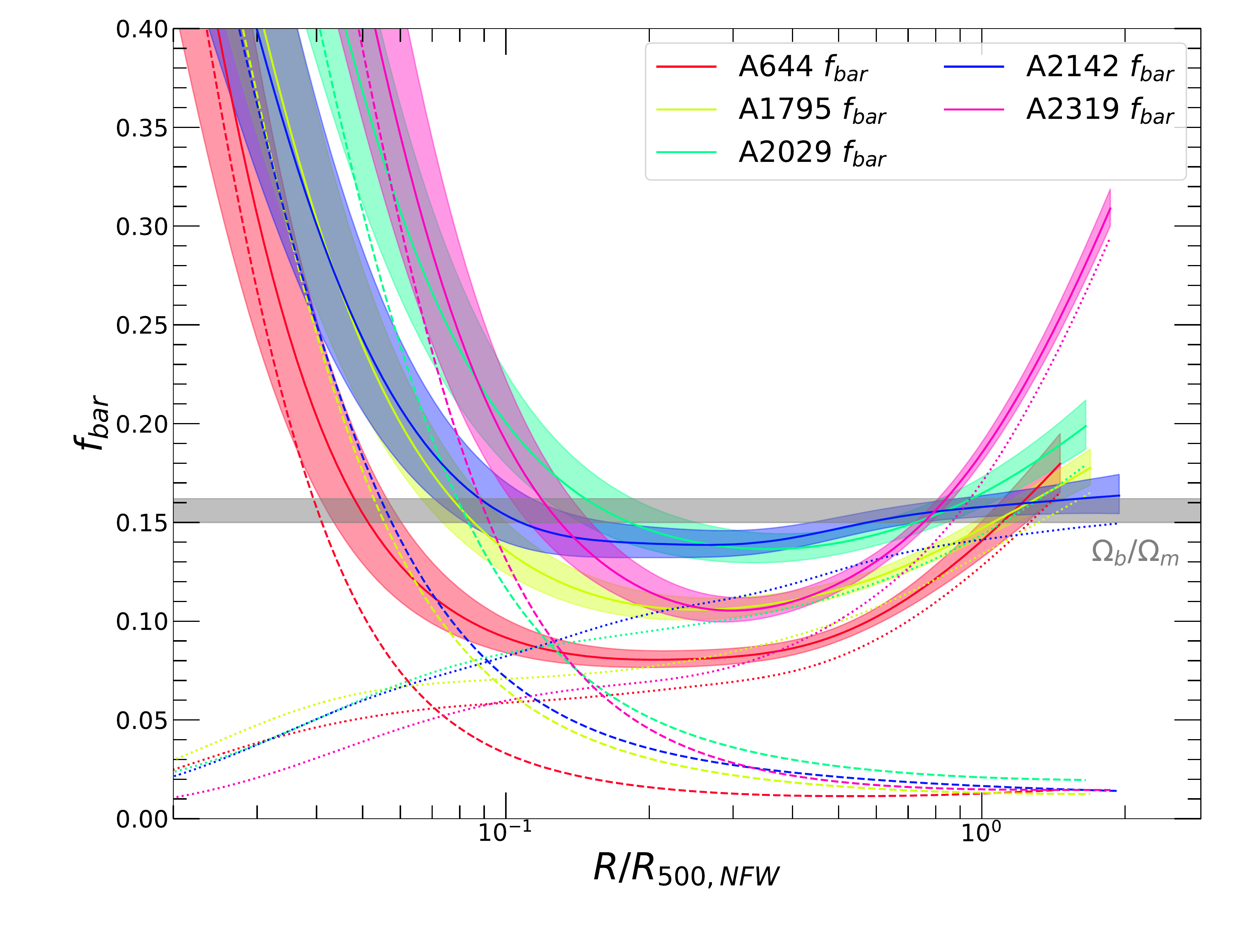}}
\caption{\label{fig:decomposition} Decomposition of the gravitational field into DM and baryonic components. \emph{Left:} Breakdown of the total differential density profile into baryonic and DM components for the five systems for which stellar mass profiles are available for both the BCG and the satellite galaxies. The baryonic components are split into the stellar mass (BCG, dashed green; satellites, dash-dotted cyan) and the gas mass (dashed red), whereas the fitted Einasto DM profiles are shown as the solid magenta curves. \emph{Right:} Total cumulative baryon fraction as a function of radius. The total baryon fraction and its uncertainty are shown as the solid curves and shaded areas, where the dashed and dotted lines show the relative contribution of gas and stars, respectively. }
\end{figure*}

As stated in Sect. \ref{sec:sample}, for a subset of systems (A644, A1795, A2029, A2142, and A2319) we have access to precise measurements of the stellar mass profiles of the cluster's BCG \citep{loubser20} and of the cumulative stellar mass of satellite galaxies \citep{vdb15}. Therefore, with the exception of a potential contribution of intracluster light (ICL) we have direct measurements of the mass profiles of all the baryonic components (gas, BCG, and satellites). For these systems, we conducted an additional analysis where the cumulative mass of the baryonic components is added as a fixed component to the mass profile at every radii. The remaining mass is then modeled as an additional model component (NFW or Einasto) to constrain the shape of the gravitational field induced by DM only. This procedure is described in detail in Sect. \ref{sec:massmod}. 

Given that the stellar mass profile of the BCG can be estimated only within a limited radial range out to $\sim50$ kpc, we fitted the BCG stellar mass using a S\'ersic functional form, which was found to provide a good description of the data,
\begin{equation}
    \rho_{BCG}(r) = \rho_0 \exp\left\{ -b_n \left((r/r_e)^{1/n}-1 \right)\right\}.
\end{equation}
Similarly, the cumulative stellar mass profiles of the satellite galaxies are quite noisy and not smooth, which is difficult to treat in the context of this analysis. \citet{vdb15} found that the stellar mass density profiles of satellite galaxies can be well described by a generalized NFW profile (see Eq. \ref{eq:gnfw}). Therefore, we fitted the stellar mass profiles beyond the BCG ($R>100$ kpc) with a gNFW profile with shape parameters fixed to \citet{vdb15}, and used the resulting functional form as input for our mass reconstruction code. 

The results of the mass reconstruction separating the baryonic and DM components are shown in Fig. \ref{fig:decomposition}. Here we focus on the Einasto reconstruction because of its greater flexibility; qualitatively similar results are obtained in the NFW case. The fitted NFW concentrations as well as the total masses obtained with this procedure are given in Table \ref{tab:master}. In the left-hand panel of Fig. \ref{fig:decomposition} we show the output differential mass density profiles for the BCG (green), gas mass (red), satellite galaxies (cyan) and DM (magenta). In the innermost regions ($R<0.02R_{500}$) the stellar mass of the BCG dominates because of the gravitational collapse of cold baryons at the bottom of the potential well. The DM component takes over beyond $\sim0.02R_{500}$ as the dominant mass component. The NFW concentrations obtained for the fit to the DM component only are 10-20\% smaller than what is obtained when fitting the total gravitational field (see Table \ref{tab:master}), which results from the subtraction of the sharply peaked BCG profile in cluster cores. The baryon-driven adiabatic expansion of the DM halo under the influence of active galactic nucleus (AGN) feedback results in slightly lower concentrations \citep{duffy10}, in agreement with our findings. The DM profiles look remarkably similar in the radial range [0.2-0.6]$R_{500}$, where our method is most accurate (see Sect. \ref{sec:syst}). The shape of the hot gas component is similar to that of the DM. It dominates the baryonic content of clusters beyond $\sim0.05R_{500}$ and even slightly ``catches up'' with the DM in the outskirts. In the case of A2319 and A644, the mass profile was found to be better described by a Burkert profile, which leads to the curved shape of the DM profiles observed here. Nongravitational heating processes such as AGN feedback inject energy into the gaseous atmosphere, making its profile somewhat shallower than that of the DM. Finally, satellite galaxies dominate the stellar mass content beyond $\sim0.1R_{500}$, such that the total mass of satellites within $R_{200}$ exceeds that of the BCG by more than an order of magnitude. However, their contribution to the total gravitational field never exceeds a few per cent. Intracluster light, which is not included in our analysis, may slightly change the shape of the total stellar component, as its distribution is thought to be centrally concentrated but somewhat shallower than the BCG \citep{montes14}; however, the contribution of ICL to the total baryon budget is expected to be small \citep[$<5\%$; e.g.,][]{gonzalez07}.

\subsection{Cumulative baryon fraction}
\label{sec:fbar}

\begin{table*}
    \caption{\label{tab:fbar} Relative contribution of gas and stars as well as the total baryonic fraction at several overdensity radii for the five objects with available complete stellar mass information. All the values are given as percentages.}
    \centering
\begin{tabular}{lccccccccc}
\hline
Cluster & $f_{\rm gas, 2500}$ & $f_{\star,2500}$ & $f_{\rm bar,2500}$ & $f_{\rm gas, 500}$ & $f_{\star,500}$ & $f_{\rm bar,500}$ & $f_{\rm gas, 200}$ & $f_{\star,200}$ & $f_{\rm bar,200}$ \\ 
\hline
\hline
A644 & $7.7\pm0.1$ & $1.2\pm0.4$ & $8.9\pm0.5$ &  $12.8\pm0.6$ & $1.3\pm0.5$ & $14.1\pm1.1$ &  $16.9\pm1.2$ & $1.5\pm0.6$ & $18.4\pm1.7$ \\ 
A1795 & $9.6\pm0.1$ & $1.7\pm0.6$ & $11.3\pm0.6$ &  $13.4\pm0.3$ & $1.3\pm0.5$ & $14.7\pm0.7$ &  $15.8\pm0.5$ & $1.3\pm0.4$ & $17.1\pm1.0$ \\ 
A2029 & $11.1\pm0.1$ & $2.8\pm0.9$ & $13.8\pm1.1$ &  $14.4\pm0.3$ & $2.1\pm0.7$ & $16.5\pm1.0$ &  $17.1\pm0.7$ & $2.0\pm0.7$ & $19.1\pm1.4$ \\ 
A2142 & $12.0\pm0.1$ & $2.3\pm0.8$ & $14.3\pm0.9$ &  $14.1\pm0.2$ & $1.7\pm0.6$ & $15.8\pm0.8$ &  $14.7\pm0.5$ & $1.5\pm0.5$ & $16.2\pm1.0$ \\ 
A2319 & $9.2\pm0.1$ & $2.0\pm0.7$ & $11.3\pm0.8$ &  $17.0\pm0.2$ & $1.5\pm0.5$ & $18.5\pm0.7$ &  $24.9\pm0.5$ & $1.5\pm0.5$ & $26.4\pm1.0$ \\ 
\hline
\end{tabular}
\end{table*}

The data presented here also allow us to study the total cumulative baryon fraction in our systems as a function of radius. The right-hand panel of Fig. \ref{fig:decomposition} shows the integrated baryon fraction $f_{\rm bar}(r)=M_{\rm bar}(<r)/M_{\rm tot}(<r)$ from the core to the outskirts, as well as the relative contribution of gas and stars. As already highlighted above, the stellar content of the BCG dominates  inside a few per cent of $R_{500}$ and decreases steeply with radius. The integrated mass of the gas component dominates over that of stars beyond $\sim 0.1R_{500}$, and inside $R_{500}$ the ICM accounts for about 90\% of the total baryonic content of clusters.  In Table \ref{tab:fbar} we quote the measured baryon fraction values for the five systems studied here at three different overdensity radii ($R_{2500}$, $R_{500}$, and $R_{200}$). 

We observe an overall depletion of baryons in the $0.2-0.5 R_{500}$ range, which is compensated by the somewhat shallower slope of the gas density profile with respect to the DM. Inside $R_{500}$ our estimated baryon fraction is equal to the cosmic value, and even slightly exceeds it beyond this point. Such an excess relative to the cosmic value is unlikely to be a real effect, as clusters are expected to be closed boxes for baryons within their virial radius \citep[e.g.,][]{white93}. An excess baryon fraction is more likely a trace of deviations from the HSE assumption, which are expected to bias HSE masses toward low values \citep[e.g.,][]{rasia06,nelson14}. This effect is particularly striking in the case of A2319, for which HSE masses are known to be biased low \citep[see the detailed discussion in][]{ghirardini18}. In \citet{eckert19} we related the cumulative hydrostatic gas fraction to the level of hydrostatic bias by comparing the measured gas fractions computed with the NFW model with the predictions of numerical simulations, which expect that at large radii the baryon fraction should match the cosmic value with a small depletion of $5-10\%$ \citep{planelles13}. We found that deviations from HSE are low, at the level of $6-10\%$, with the notable exception of A2319. However, the stellar content had to be assumed. Our measurements of the total baryon fraction are consistent with the results of \citet{eckert19} within $R_{500}$ but slightly higher within $R_{200}$, which can be explained by the different model used here. Indeed, the study presented in \citet{eckert19} was based on NFW fits to the total density profile, whereas here we consider the baryonic and DM components separately and model the DM profile with the more flexible Einasto functional form. At $R_{200}$ our baryon fractions exceed the cosmic value by 10-20\%, potentially indicating a stronger impact of nonthermal pressure beyond $R_{500}$. This analysis thus shows that our mass profiles are accurate at the $\lesssim 20\%$ level throughout the range considered here, which has little impact on our main conclusions.

\subsection{The galaxy cluster radial acceleration relation}
\label{sec:rar}

\begin{figure*}
\begin{minipage}[c]{0.6\textwidth}
        \centerline{\resizebox{\hsize}{!}{\includegraphics{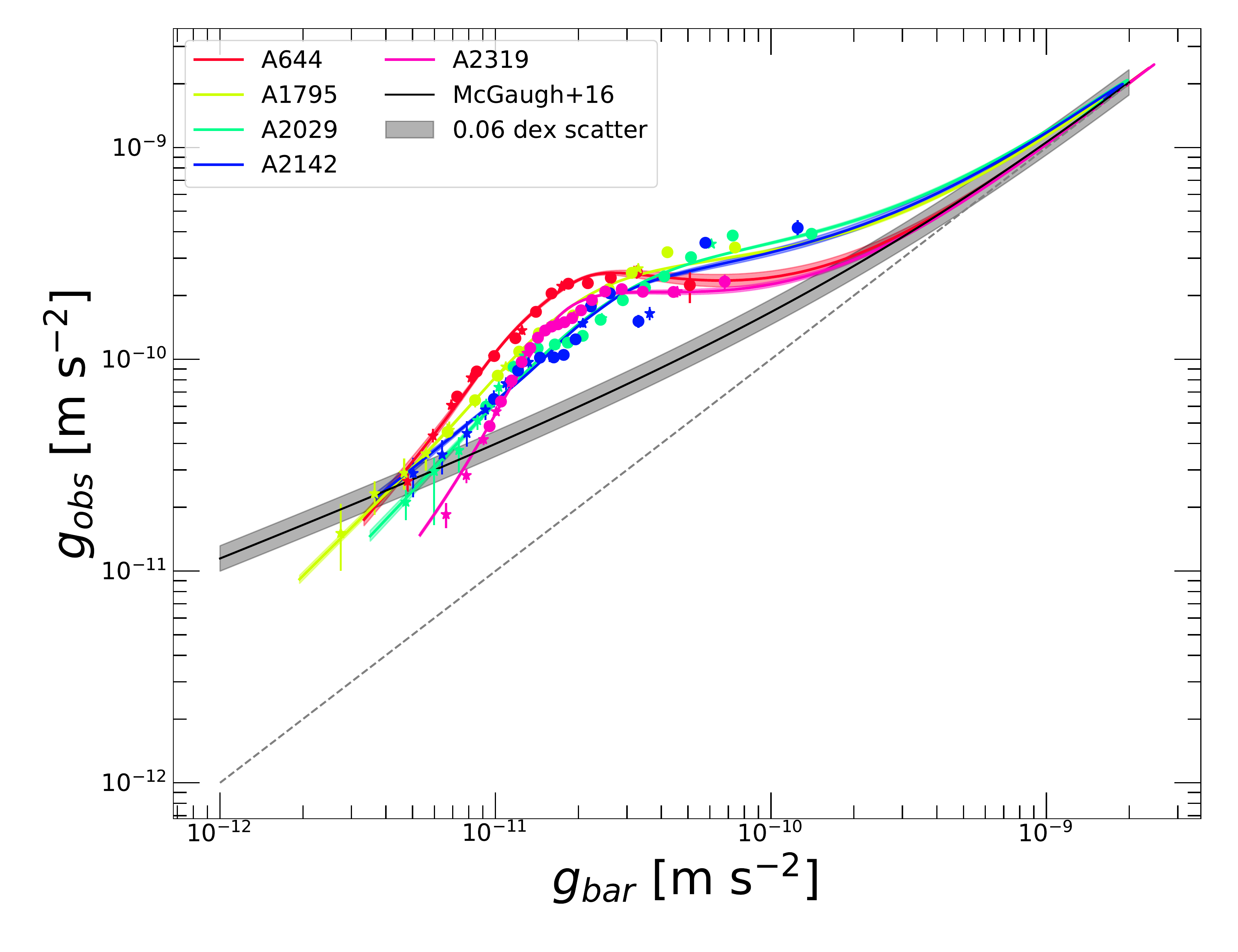}}}
\end{minipage}
\begin{minipage}[c]{0.4\textwidth}
        \caption{\label{fig:rar}Radial acceleration relation for X-COP galaxy clusters. The solid colored curves show the result of the Einasto fit to the baryons+DM model, whereas the individual data points indicate the accelerations computed using the nonparametric log-normal mixture model (NP). For comparison, the solid black curve and shaded area show the RAR obtained by \citet{mcgaugh16} for rotationally supported galaxies and the associated 0.06 dex scatter. The dashed line indicates the one-to-one relation.}
\end{minipage}
\end{figure*}

The availability of precise measurements of the gas, stellar, and DM mass components allows us to study the relation between the total, observed gravitational acceleration $g_{\rm obs}$ and the acceleration that would be expected in the absence of DM from the sum of the detected baryonic components, $g_{\rm bar}$. Studying the rotation curves of rotationally supported galaxies, \citet{mcgaugh16} showed that the relation between $g_{\rm bar}$ and $g_{\rm obs}$ is nearly universal across the considered sample and smoothly deviates from the one-to-one relation when moving toward the external regions where the gravitational force is low. If this relation is found to be universal across all gravitationally bound halos, it could become the smoking gun for modified gravity theories such as MOND \citep{milgrom83} and its developments. The RAR was found to hold over a wide range in stellar mass \citep{lelli17} with a low scatter of less than 0.13 dex and possibly even less than 0.06 dex \citep{li18}. Studies of the RAR in galaxy clusters found that the normalization of the observed acceleration is higher than what was found in rotationally supported galaxies \citep{chan20,tian20,pradyumna21}, although the uncertainties were too large to study in detail the shape of the RAR. 

In Fig. \ref{fig:rar} we show the RAR in X-COP galaxy clusters. The baryonic and observed acceleration were computed locally from the reconstructed profiles,
\begin{equation}
g_{\rm obs}(r) = \frac{GM_{\rm HSE}(<r)}{r^2}, 
\end{equation}
\begin{equation}
g_{\rm bar}(r)=\frac{G(M_{\rm gas}(<r) + M_{\rm \star,BCG}(<r)+M_{\rm \star,sat}(<r))}{r^2}.   
\end{equation}
For the observed acceleration we consider both the nonparametric reconstruction and the results of the Einasto fit to the DM component only. As can be seen in Fig. \ref{fig:rar}, the two methods provide consistent results throughout the entire range. The data are compared with the RAR relation determined by \citet{mcgaugh16} in rotationally supported galaxies,

\begin{equation}
g_{\rm obs} = \frac{g_{\rm bar}}{1-\exp(-\sqrt{g_{\rm bar}/g_{\dagger}})} \equiv \mathcal{F}(g_{\rm bar} )
\label{eq:mcgaugh}
,\end{equation}

\noindent with $g_{\dagger}=1.2\times10^{-10}$ m s$^{-2}$. At high acceleration ($g>10^{-10}$ m s$^{-2}$), the baryonic acceleration induced by the stellar mass of the BCG dominates, as expected from the breakdown of the density profiles shown in Fig. \ref{fig:decomposition}. Beyond this point, the observed gravitational acceleration starts to exceed the expectation of the RAR in spiral galaxies. In this regime, the baryonic acceleration rapidly decreases, whereas the total acceleration remains almost flat. This corresponds to the $[0.05-0.2]R_{500}$ radial range in Fig. \ref{fig:decomposition} where the stellar mass of the BCG falls off rapidly and the DM component largely dominates. The behavior of the relation is inconsistent with the \citet{mcgaugh16} relation at high statistical significance, even when considering the nonparametric reconstruction only. The observed acceleration is up to 5 times higher than expected, which cannot be explained by a potential bias in our mass reconstruction, since the HSE method rather tends to {underestimate} the true mass (see Sect. \ref{sec:syst} for a full discussion). 

Around $g_{\rm bar}\sim2\times 10^{-11}$ m s$^{-2}$ we observe another change of regime, where both the observed and the baryonic components decrease with a slope that is almost parallel to the one-to-one relation. This corresponds to the regime where the baryonic component is dominated by the ICM mass and where the gas density profiles are slightly shallower than the DM profiles. Our systems eventually catch up with the \citet{mcgaugh16} RAR in the outermost regions, although the slope of the relation is widely different. The complex shape observed here indicates that the RAR in rotationally supported galaxy is not a universal property of gravity, since it fails to reproduce a gravitational field of similar strength in massive halos.

\subsection{Modeling the galaxy cluster RAR}
\label{sec:model_rar}

\begin{figure*}
        \centerline{\resizebox{\hsize}{!}{\includegraphics{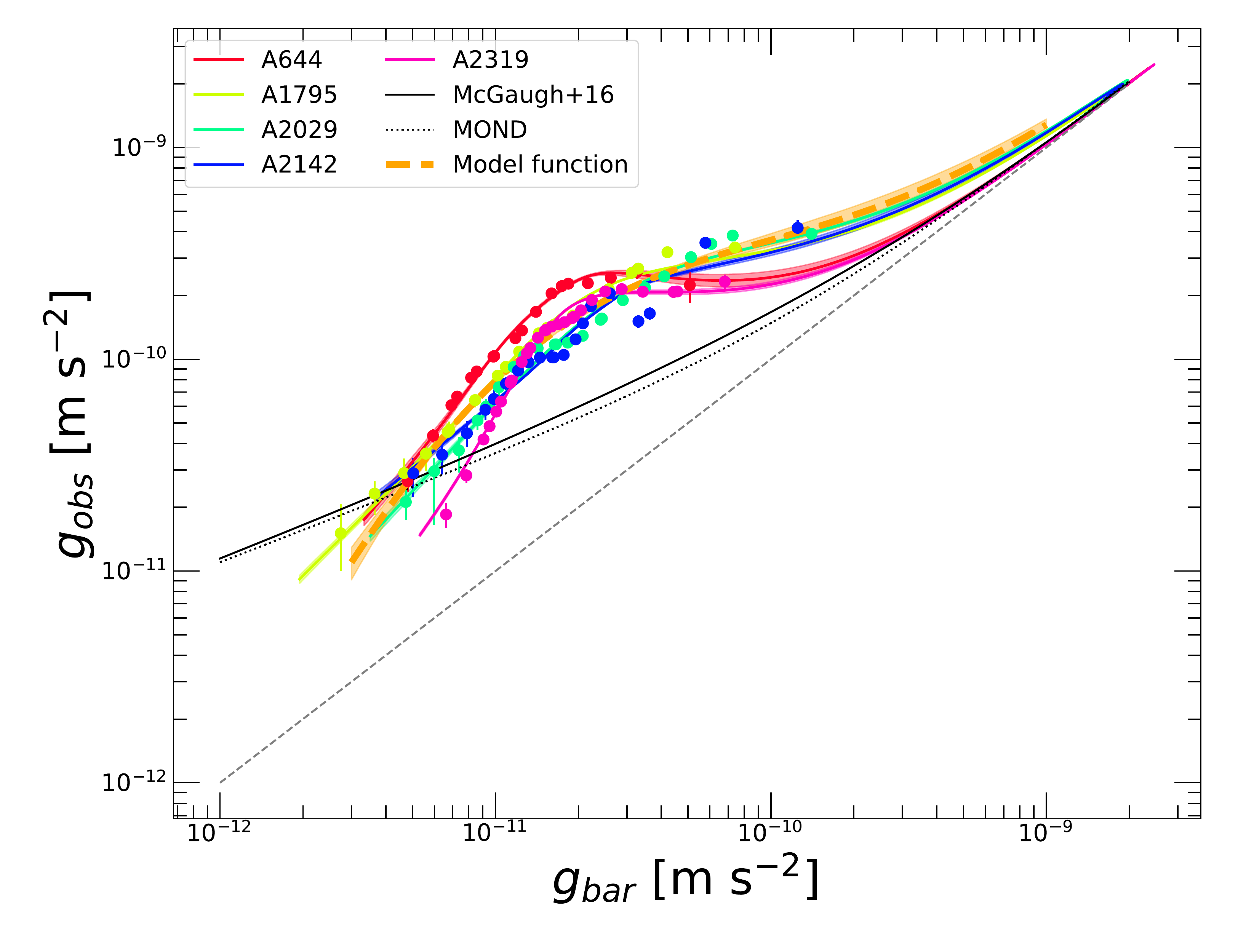}\includegraphics{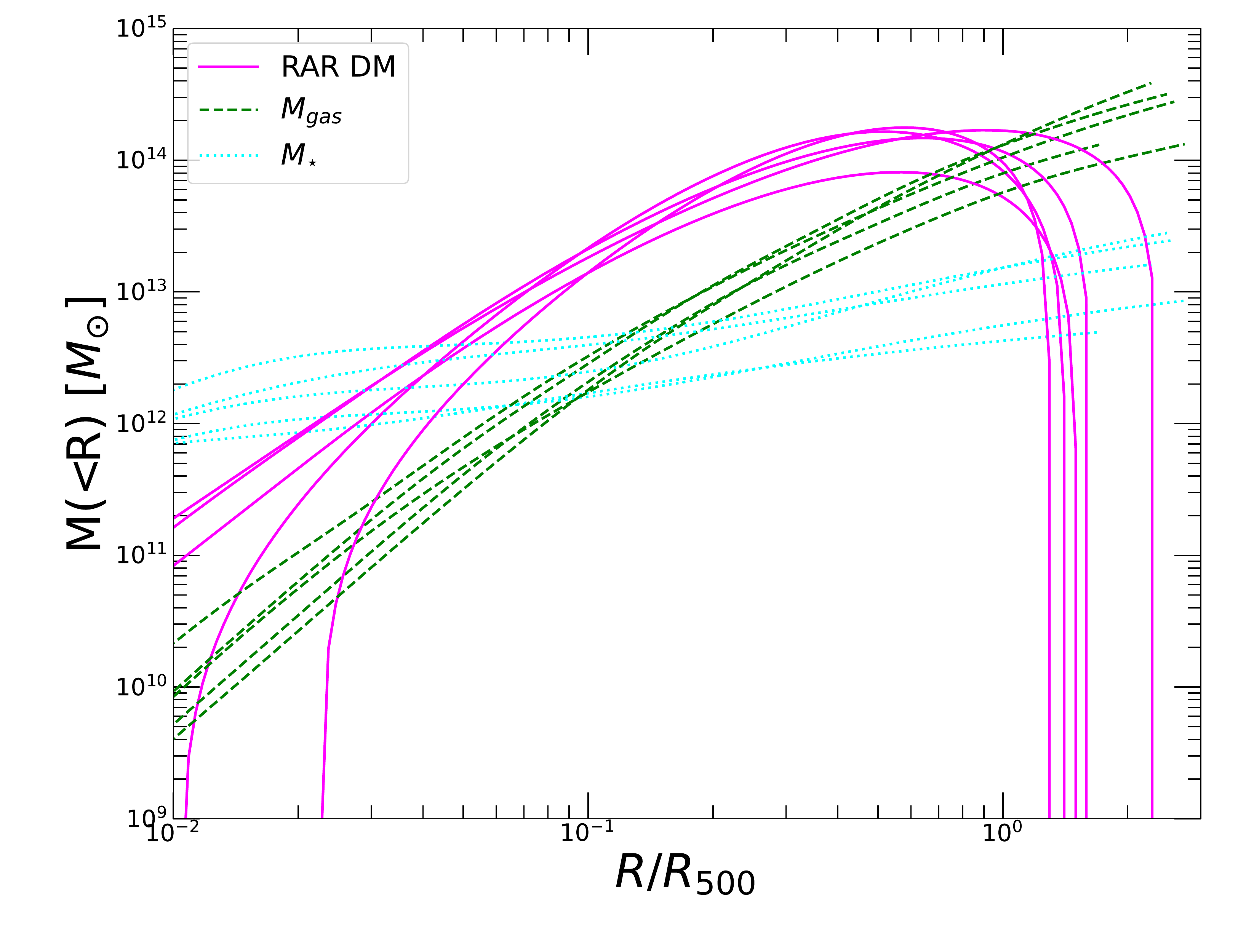}}}
        \caption{\label{fig:model_rar} Modeling of the RAR in galaxy clusters. \emph{Left:} Same as in Fig. \ref{fig:rar}. The thick dashed orange curve and shaded area show the best fit to the NP data with the model that describes two transitions (Eq. \ref{eq:model_rar}) and its corresponding $1\sigma$ error envelope. The dotted line shows the classical MOND relation. \emph{Right:} Breakdown of the various mass components in the RAR paradigm for the five systems with complete descriptions of the baryonic components assuming that the \citet{mcgaugh16} relation is universal. The curves show the cumulative radial profiles of the gas mass (dashed green), stellar mass (dotted cyan), and the required missing $M_{DM,RAR}$ computed using Eq. \ref{eq:mond_dm} (solid magenta).}
\end{figure*}

While our data unequivocally confirm that galaxy clusters do not follow an acceleration relation similar to that of disk galaxies, Fig. \ref{fig:rar} shows that galaxy clusters define their own relation in the $g_{\rm obs}-g_{\rm bar}$ plane, which we can attempt to quantify. The complex shape of the relation implies that it cannot be described simply with a single acceleration scale. To model the observed relation, we introduce a modification of the MOND law with a second acceleration scale as

\begin{equation}
g_{\rm obs} = g_{\rm bar} \left(1+\frac{g_1}{g_{\rm bar}} \right)^{\alpha_1}  \left(1+\frac{g_2}{g_{\rm bar}} \right)^{\alpha_2-\alpha_1} 
\label{eq:model_rar}
,\end{equation}

\noindent with $g_1, g_2$ the two characteristic acceleration scales describing the transition between the three regimes, and $\alpha_1, \alpha_2$ the slopes of the correction term in the regime where the corresponding term dominates. This functional forms models a correction that is proportional to $g_{\rm bar}^{-\alpha_2}$ when $g_{\rm bar}\ll g_{2}$, transitioning to $g_{\rm bar}^{-\alpha_1}$ when $g_2<g_{\rm bar}<g_1$ and converging to 1 when $g_{\rm bar}\gg g_1$. Our empirical formula approaches the MOND formalism at low acceleration when $\alpha_1=\alpha_2=0.5$ and $g_1$ is the classical MOND acceleration $a_0\approx10^{-10}$ m/s$^2$. 

We fitted the nonparametric points on the $g_{\rm obs}-g_{\rm bar}$ plane with the model described in Eq. \ref{eq:model_rar} and attempted to quantify the parameters of the model and the intrinsic scatter. The model provides an adequate description of the data at hand, with an intrinsic scatter $\sigma_{\ln g_{\rm obs}}=0.21\pm0.02$. The best-fit curve is shown in the left-hand panel of Fig. \ref{fig:model_rar} where the functional form and its uncertainty are compared to the data. For the two acceleration scales we obtain $g_{1} = 2.3_{-0.8}^{+1.2}\times 10^{-10}$ m s$^{-2}$ and $g_2 = 1.2_{-0.7}^{+0.9}\times 10^{-11}$ m s$^{-2}$, in qualitative agreement with the above discussion. The fitted slopes yield $\alpha_1=1.4_{-0.5}^{+0.7}$ and $\alpha_2=-1.2_{-0.7}^{+0.9}$. The scale $g_1$ is of the same order as the classical MOND acceleration, albeit slightly higher. However, the fitted slope $\alpha_1$ is considerably larger than its corresponding MOND counterpart, which indicates that the transition away from the baryon-dominated regime is much faster than in disk galaxies. Below the second characteristic scale $g_2$ the $g_{\rm obs}-g_{\rm bar}$ relation is much steeper than expected in MOND.

As previously recognized by several authors \citep[e.g.,][]{sanders99,sanders03}, an extra mass component is required to reconcile the gravitational field of galaxy clusters with the MOND paradigm. In E19 we showed that the total baryonic mass in X-COP clusters within $R_{500}$ should exceed the measured baryonic mass by about a factor of 2. Our data can be used to determine the mass profile of the missing component assuming that the \citet{mcgaugh16} relation (Eq. \ref{eq:mcgaugh}) applies. Specifically, for each value of $g_{\rm obs}$ we can determine the missing enclosed mass $M_{DM,RAR}(<r)$ such that the expected acceleration matches the observed one. In other terms, for any given value of $g_{\rm obs}$ and $g_{\rm bar}$ we numerically search for the value of $M_{DM,RAR}$ that satisfies the condition
\begin{equation}
g_{\rm obs} = \mathcal{F}\left(g_{\rm bar} + \frac{GM_{DM,RAR}(<r)}{r^2}\right).
\label{eq:mond_dm}
\end{equation}
We used the Einasto reconstructions to determine the mass profiles of $M_{DM,RAR}$. In the right-hand panel of Fig. \ref{fig:model_rar} we show the enclosed mass profiles of the missing component assuming that Eq. \ref{eq:mcgaugh} holds.  The missing component dominates the mass profile beyond $\sim0.05R_{500}$, similar to the CDM case (Fig. \ref{fig:decomposition}). The required enclosed mass exceeds the gas mass and reaches a maximum around $\sim0.5R_{500}$ ($\sim600-700$ kpc). This corresponds to the second transition scale $g_2$, beyond which the slope of the relation is steeper than the expectation. As described in Sect. \ref{sec:rar}, the galaxy cluster data eventually catch up with the \citet{mcgaugh16} relation in the outskirts, implying that the required correction is negligible and the enclosed mass profile {decreases}. Since the cumulative mass cannot decrease, this analysis shows that the required mass distribution of the missing mass in MOND is unphysical, unless our analysis is hampered by substantial systematic uncertainties (see Sect. \ref{sec:syst}).

\section{Discussion}

\subsection{Systematic uncertainties}
\label{sec:syst}

A key issue to be addressed is the potential impact of systematic uncertainties. While our tests with mock data demonstrate that our analysis pipeline introduces minimal biases, from spectral fitting to hydrostatic mass reconstruction (see Sect. \ref{sec:mock}), a number of other sources of systematic uncertainty need to be addressed. Here we discuss the main sources of uncertainty: hydrostatic bias, gas inhomogeneities, and effective area calibration. 

\textbf{Hydrostatic bias}: In the presence of mergers and nongravitational energy input, such as AGN feedback, residual gas motions can act as an additional source of pressure support on top of thermal pressure \citep[e.g.,][]{rasia04,lau09,nelson14,biffi16}. In the presence of a substantial level of residual gas motions, mass estimates derived under the hydrostatic assumption are known to be biased toward low values, an effect that various numerical simulations predict to be in the range $5-30\%$ at $R_{500}$ \citep{rasia04,nagai07,angelinelli20,barnes21,bennett21}. Through a direct comparison between hydrostatic masses and masses obtained using alternative methods \citep{ettori19}, we found that X-COP hydrostatic masses are 10-15\% lower than weak lensing estimates \citep{herbonnet2020}. In \citet{eckert19} we used the integrated gas fraction as an anchor assuming that the true gas fraction can be robustly predicted by the $\Lambda$CDM framework to estimate the level of hydrostatic bias, which was found to be low (7\% at $R_{500}$, 10\% at $R_{200}$). We find similar results in this analysis, as described in Sect. \ref{sec:fbar}, where we can see that, with the exception of a single system that is known to be strongly biased \citep[A2319;][]{ghirardini18}, the measured baryon fractions agree with the expectations at $R_{500}$ and slightly exceed the cosmic baryon fraction beyond this point, indicating a mild level of hydrostatic bias. Since the relative level of nonthermal pressure support is expected to increase with distance to the cluster core \citep[e.g.,][]{nelson14}, we conclude that our measurements are mildly biased ($<10\%$) out to $R_{500}$. The HSE assumption is likely to be a poorer approximation beyond this point. 
    
    \textbf{Gas inhomogeneities:} Inhomogeneities in the gas distribution can potentially impact the recovered thermodynamic profiles, both for the gas density \citep{mathiesen99,nagai,vazza13} and the temperature profiles \citep{rasia14}. In the presence of overdense, cool substructures that have not yet mixed with the surrounding plasma, the recovered X-ray densities and temperatures are expected to be biased toward high and low values, respectively, compared to the mass-weighted quantities. The multi-temperature structure may also be important in the innermost regions ($<10$ kpc) where the hot ICM may mix with the cooler gas content of the BCG. Several works based on numerical simulations \citep[e.g.,][]{rasia12,pearce20} predict that temperature inhomogeneities would bias the X-ray derived temperatures, which might greatly increase the bias in the recovered HSE masses. For instance, \citet{pearce20} predict that masses based on X-ray data only may be largely biased low (up to 50\% in the X-COP mass range), whereas HSE masses using SZ-derived pressure profiles should be closer to the true, mass-weighted hydrostatic masses. However, we note that our gas density profiles are estimated using the azimuthal median technique \citep{eckert15}, which excises the regions of enhanced surface brightness and returns gas density profiles that are free from the clumping effect. The regions corresponding with obvious substructures are masked during the spectral extraction procedure, such that their impact on the recovered temperatures should be limited. Our masses estimated from X-ray data only are consistent with the results obtained in combination with SZ data \citep{ettori19} and the pressure profiles determined independently from X-ray and SZ data are in excellent agreement \citep{ghirardini19}. Therefore, we conclude that the impact of gas inhomogeneities is small ($<10\%$) and certainly not as severe as predicted by, for example, \citet{pearce20}. The difference could be explained either by the different procedures used in the simulations and observations or by an incomplete mixing in the corresponding simulations. To investigate the origin of the difference, it would be necessary to apply our pipeline on mock observations of simulated clusters, similar to the analysis presented in Sect. \ref{sec:mock}, which is beyond the scope of this paper.
    
    \textbf{Effective area calibration:} The calibration of the \emph{XMM-Newton} effective area affects the measured spectroscopic temperatures, as the spectral model needs to be folded through the instrumental response to fit the X-ray spectra. There is a known inconsistency between the temperatures determined by \emph{XMM-Newton}/EPIC and by \emph{Chandra}/ACIS \citep{nevalainen10,schellenberger15}, with ACIS returning systematically higher temperatures than EPIC when fitting over the entire 0.5-10 keV energy range. The discrepancy increases with plasma temperature, and reaches $\sim20\%$ for 10 keV plasma. It is still unclear whether the \emph{XMM-Newton} temperatures derived here are accurately reproducing the true spectroscopic temperatures. If the \emph{Chandra} temperatures are correct, our masses at fixed radii would be underestimated by a similar amount, given that to first order the HSE mass is proportional to the fitted temperature. Our reconstruction of the gravitational field is based on a combination of \emph{XMM-Newton} and \emph{Planck} data, the latter providing the dominant contribution to the temperature profiles at large radii ($\geq R_{500}$). Therefore, the uncertainty in the effective area calibration mostly affects the regions located inside $0.5 R_{500}$, where our temperatures are constrained primarily by the spectroscopic X-ray measurements.
    
Overall, we estimate that the systematic uncertainties associated with each of the effects discussed here amount to $10-20\%$, which should not affect the conclusions of this paper with the exception of the results on the baryon fraction (Table \ref{tab:fbar}). 

\subsection{Consistency with the $\Lambda$CDM framework}

All of the results presented in this paper agree with the predictions of the $\Lambda$CDM framework and the bottom-up structure formation paradigm. The average mass profile of the X-COP clusters is well represented by an NFW model, with deviations of at most 10\% over the entire radial range of interest ($[0.01-2]R_{500}$; see the left-hand panel of Fig. \ref{fig:c-m}). The average NFW concentration of the sample matches perfectly the $\Lambda$CDM predictions (right-hand panel of Fig. \ref{fig:c-m}), and N-body simulations also reproduce accurately the observed scatter in the mass-concentration relation at $M_{200}\sim10^{15}M_\odot$. Finally, with the exception of a system known for its high level of nonthermal pressure support (A2319), the total cumulative baryon budget (gas + stars) is close to the cosmic baryon fraction $\Omega_b/\Omega_m$, with a slight excess of 10-20\% that can most likely be attributed to a hydrostatic bias (see Sect. \ref{sec:fbar}). Numerical simulations including different gas physics and hydrodynamic solvers generically predict that in the most massive halos the gravitational binding energy largely exceeds nongravitational energy input, for example from AGN and supernovae. As a result, the baryon fraction enclosed within the virial radius should closely match the universal value \citep{white93,eke98,ettori09,planelles13,eckert19,mantz22}, in agreement with the findings presented here. We note however that given the uncertain impact of hydrostatic bias, our baryon fractions only provide a weak test of $\Lambda$CDM, and more accurate mass reconstructions combined, for example, with weak lensing data are required to determine the exact baryon fraction of galaxy clusters within $R_{200}$.

While the NFW model provides an adequate representation of the average mass profiles, we do observe deviations from the predicted universal shape, both in the innermost ($R<0.1R_{500}$) and outermost ($R>R_{500}$) regions covered in our study. The profiles estimated through our nonparametric technique appear to show a wider variety of shapes than can be described by the NFW parametric form (see Fig. \ref{fig:mod_vs_gp}). Conversely, the additional degree of freedom afforded by the Einasto parametric form allows the observed behavior to be reproduced more closely in each individual case. Some systems (e.g., A2255 and A644) show a larger amount of curvature in their mass profiles; the same systems were also found in E19 to be better described by models including a central core (Burkert, isothermal sphere). These systems exhibit a somewhat disturbed X-ray morphology as indicated by their large centroid shifts (see Paper \rom{2}); thus, it is possible that the X-ray peak does not trace exactly the bottom of the potential well. This effect is likely to be present in A2255, which exhibits a very flat central gas density profile and a substantial offset between the X-ray peak and the position of the BCG \citep[123 kpc,][]{rossetti16}. The apparent flat mass distribution in the core of this system may thus be explained by miscentering. On top of that, while the NFW profile is expected to provide an excellent description of the average mass profile extracted from CDM simulations, individual profiles show a scatter of $\sim0.1$ dex around the mean \citep[see, e.g.,][]{bhattacharya13,ludlow16}, in agreement with our observations (see Fig.~\ref{fig:c-m}). In conclusion, the study presented here does not show any significant tension with $\Lambda$CDM predictions, and extensions of the paradigm, either in the form of a different DM candidate (e.g., WDM or SIDM) or modifications of the theory of gravity, are not required by our data. 

\subsection{The RAR is not universal}

Through our decomposition of the measured gravitational field into DM and baryonic components (gas, BCG, and satellite galaxies), we provided a detailed reconstruction of the RAR in a subset of five systems (see Fig. \ref{fig:rar}). Our study improves over previous attempts to test the RAR in the galaxy cluster regime \citep{chan20,tian20,pradyumna21} as it combines our measurements of the gravitational field over two decades in radius with a full characterization of the various baryonic components (see Fig. \ref{fig:decomposition}). The relation derived by \citet{mcgaugh16} for disk galaxies fails to reproduce the relation between observed and baryonic accelerations in X-COP clusters at very high significance. As described in Sect. \ref{sec:rar}, we observe three distinct regimes in the $g_{\rm obs}-g_{\rm bar}$ plane: at high acceleration ($g_{\rm bar}>10^{-10}$ m s$^{-2}$), the gravitational field is dominated by the stellar content of the BCG. Below this threshold, the relation departs from the \citet{mcgaugh16} relation and the observed acceleration becomes up to 5 times higher than that expected from the RAR in spiral galaxies. Below a second characteristic scale, $g_{\rm bar}\sim2\times 10^{-11}$ m s$^{-2}$, the relation steepens again and eventually catches up with the \citet{mcgaugh16} relation. The discrepancy cannot be attributed to potential systematics in our reconstruction such as hydrostatic bias (see Sect. \ref{sec:syst}) as they would most likely lead to an underestimation of the gravitational acceleration and increase the gap. Therefore, the behavior of the relation in X-COP clusters vastly differs from that in spiral galaxies, both qualitatively and quantitatively, indicating that the RAR is apparently not universal.

If the tight relation observed in spiral galaxies results from a modification of the theory of gravity at large scales (e.g., MOND), the observed behavior shows that an additional source of acceleration is required on top of the known baryonic components \citep[see also][]{ettori19}. A possible way of alleviating the tensions with the MOND paradigm in the cluster regime is to invoke an additional matter component such as a light sterile neutrino with a mass in the range $1-10$ eV \citep[e.g.,][]{sanders07,angus08,nieuw13}, which may be a natural candidate for a minimal extension of the standard model of particle physics. The missing component would dominate the matter content at intermediate radii where the discrepancy with MOND is large. \citet{angus10} argue that in the case of a sterile neutrino with a mass of 11 eV the gravitational field of galaxy clusters would be deep enough to retain the sterile neutrinos, while at galaxy scales the neutrinos would free stream from the halo. However, we note that invoking a substantial amount of hot DM affects the formation of structures in the Universe, which renders the expected halo mass function inconsistent with the observations \citep{angus13}. This scenario also invokes at the same time a modification of the gravitational law and a (less abundant) DM component, making it intrinsically more complex. Our analysis of the required missing mass in the standard RAR calibrated on disk galaxies indicates that this component should dominate the mass budget in the inner regions and its total mass should be comparable to the total baryonic mass (see Fig. \ref{fig:model_rar}). However, the mass distribution of the missing component should exhibit an unphysical decreasing trend beyond $\sim0.5R_{500}$, since our data at large radii eventually catch up with the \citet{mcgaugh16} relation. If the mass density of the MOND DM is set to exactly zero beyond the radius where the missing component peaks, our observed gravitational acceleration would need to be biased low by about a factor of 2 at $R_{200}$ to match the predicted acceleration. While not totally impossible, this discrepancy is quite a bit larger than our assessment of systematic uncertainties (see Sect. \ref{sec:syst}) and thus this scenario is disfavored by our data. 

Alternatively, it has been claimed that the MOND characteristic scale $a_0$ may be mass dependent \citep{hodson17}. While our clusters indeed define their own RAR with a scatter $\sigma_{\ln g}=0.21$, the behavior of the recovered RAR in X-COP clusters is quite complex and our model requires two characteristic acceleration scales, $g_1$ and $g_2$ (see Eq. \ref{eq:model_rar}). The fitted value of $g_1$ is close to $a_0$, whereas the second characteristic scale $g_2$ is about an order of magnitude smaller. The $g_{\rm bar}-g_{\rm obs}$ relation steepens again below $g_2$, which cannot be easily explained by a mass dependence of $a_0$.

Overall, our results can be more easily explained in the $\Lambda$CDM scenario. The decomposition of the gravitational field into its various components can be well explained: the prevalence of the stellar content in the innermost regions reflects the collapse of cold baryons to the bottom of the potential well, whereas the somewhat flatter distribution of hot gas with respect to the DM profile can be explained by AGN feedback, which is known to inject nongravitational energy into the surrounding medium \citep[e.g.,][]{lebrun14}. Moreover, a few galaxy evolution models are also able to reproduce the shape of the RAR in rotationally supported galaxies in the $\Lambda$CDM context \citep{navarro17,ludlow17b,dutton19}. Therefore, it is conceivable that the RAR may arise as a byproduct of the galaxy formation process as an interplay between baryonic and total acceleration.

\section{Conclusions}

In this paper we have performed a detailed analysis of the gravitational field of X-COP galaxy clusters over the radial range $[0.01-2]R_{500}$ from a combination of deep \emph{XMM-Newton} and \emph{Planck} data. Our results can be summarized as follows:

\begin{itemize}
    \item We introduced a novel framework for the reconstruction of hydrostatic mass profiles from X-ray and/or SZ data, which we distribute in the form of the public Python package \texttt{hydromass}. Hydrostatic equilibrium profiles can be reconstructed assuming a mass model, from a parametric pressure profile, or in a nonparametric way as a linear combination of log-normal functions. Our framework improves over our previous works (\citealt{ettori19}; the differences in the mass values are less than 10 per cent; see Table \ref{tab:mass_ratio}) as it implements PSF deconvolution, affecting in particular the inner regions, and fits with the Einasto mass model, where the parameter $\alpha$ is left free to vary. The various methods discussed here are integrated within a common framework that includes an efficient Bayesian optimization scheme.
    
    \item We validated our method extensively using a set of mock \emph{XMM-Newton} observations of a synthetic NFW cluster in HSE. Our mock observations include a wide range of instrumental effects, such as effective area, energy redistribution, vignetting, PSF convolution, and a sophisticated background model. The mock observations were analyzed using the same pipeline as the actual observations. Our code was found to reproduce the input profile very accurately, both for the 3D temperature profile and the hydrostatic mass (see Fig. \ref{fig:mock}).
    
    \item Applying our method to the 12 X-COP galaxy clusters, we find that the NFW and Einasto profiles both provide a very good representation of the average mass profiles over the entire radial range, in agreement with the predictions of the $\Lambda$CDM paradigm. The additional flexibility afforded by the Einasto model allows it to trace the individual profiles more closely than the NFW in the innermost and outermost regions.
    
    \item The very high statistical quality of our data allows us to measure the NFW concentration of individual systems with typical uncertainties of a few per cent (see Fig. \ref{fig:c-m}). Modeling the relation between concentration and mass in our sample, we find an average concentration $c_{200}=3.69\pm0.39$ at $M_{200}\approx10^{15}M_\odot$, in remarkable agreement with $\Lambda$CDM predictions. The intrinsic scatter of the mass-concentration relation was found to be $\sigma_{\ln c_{200}} = 0.37_{-0.07}^{+0.11}$, which again is consistent with the expectations of N-body simulations of structure formation.
    
    \item For a subset of five systems, we decomposed the gravitational field into its baryonic and DM components. We found that the stellar content of the BCG dominates the gravitational field in the innermost regions ($<0.02R_{500}$), which can be explained by the collapse of cold baryons to the bottom of the potential well. The contribution of the stellar mass decreases sharply with radius, such that we observe a depletion of baryons in the range $[0.1-0.5]R_{500}$. The hot ICM dominates the baryonic mass beyond $\sim0.1R_{500}$ and exhibits a somewhat shallower profile than the DM. Our systems reach the cosmic baryon fraction at the virial radius. We note a slight excess at large radii with respect to the cosmic baryon fraction, which we interpret as evidence for a mild contribution of nonthermal pressure ($10-20\%$) at $R_{200}$. 
    
    \item We studied in detail the relation between observed and baryonic acceleration (RAR) in the five systems for which the baryonic content can be completely characterized (Fig. \ref{fig:rar}). The recovered relation deviates from the relation observed in spiral galaxies \citep{mcgaugh16} at high significance, with the observed acceleration exceeding the expected value by a factor of $\sim5$ around $g_{\rm bar}=2\times10^{-11}$ m s$^{-2}$. We found that the RAR in galaxy clusters exhibits a complex shape with three distinct regimes delimited by two characteristic scales, which makes it difficult to reconcile with the MOND paradigm.
\end{itemize}

Application of the presented framework to larger samples \citep[e.g., CHEX-MATE,][]{chex-mate} will allow us to study in detail the shape of galaxy cluster mass profiles and test the validity of the HSE assumption. 

\begin{acknowledgements}
SE acknowledges financial contribution from the contracts ASI-INAF Athena 2019-27-HH.0,
``Attivit\`a di Studio per la comunit\`a scientifica di Astrofisica delle Alte Energie e Fisica Astroparticellare''
(Accordo Attuativo ASI-INAF n. 2017-14-H.0), INAF mainstream project 1.05.01.86.10, and
from the European Union's Horizon 2020 Programme under the AHEAD2020 project (grant agreement n. 871158). SIL is supported in part by the National Research Foundation of South Africa (NRF Grant Number: 120850). Opinions, findings and conclusions or recommendations expressed in this publication is that of the author(s), and the NRF accepts no liability whatsoever in this regard.
\end{acknowledgements}

\bibliographystyle{aa}
\bibliography{hydromass}

\appendix

\section{Fitting results for individual clusters}

\begin{figure*}
        \centerline{\resizebox{\hsize}{!}{\vbox{
                                \includegraphics[width=0.45\textwidth]{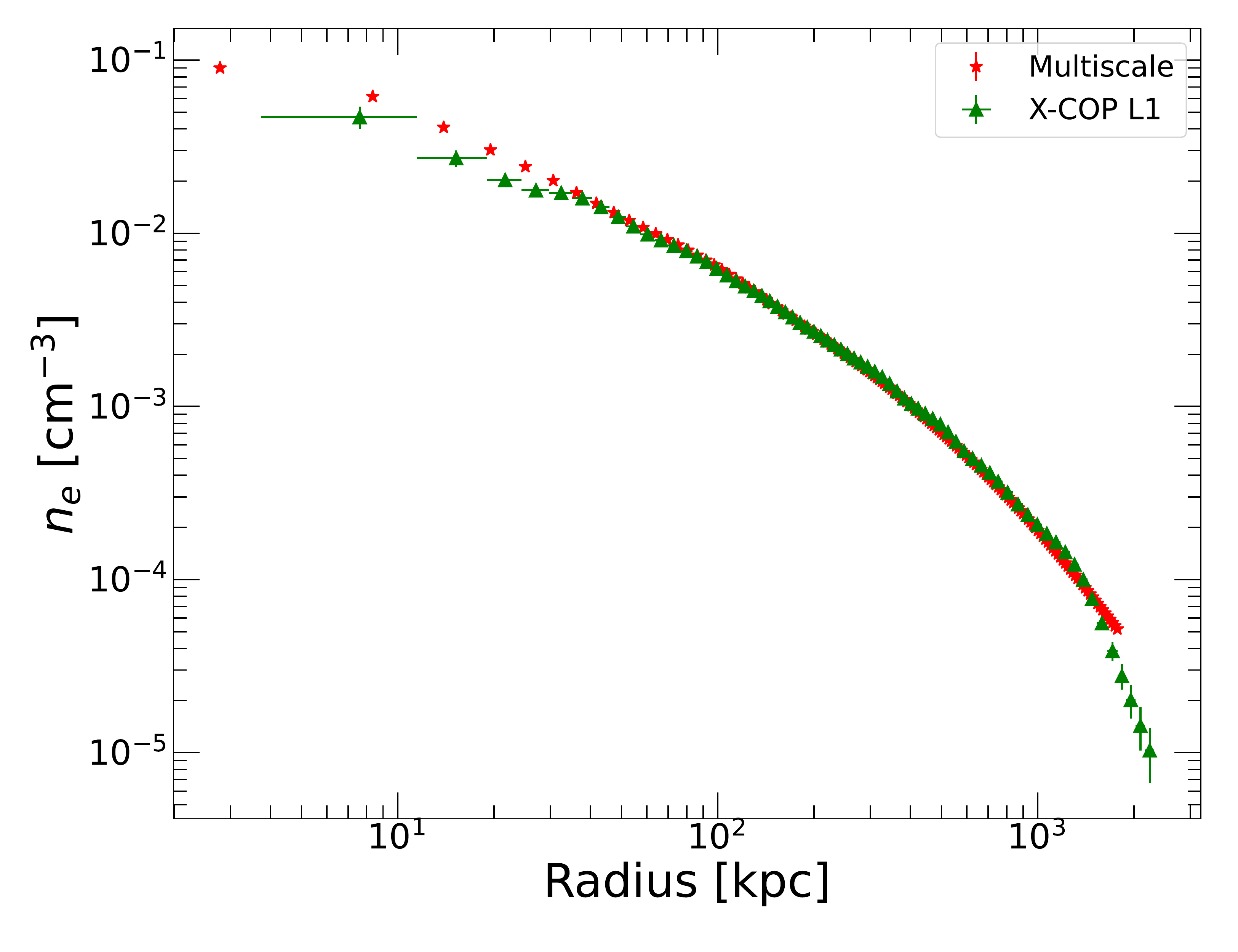}
                                \includegraphics[width=0.45\textwidth]{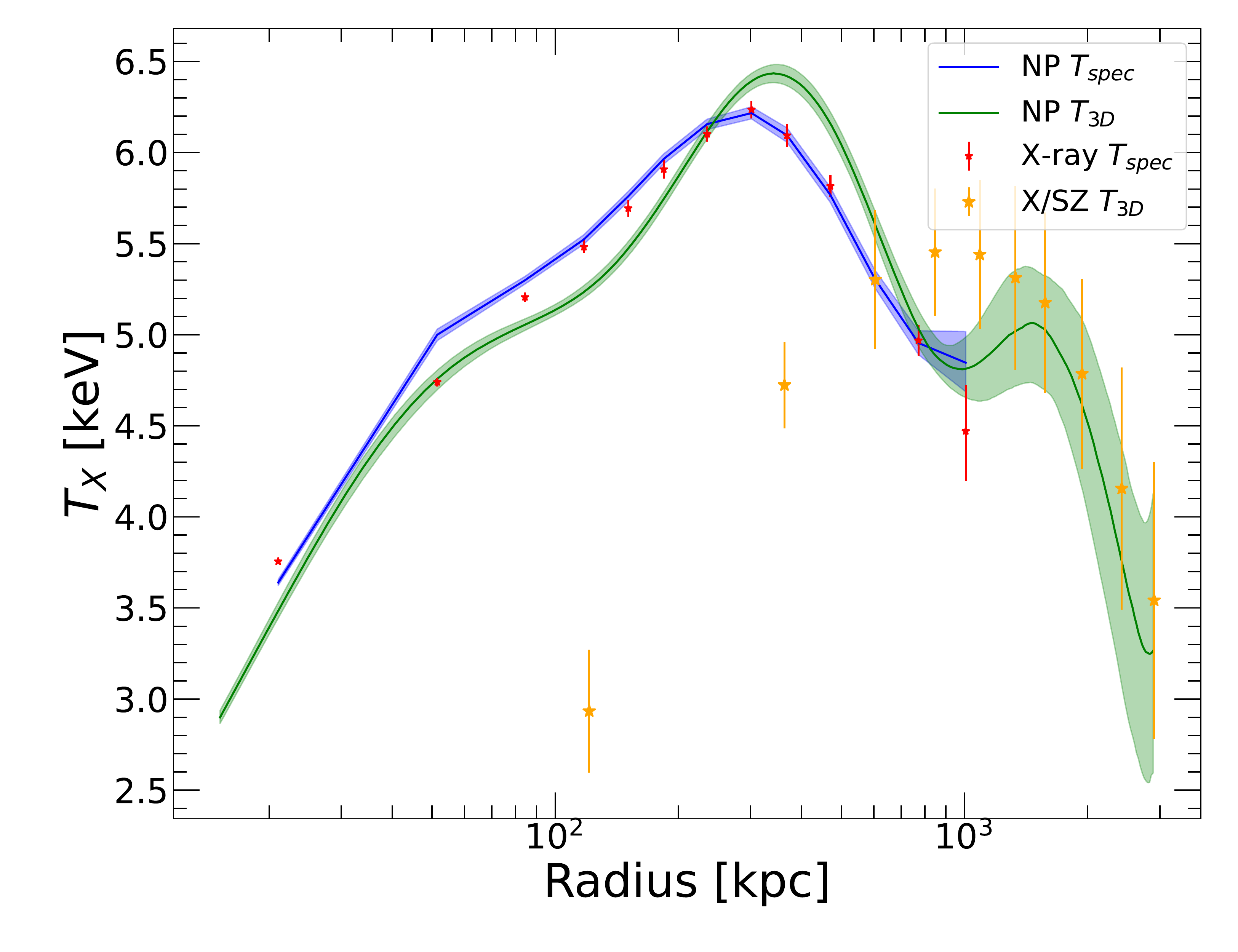}\\
                                
                                \includegraphics[width=0.45\textwidth]{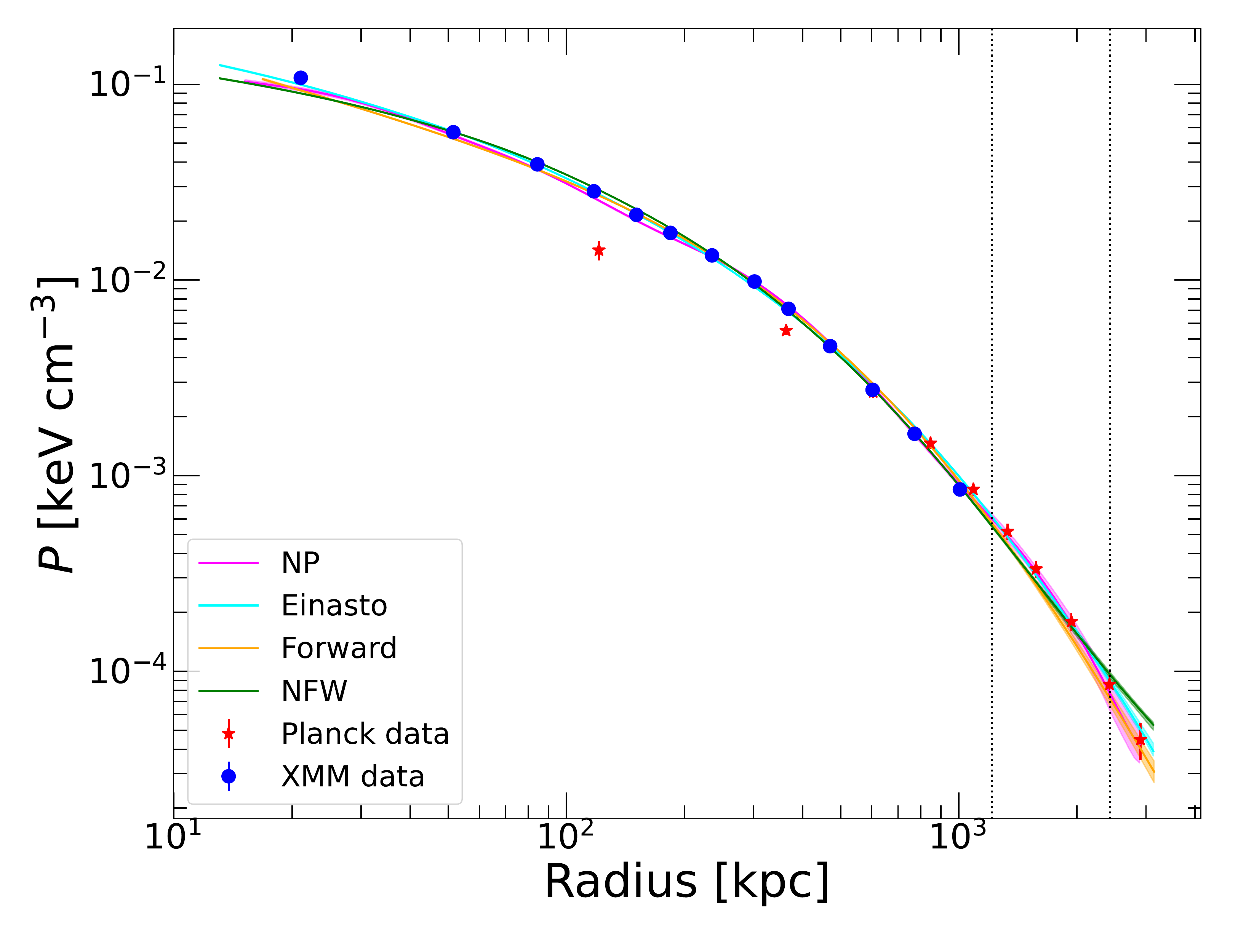}
                                \includegraphics[width=0.45\textwidth]{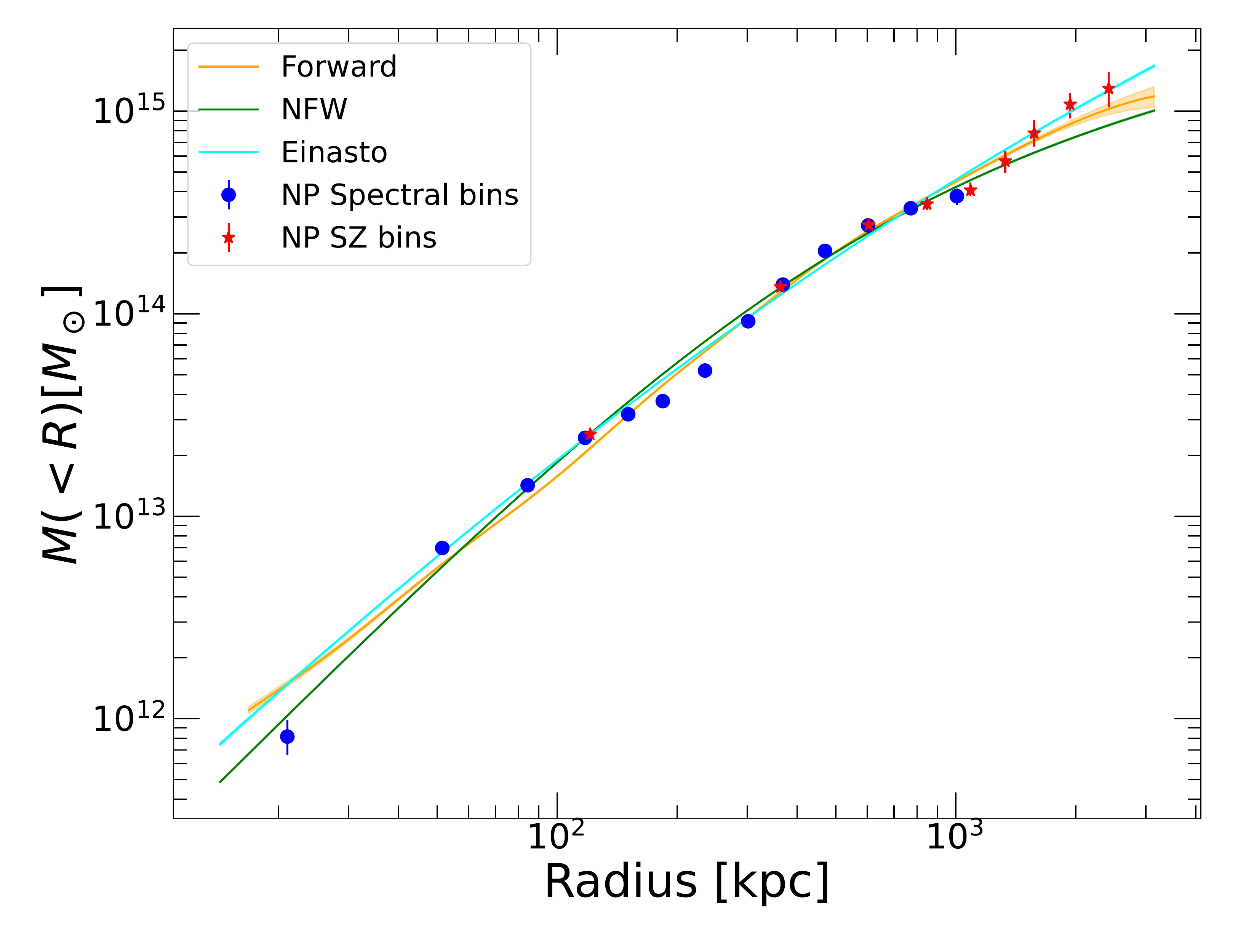}\\
                                
                                \includegraphics[width=0.45\textwidth]{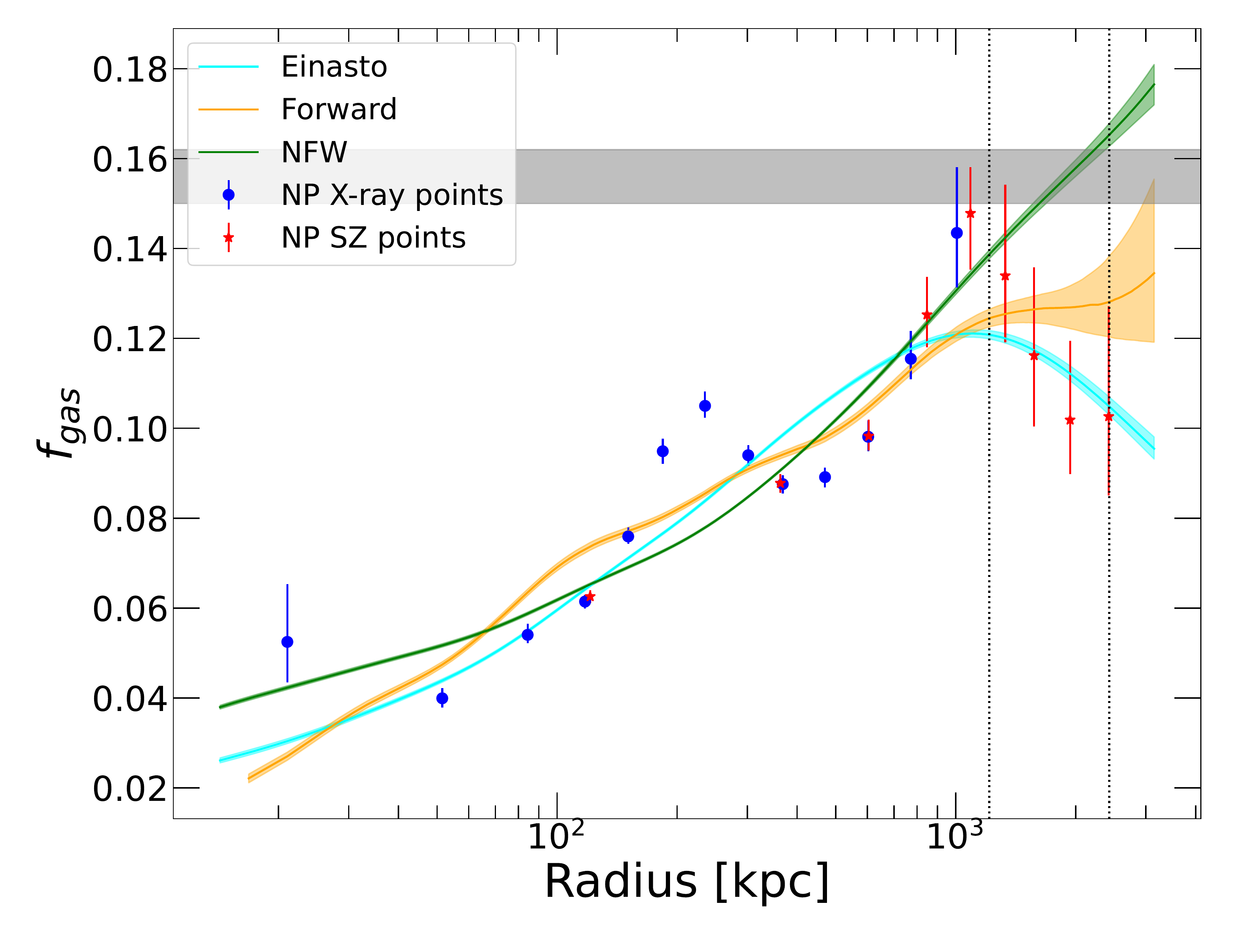}
                                \hspace{0.5cm}
                                \includegraphics[width=0.37\textwidth]{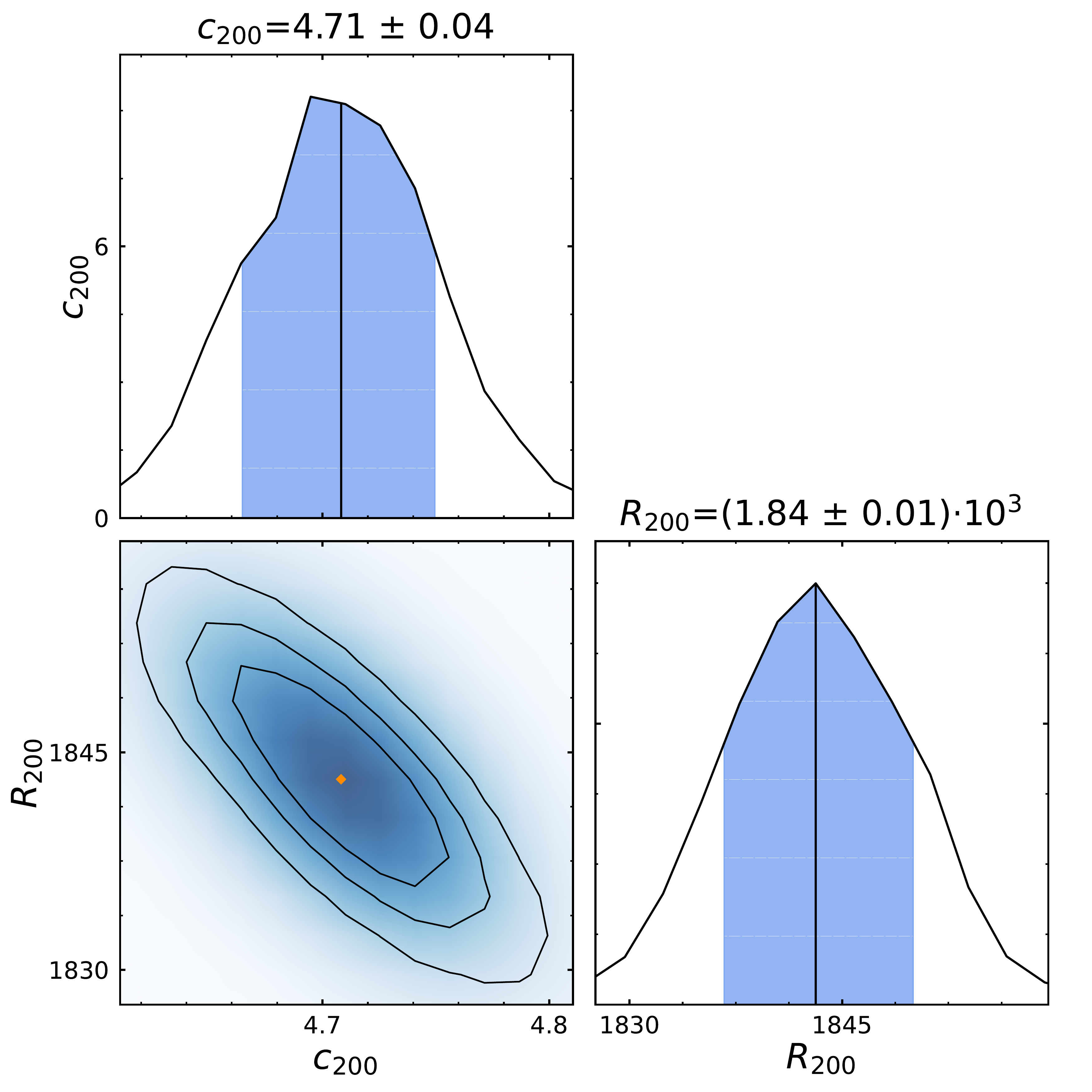}
        }}}
        \caption{Same as Fig. \ref{fig:a1795} but for A85. } 
\end{figure*}

\begin{figure*}
        \centerline{\resizebox{\hsize}{!}{\vbox{
                                \includegraphics[width=0.45\textwidth]{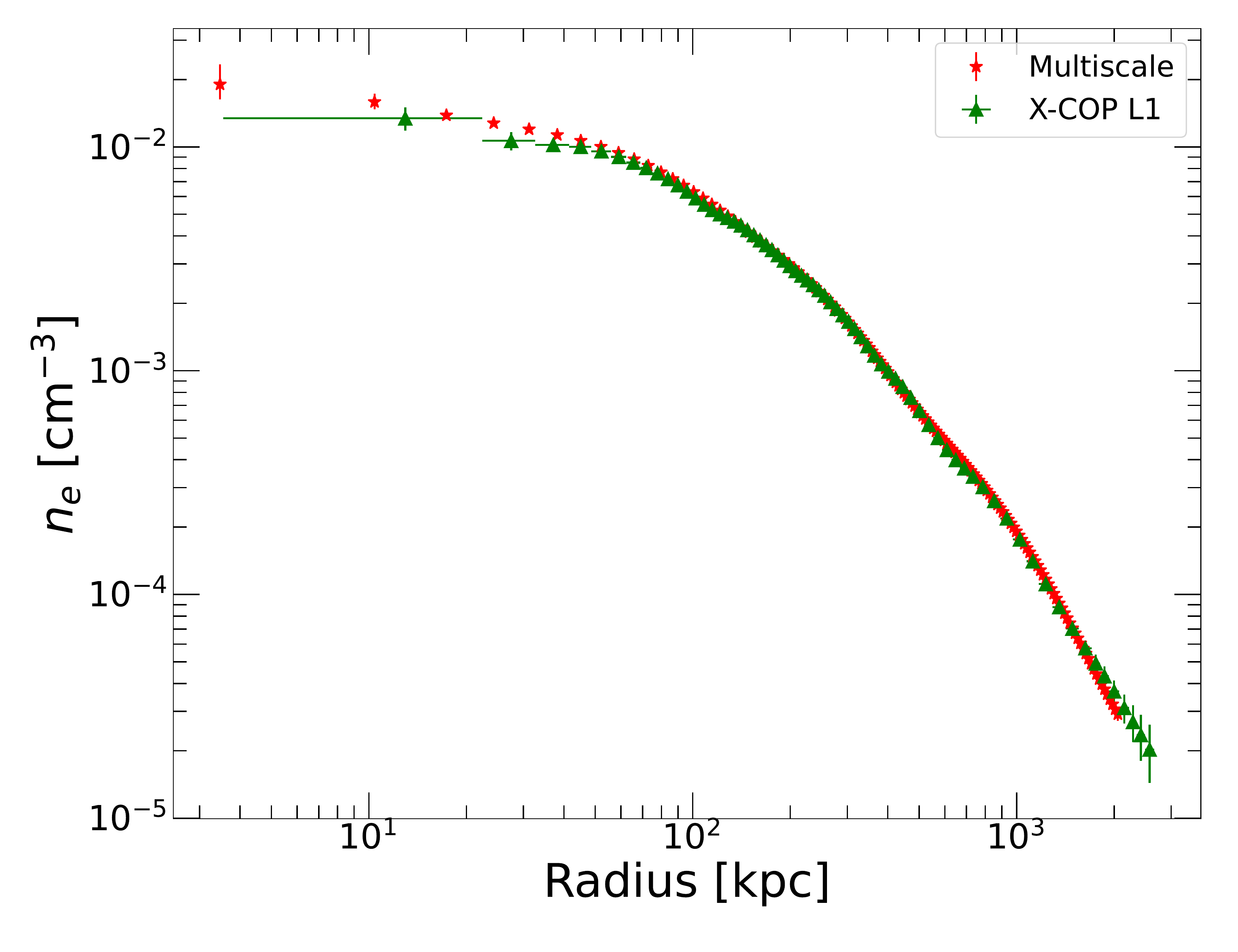}
                                \includegraphics[width=0.45\textwidth]{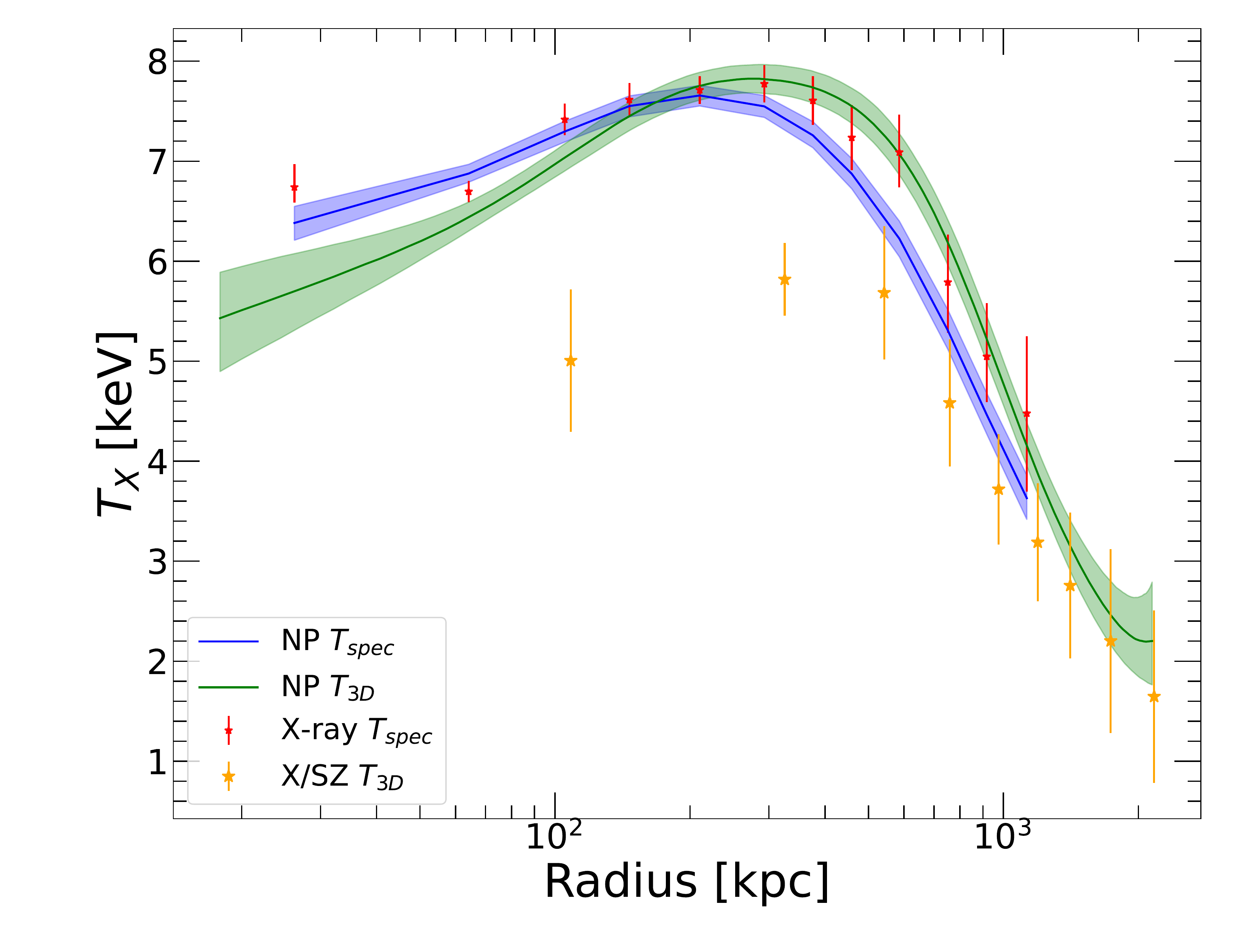}\\
                                
                                \includegraphics[width=0.45\textwidth]{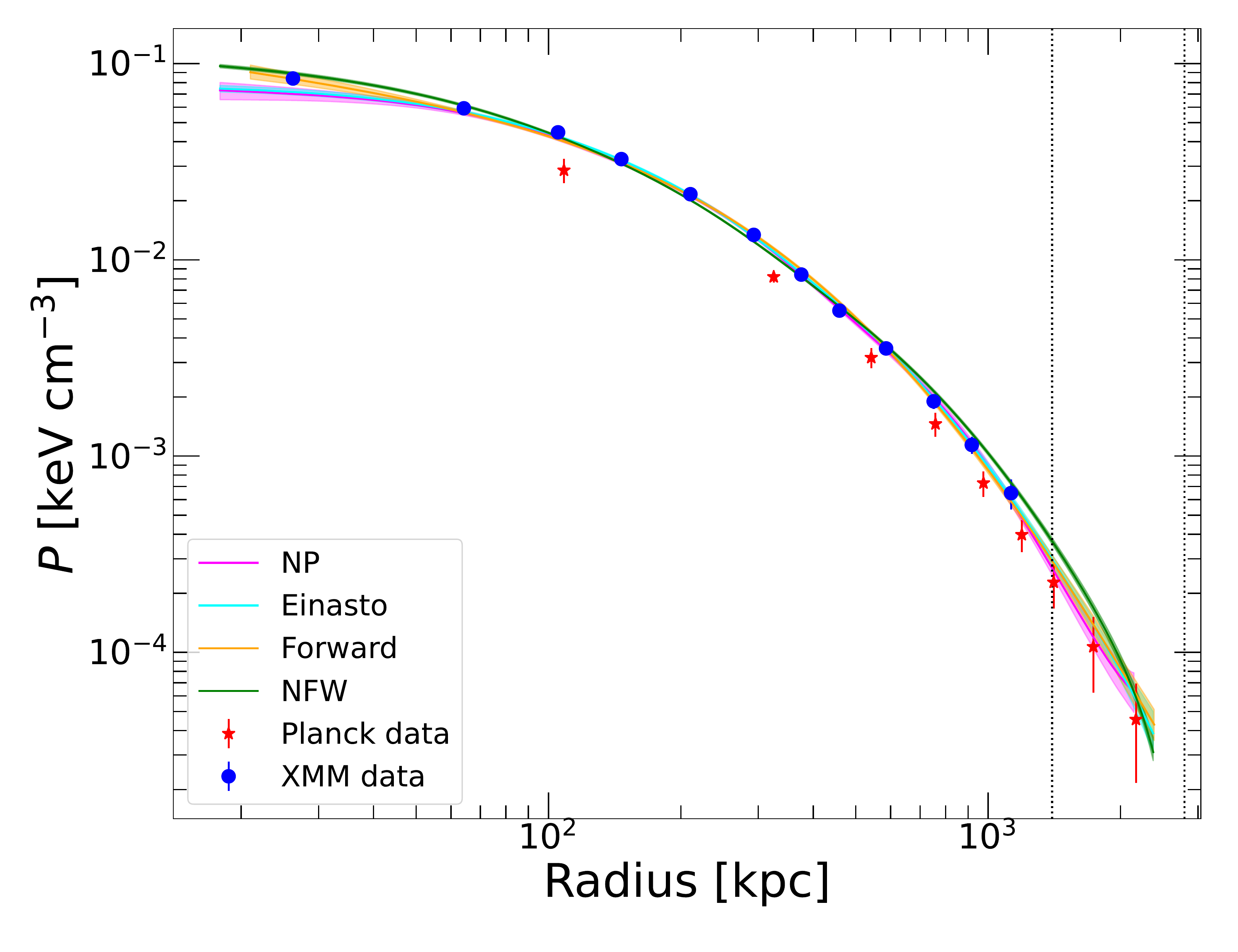}
                                \includegraphics[width=0.45\textwidth]{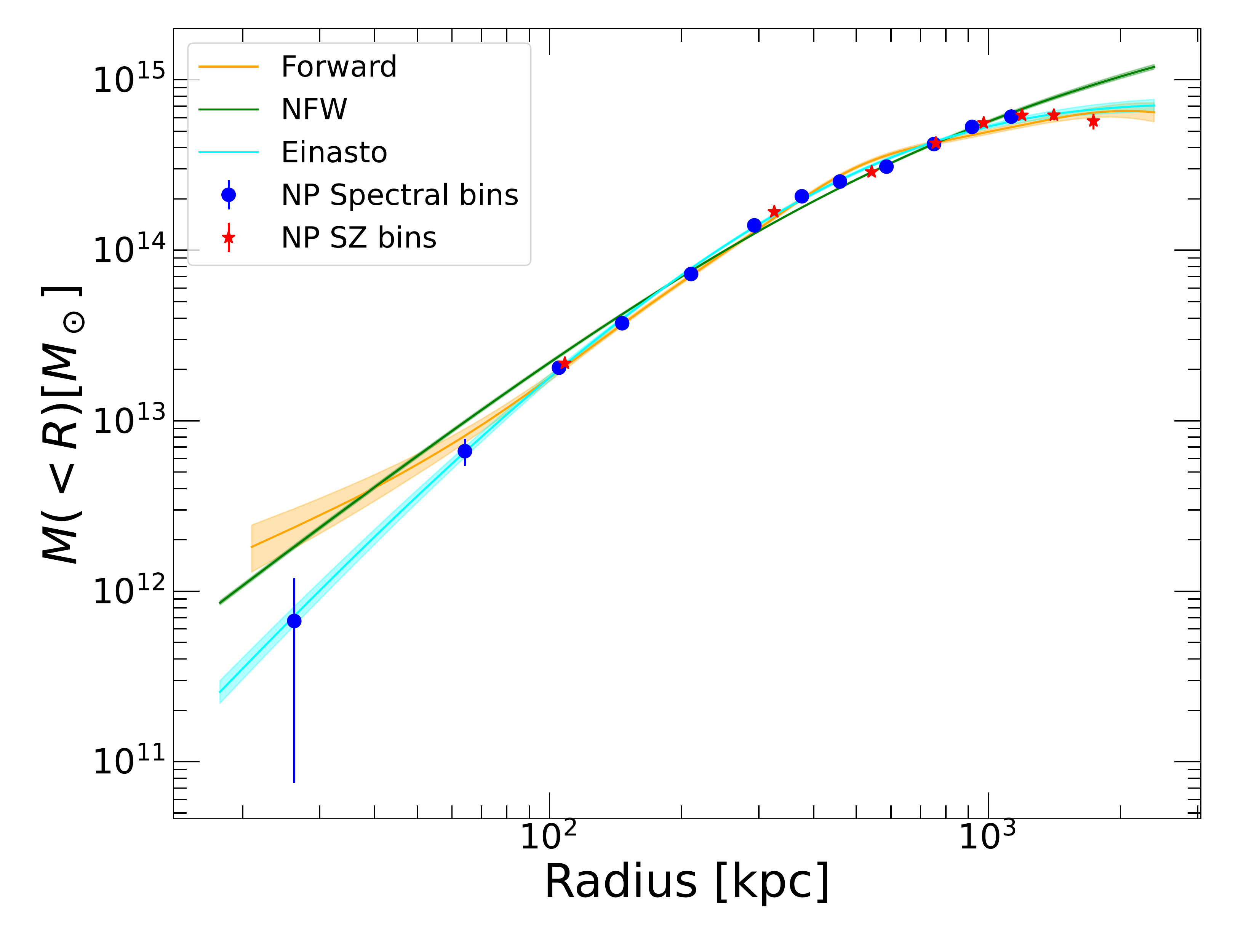}\\
                                
                                \includegraphics[width=0.45\textwidth]{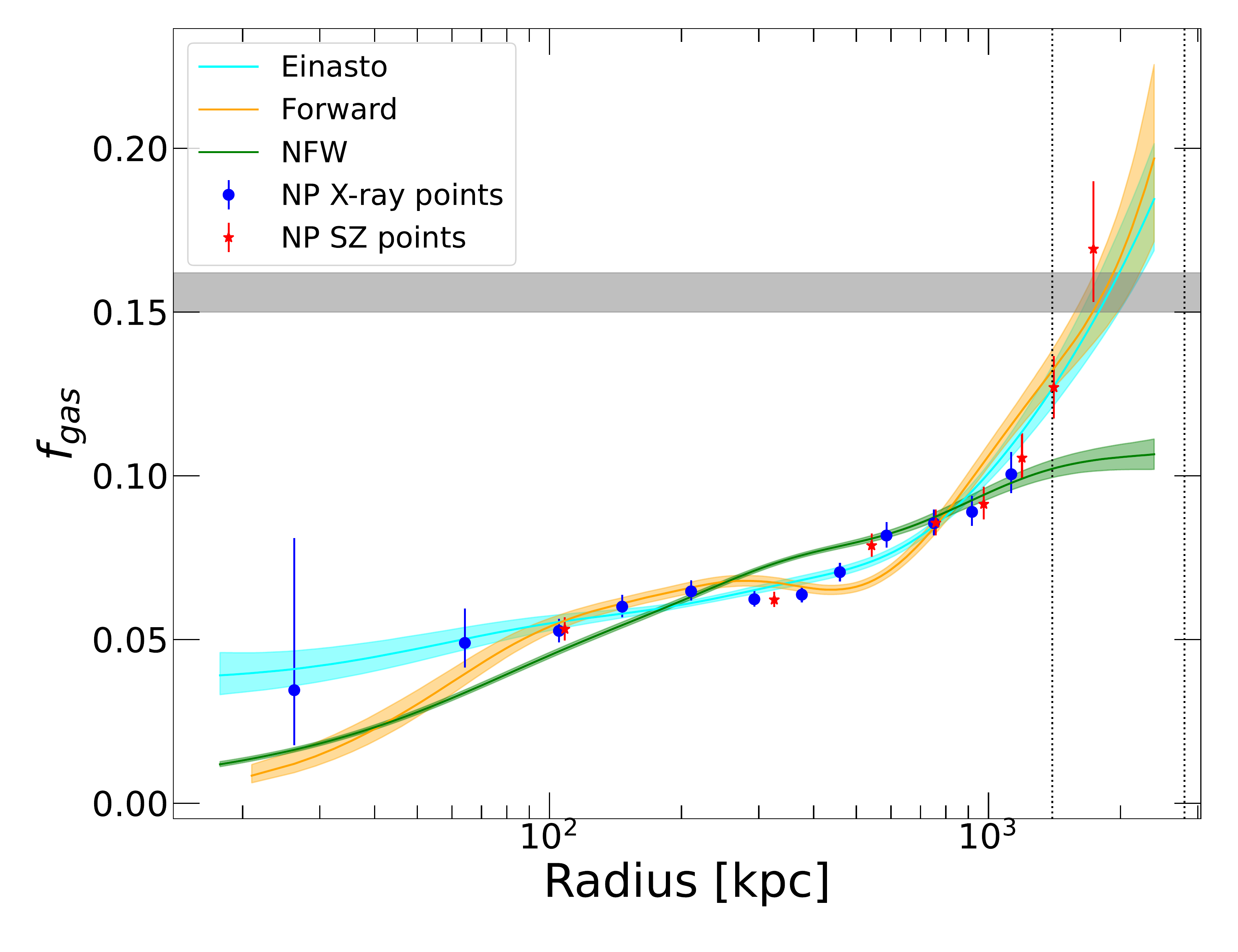}
                                \hspace{0.5cm}
                                \includegraphics[width=0.37\textwidth]{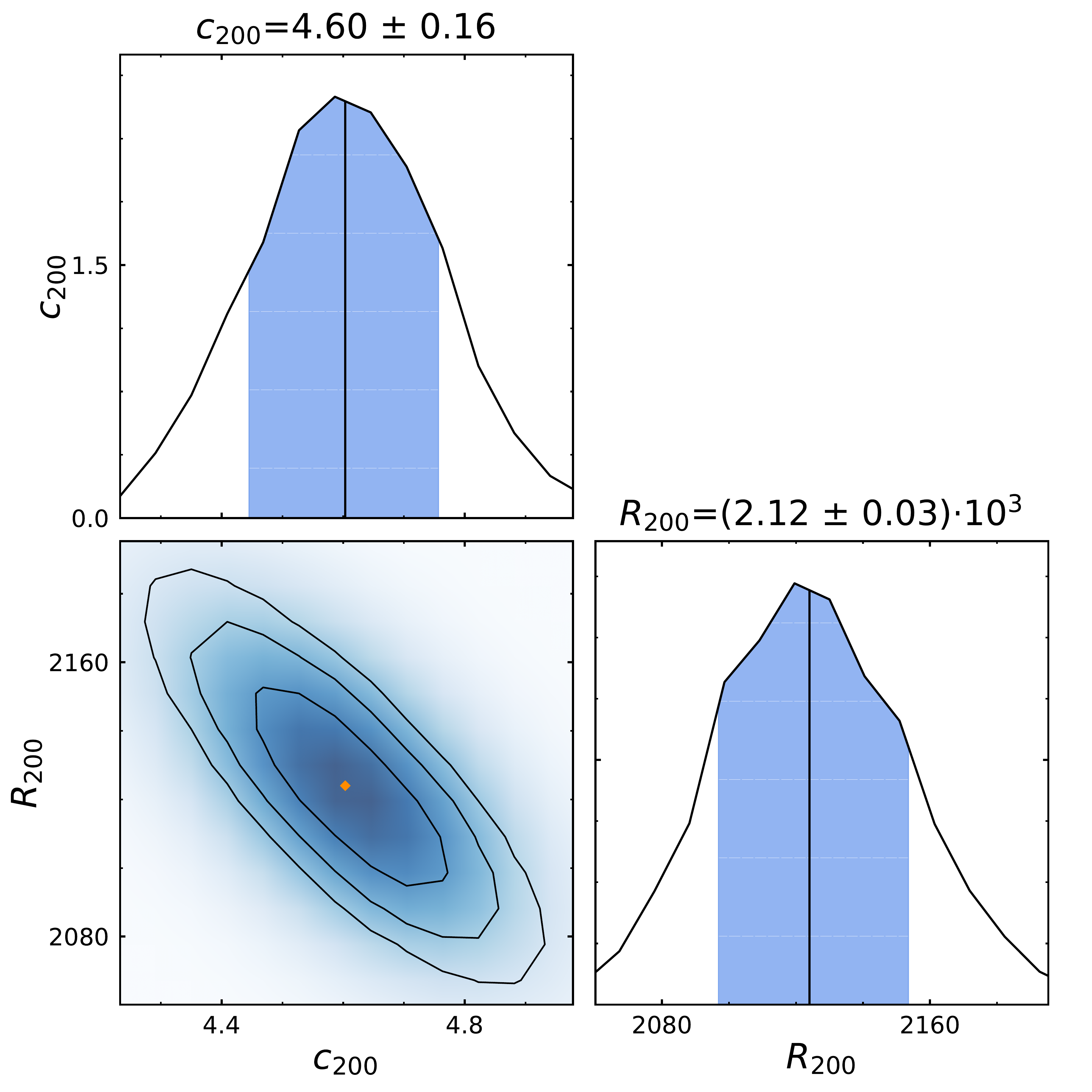}
        }}}
        \caption{Same as Fig. \ref{fig:a1795} but for A644. } 
\end{figure*}

\begin{figure*}
        \centerline{\resizebox{\hsize}{!}{\vbox{
                                \includegraphics[width=0.45\textwidth]{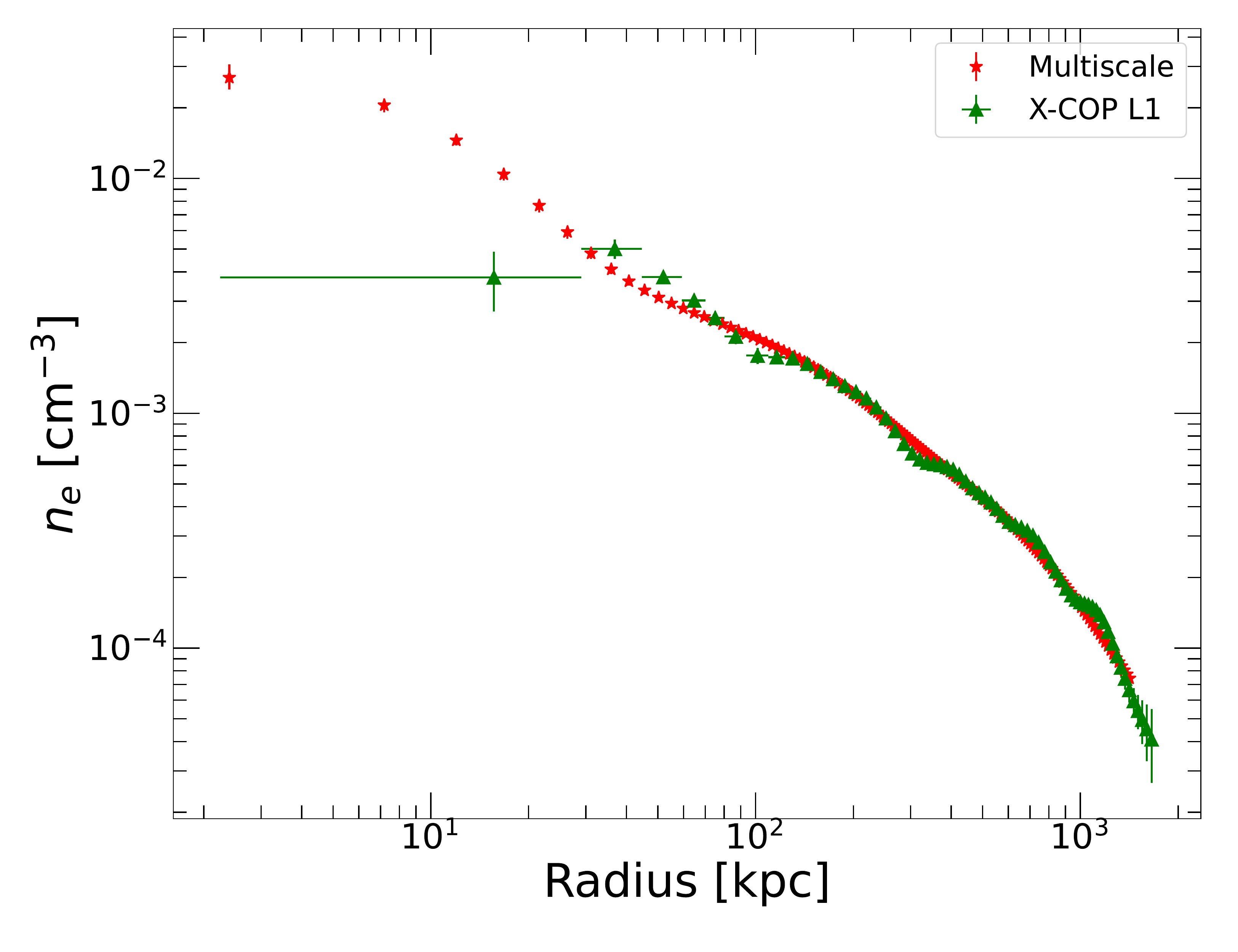}
                                \includegraphics[width=0.45\textwidth]{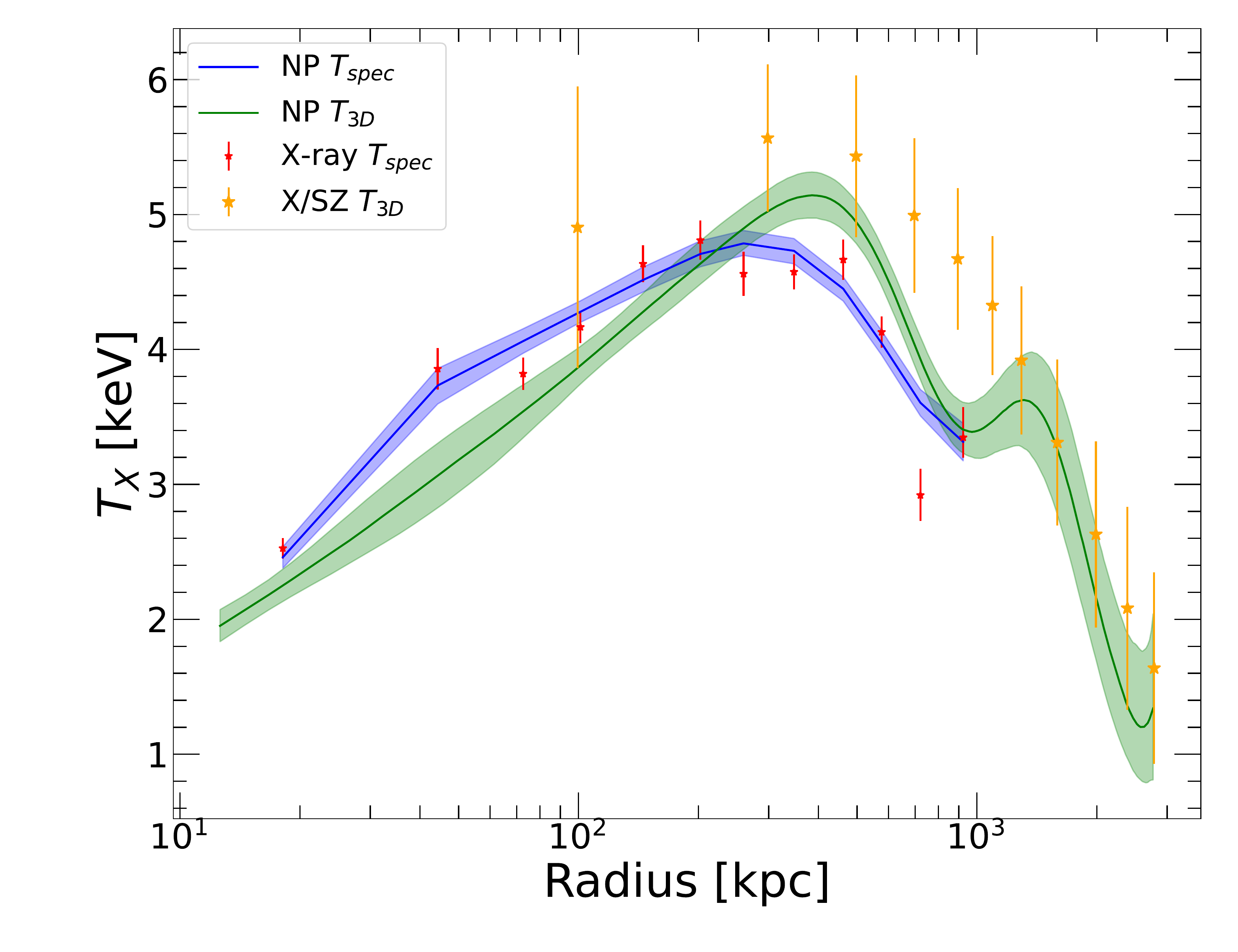}\\
                                
                                \includegraphics[width=0.45\textwidth]{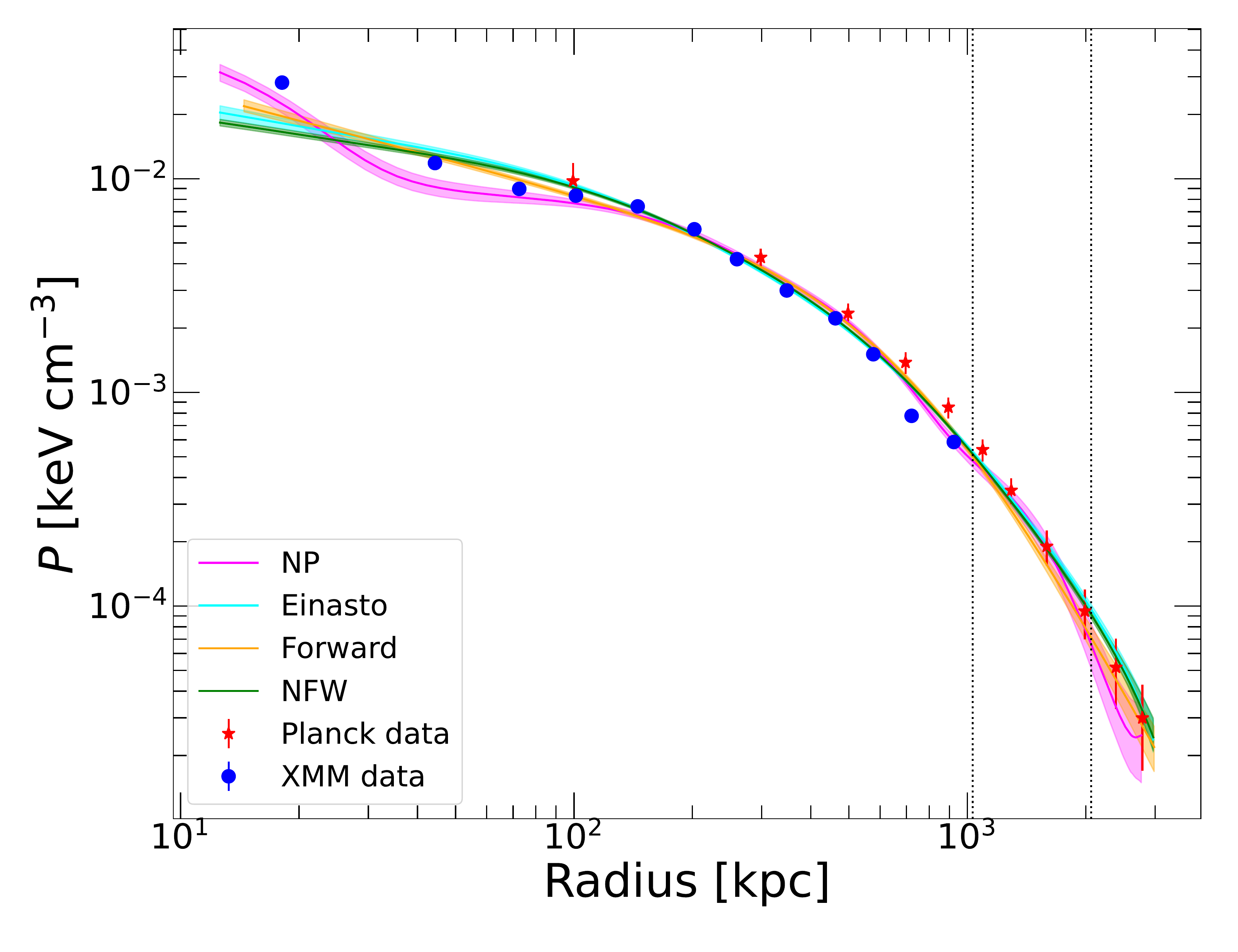}
                                \includegraphics[width=0.45\textwidth]{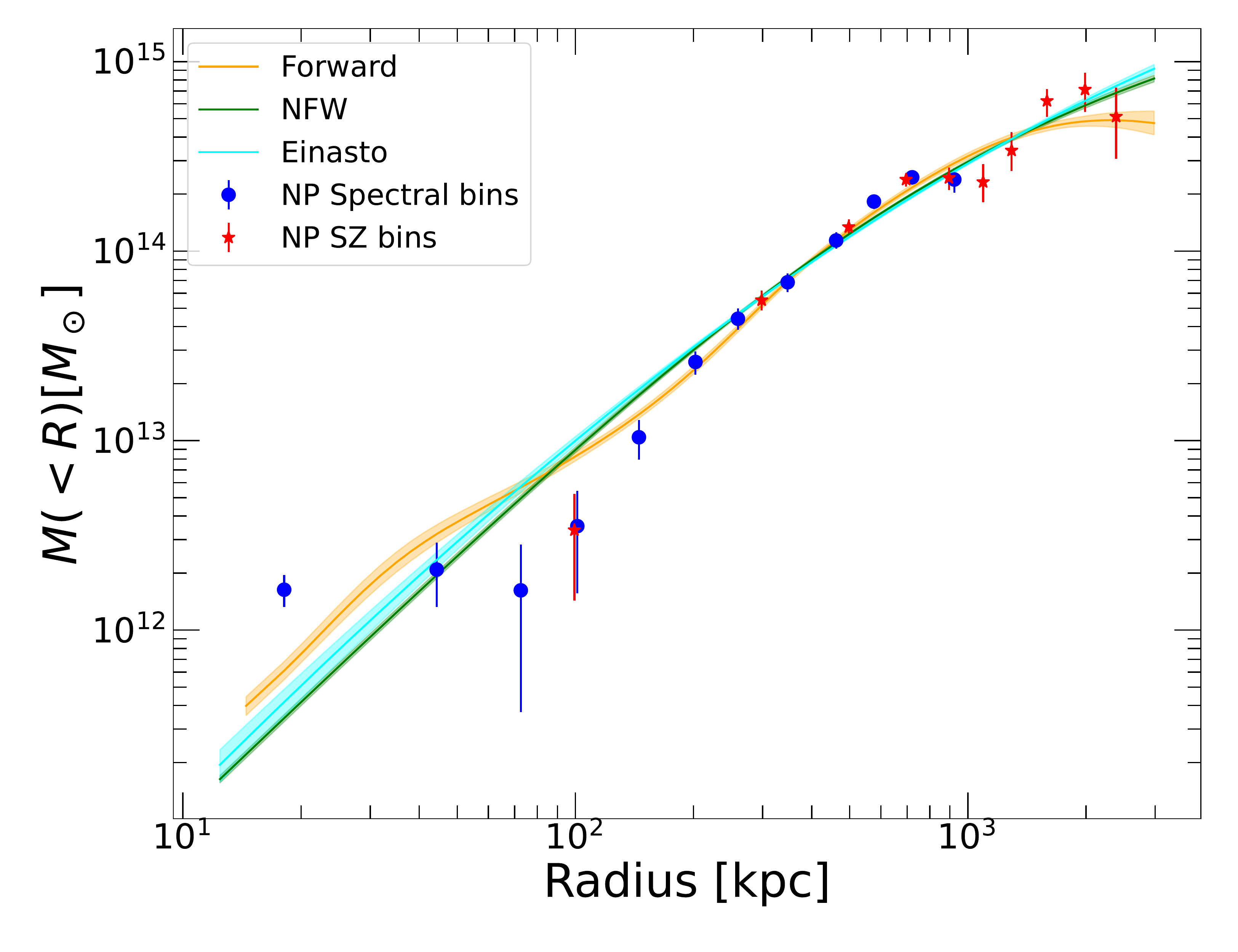}\\
                                
                                \includegraphics[width=0.45\textwidth]{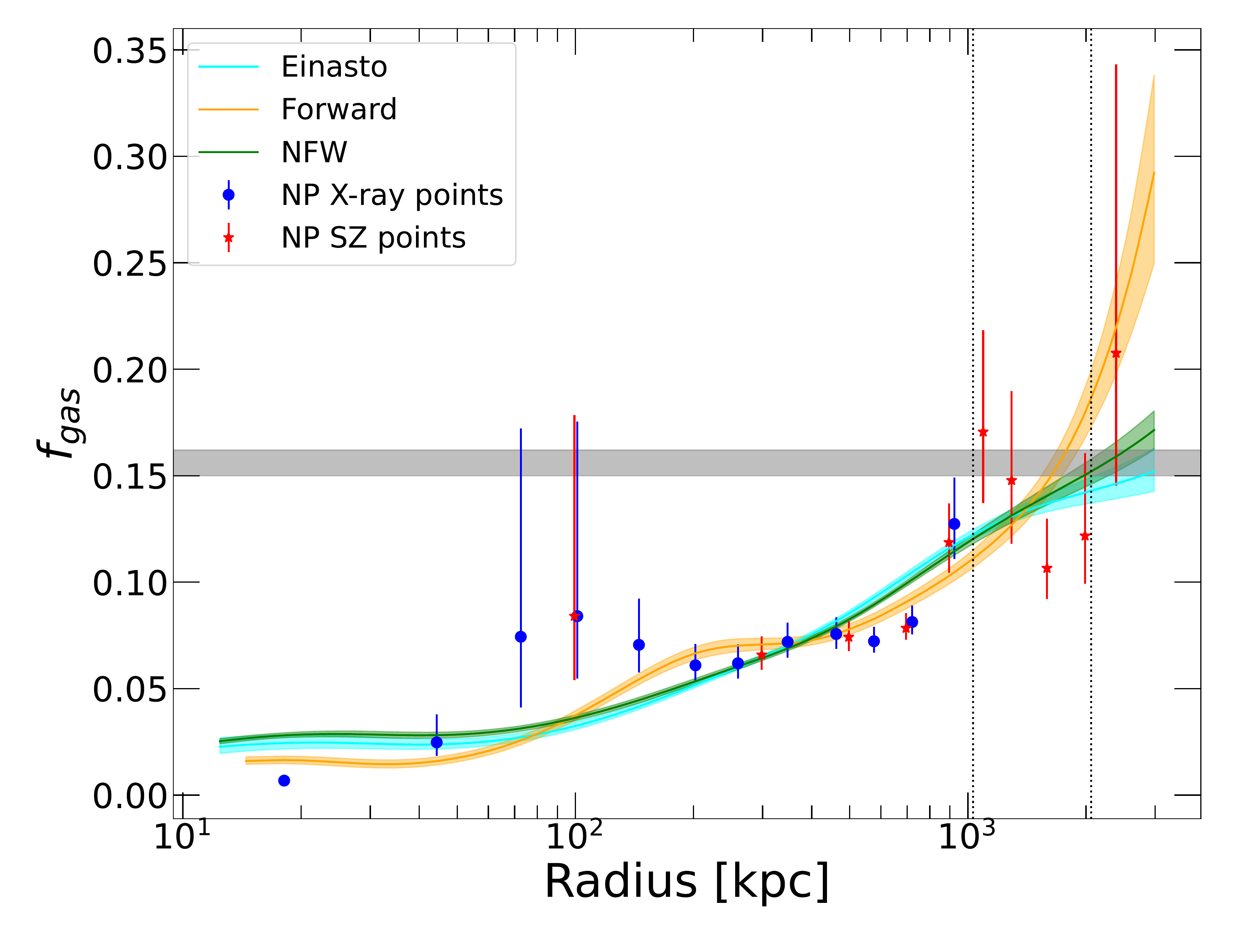}
                                \hspace{0.5cm}
                                \includegraphics[width=0.37\textwidth]{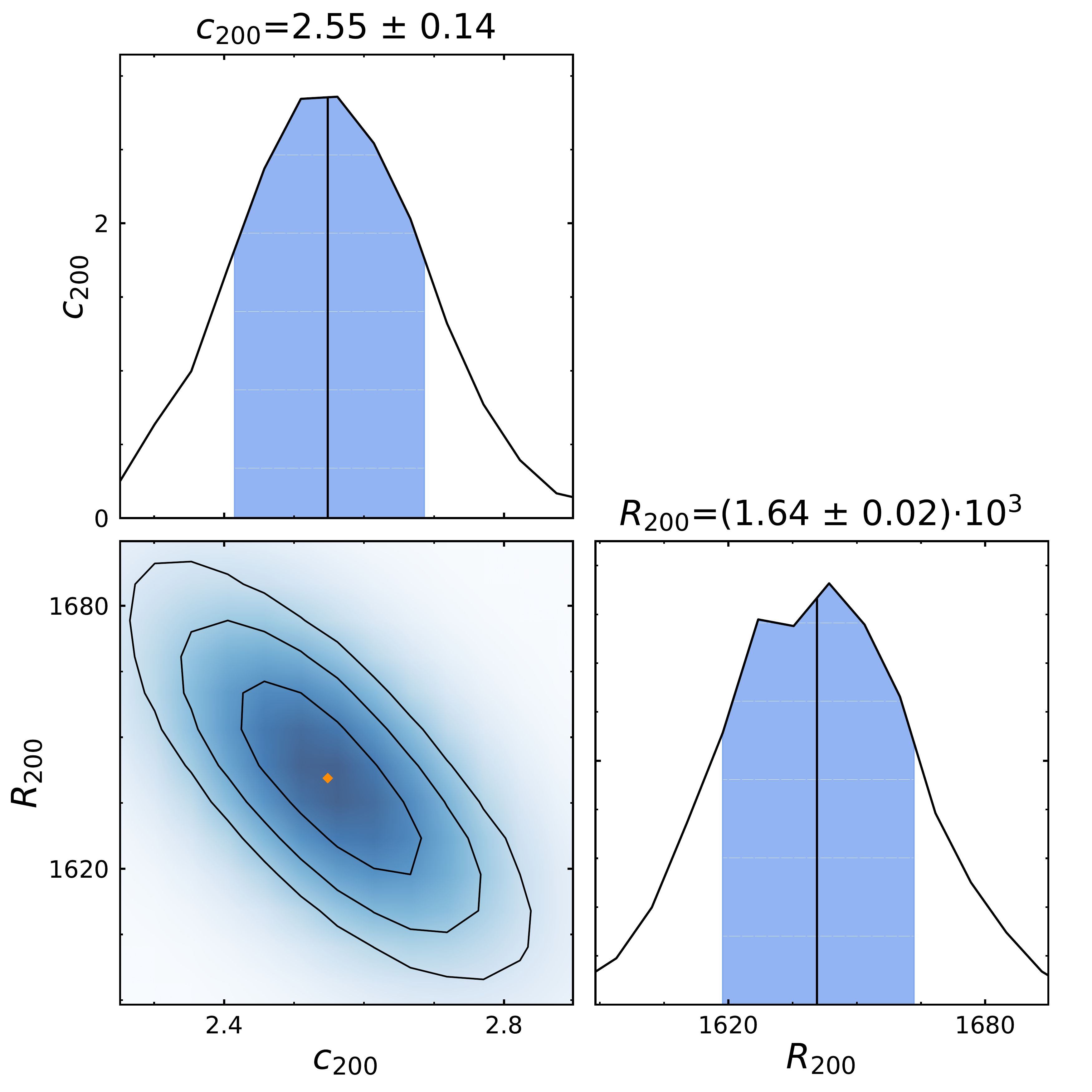}
        }}}
        \caption{Same as Fig. \ref{fig:a1795} but for A1644. } 
\end{figure*}

\begin{figure*}
        \centerline{\resizebox{\hsize}{!}{\vbox{
                                \includegraphics[width=0.45\textwidth]{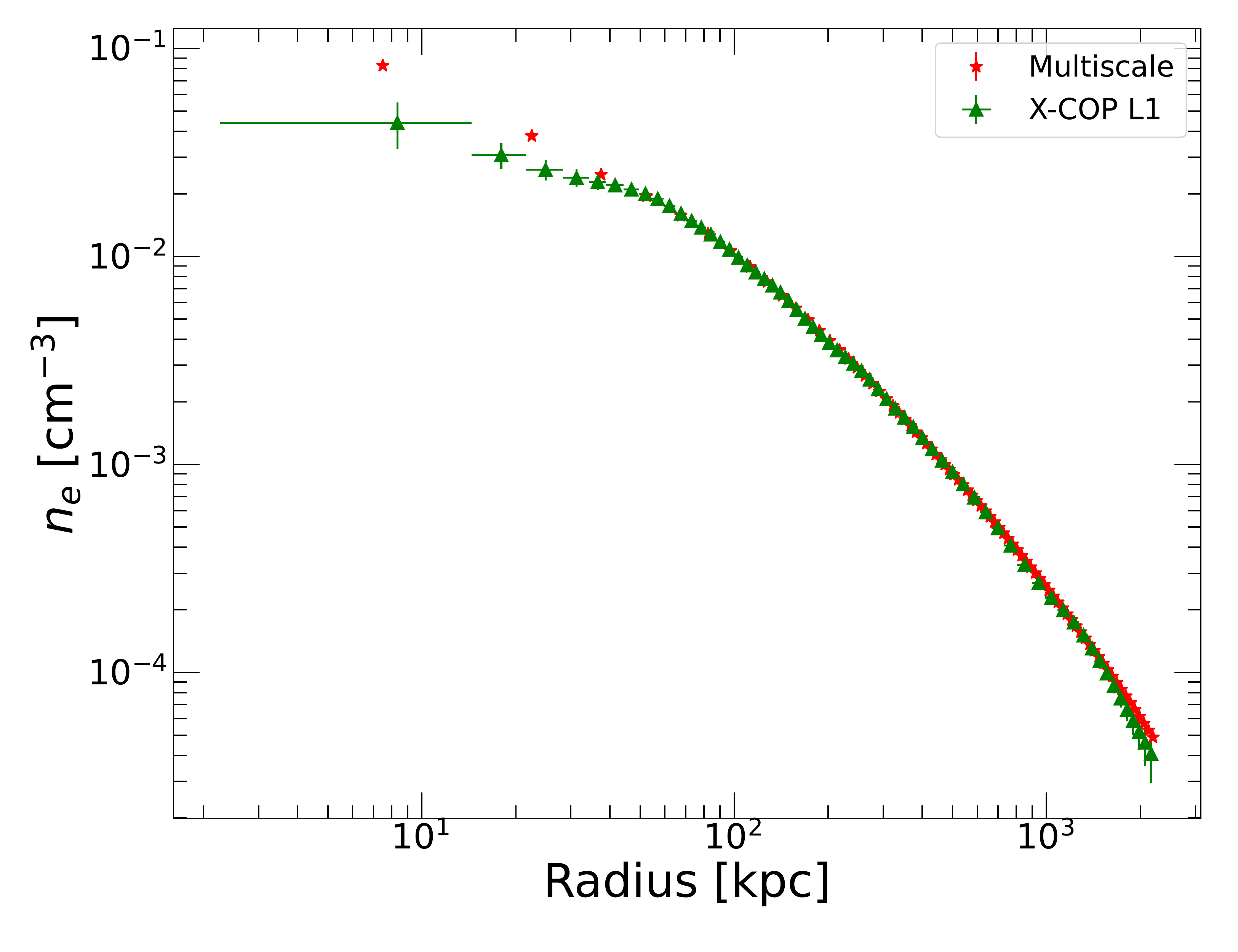}
                                \includegraphics[width=0.45\textwidth]{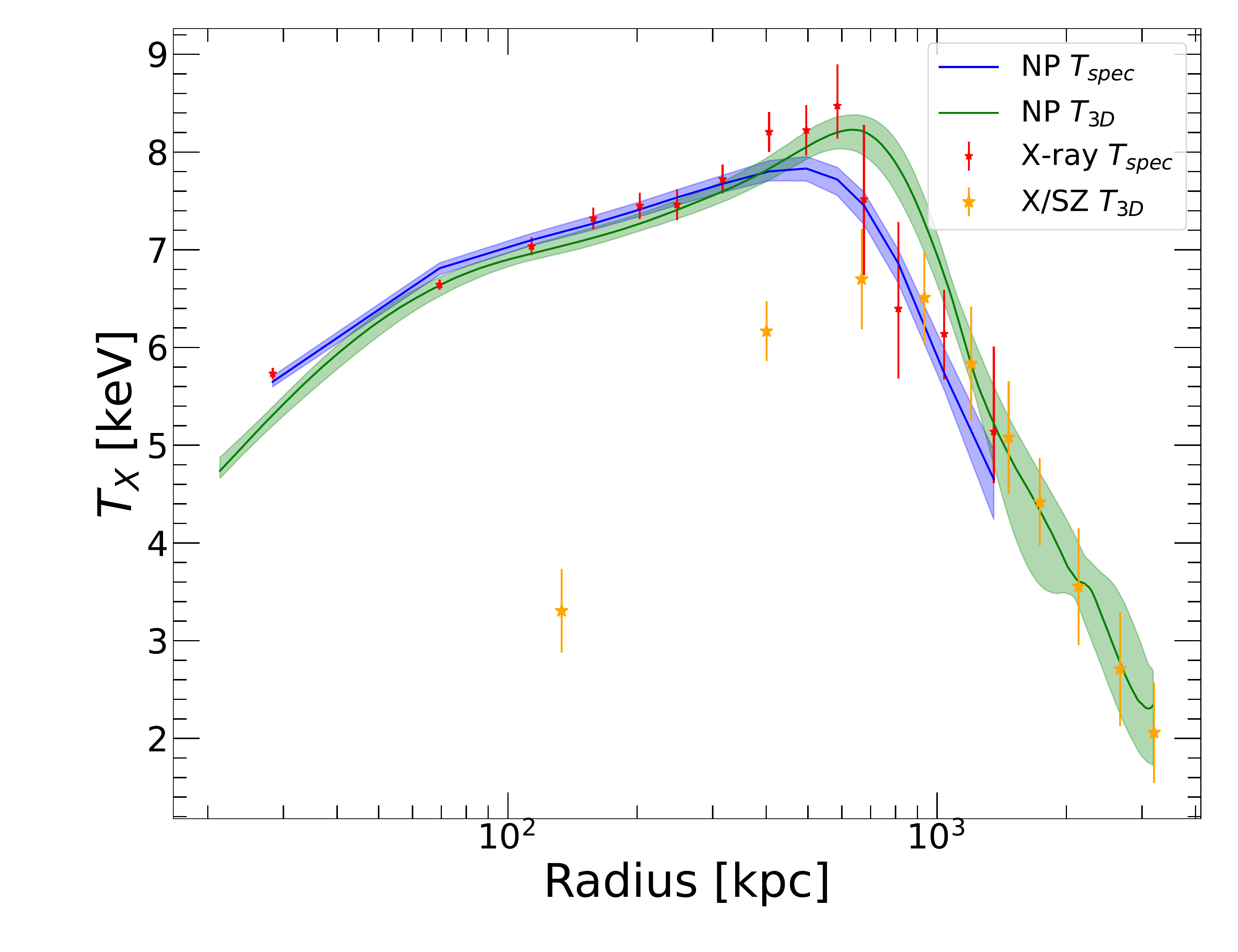}\\
                                
                                \includegraphics[width=0.45\textwidth]{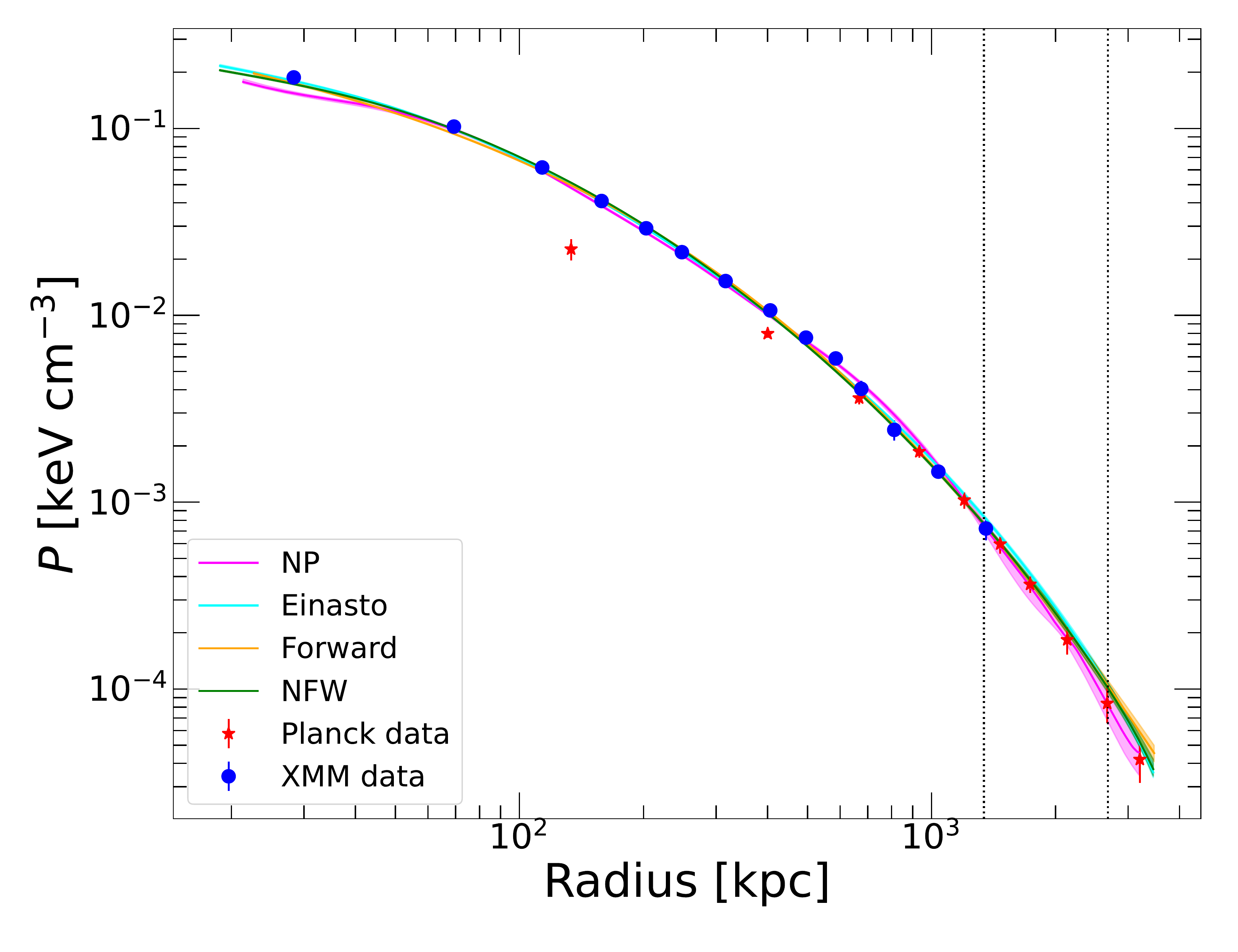}
                                \includegraphics[width=0.45\textwidth]{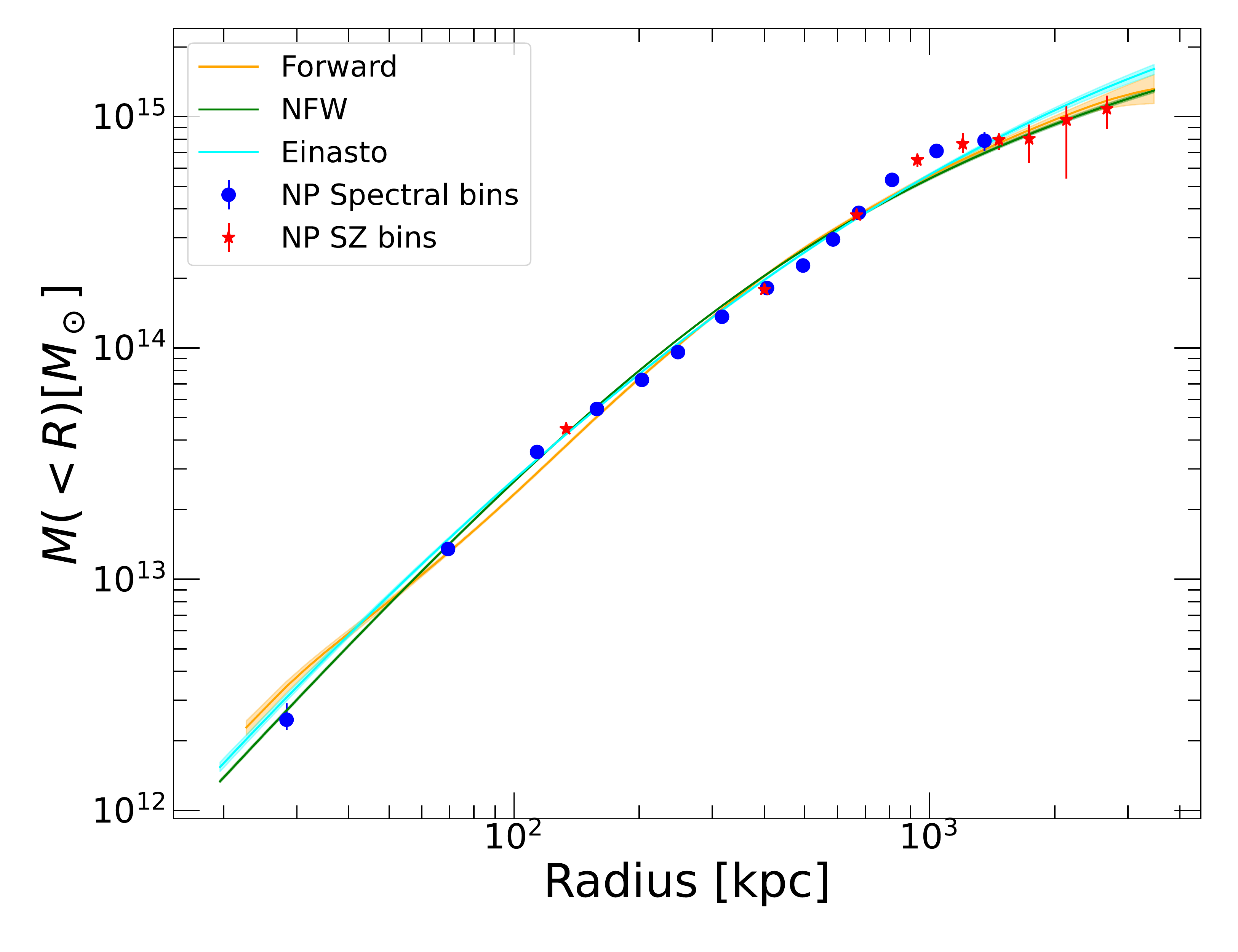}\\
                                
                                \includegraphics[width=0.45\textwidth]{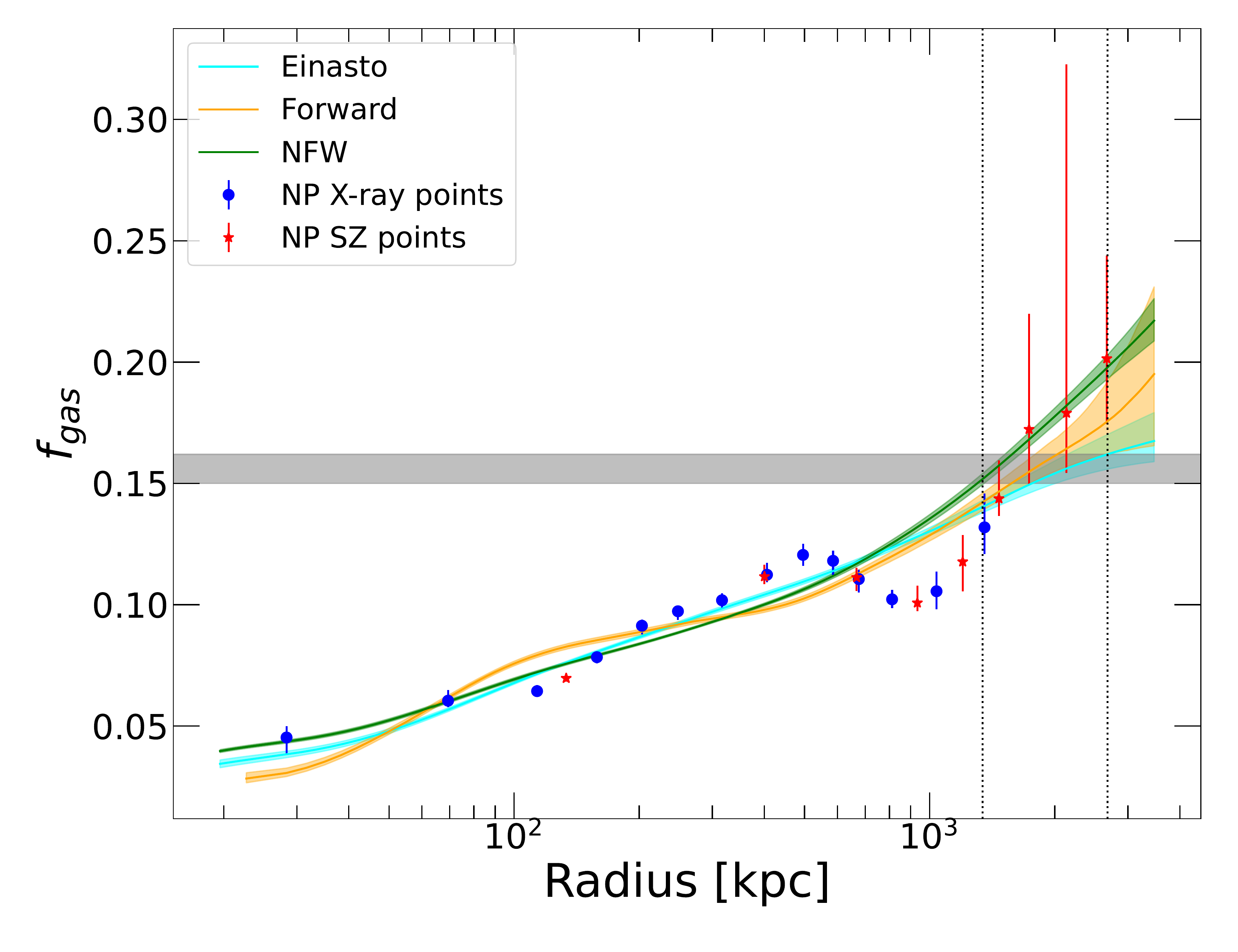}
                                \hspace{0.5cm}
                                \includegraphics[width=0.37\textwidth]{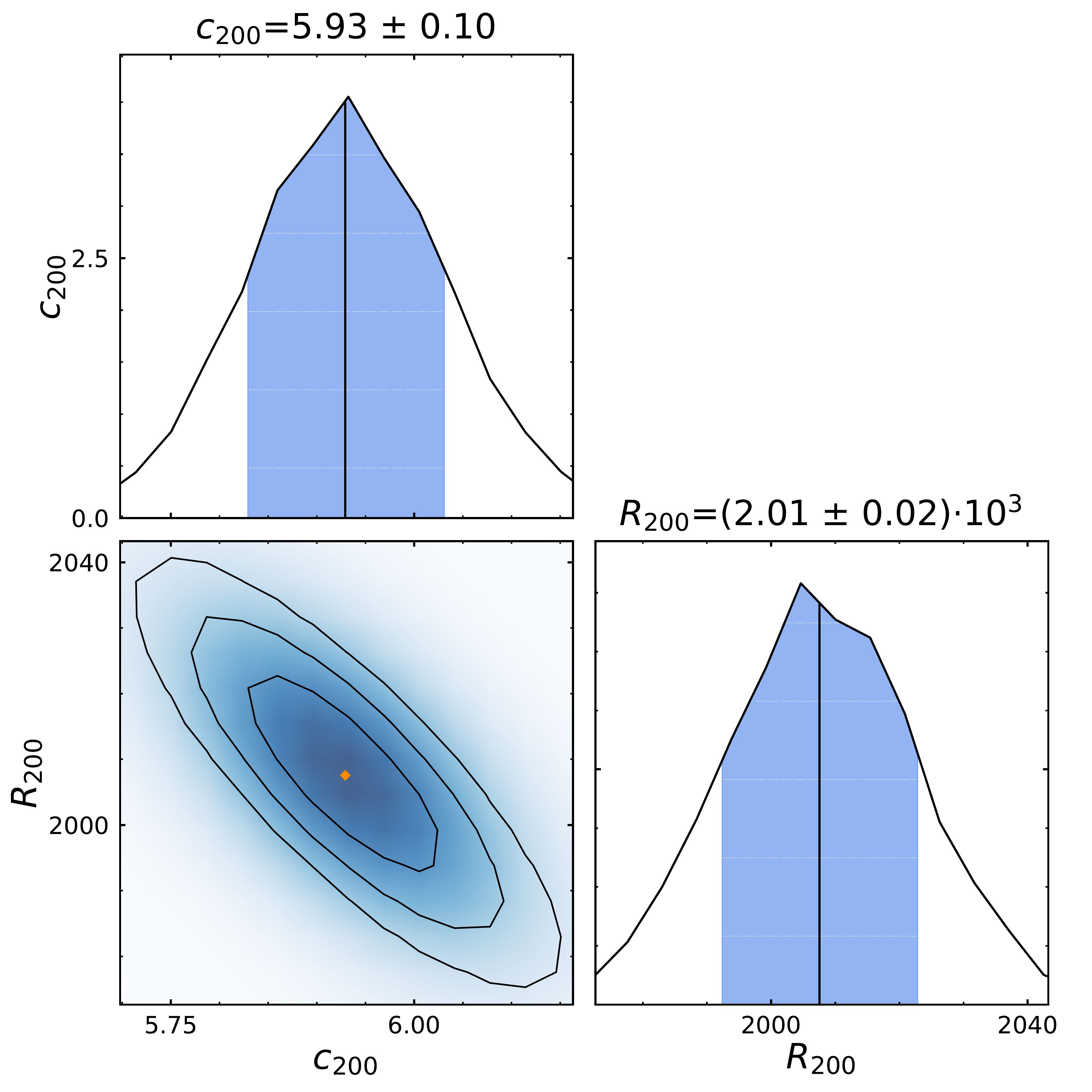}
        }}}
        \caption{Same as Fig. \ref{fig:a1795} but for A2029. } 
\end{figure*}

\begin{figure*}
        \centerline{\resizebox{\hsize}{!}{\vbox{
                                \includegraphics[width=0.45\textwidth]{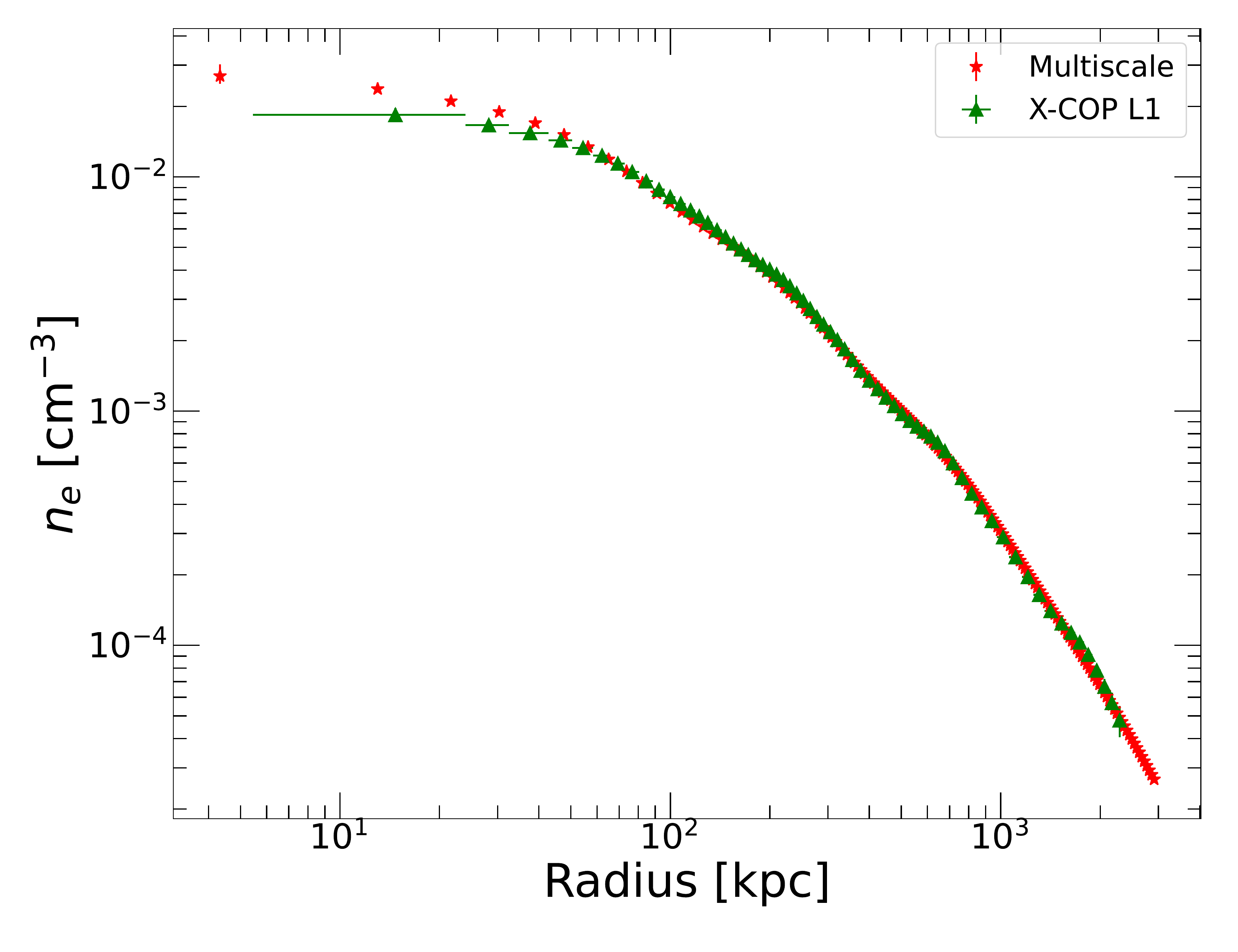}
                                \includegraphics[width=0.45\textwidth]{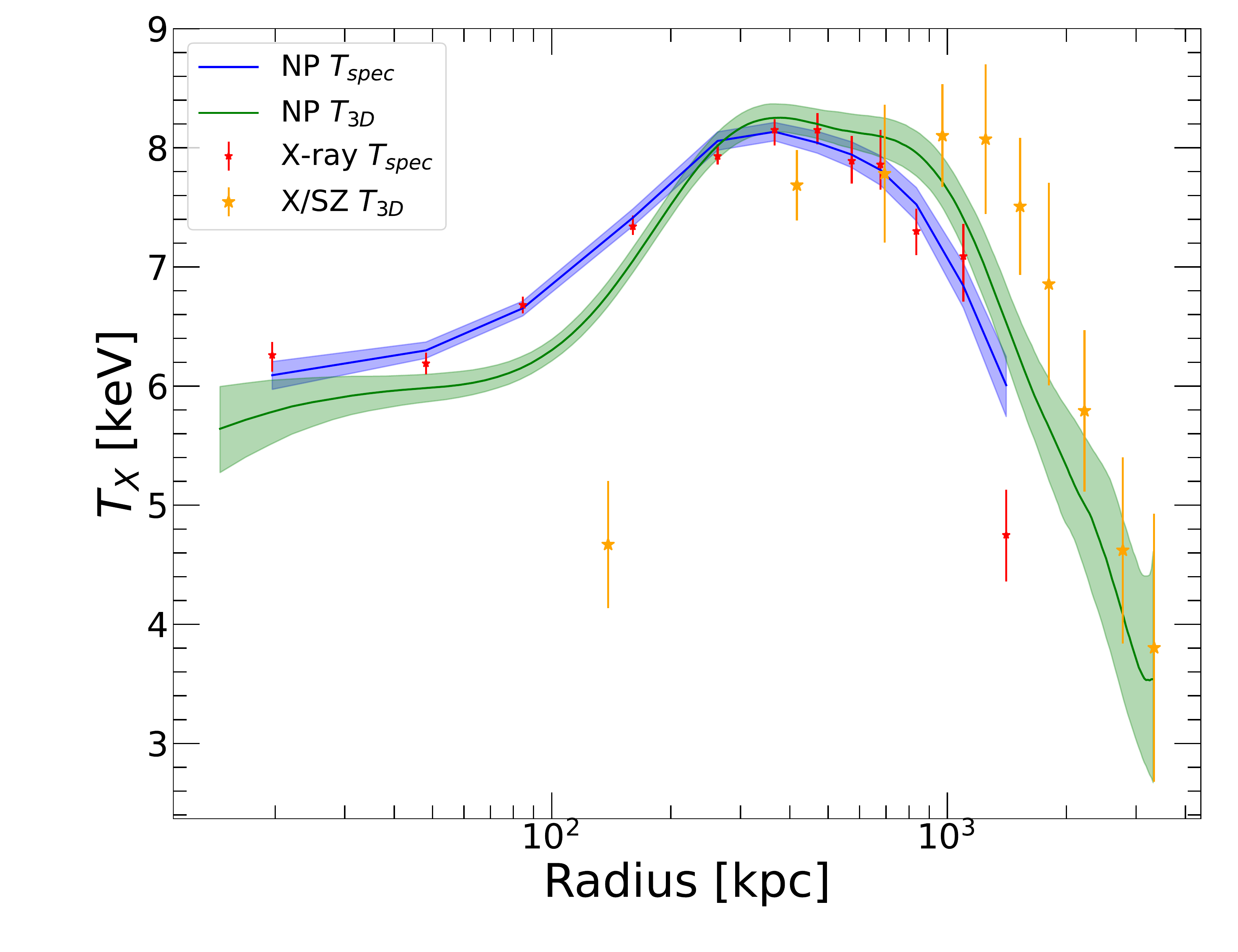}\\
                                
                                \includegraphics[width=0.45\textwidth]{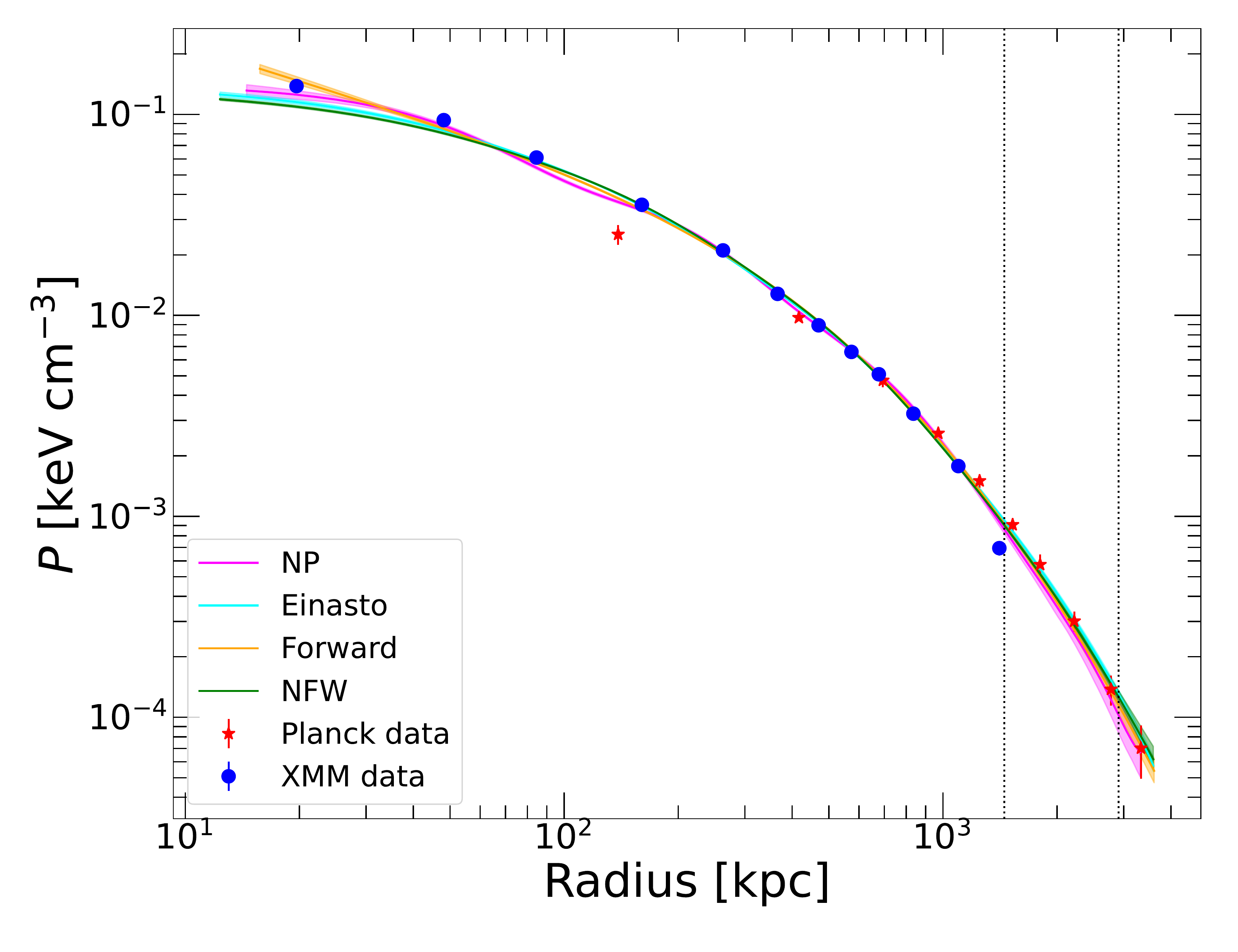}
                                \includegraphics[width=0.45\textwidth]{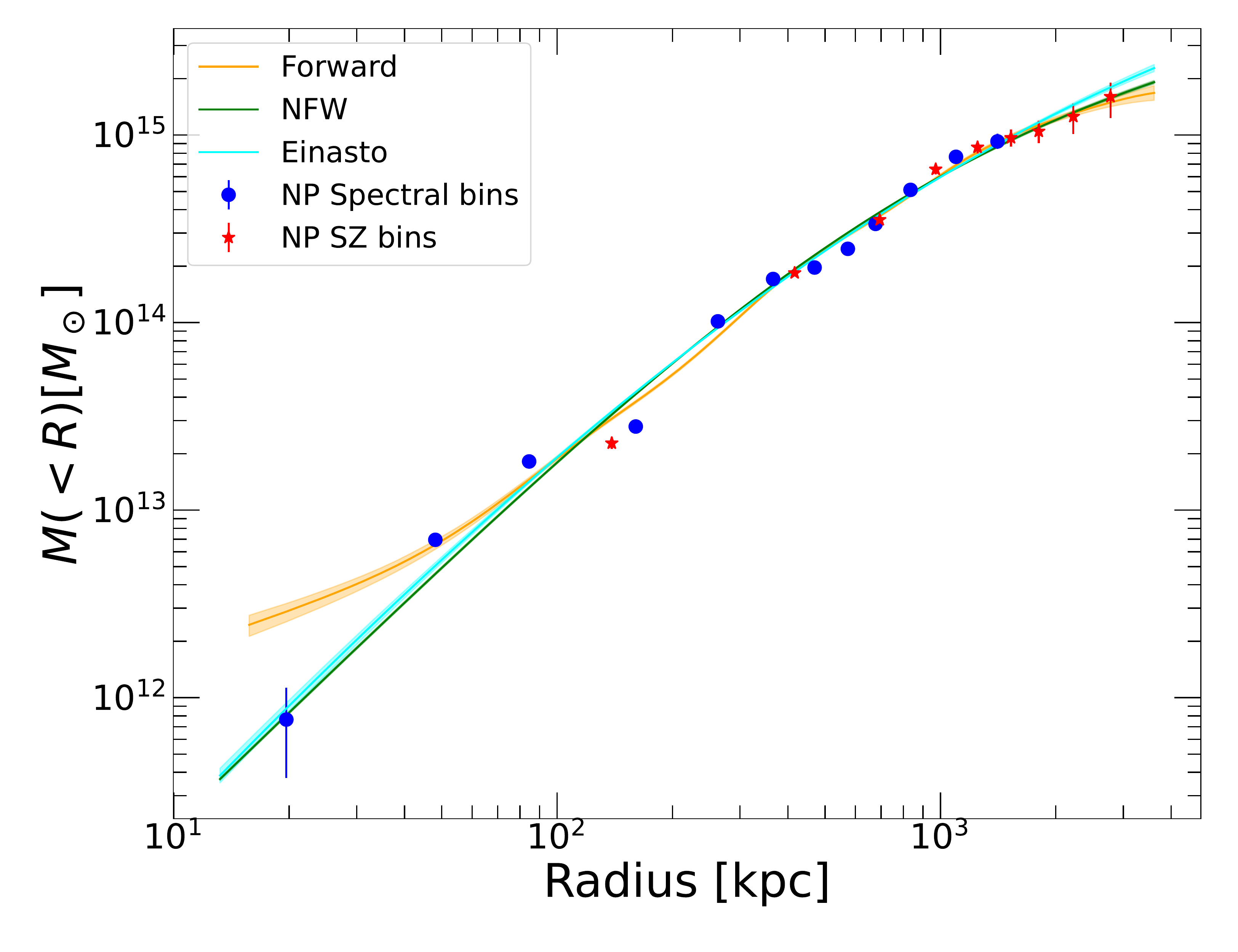}\\
                                
                                \includegraphics[width=0.45\textwidth]{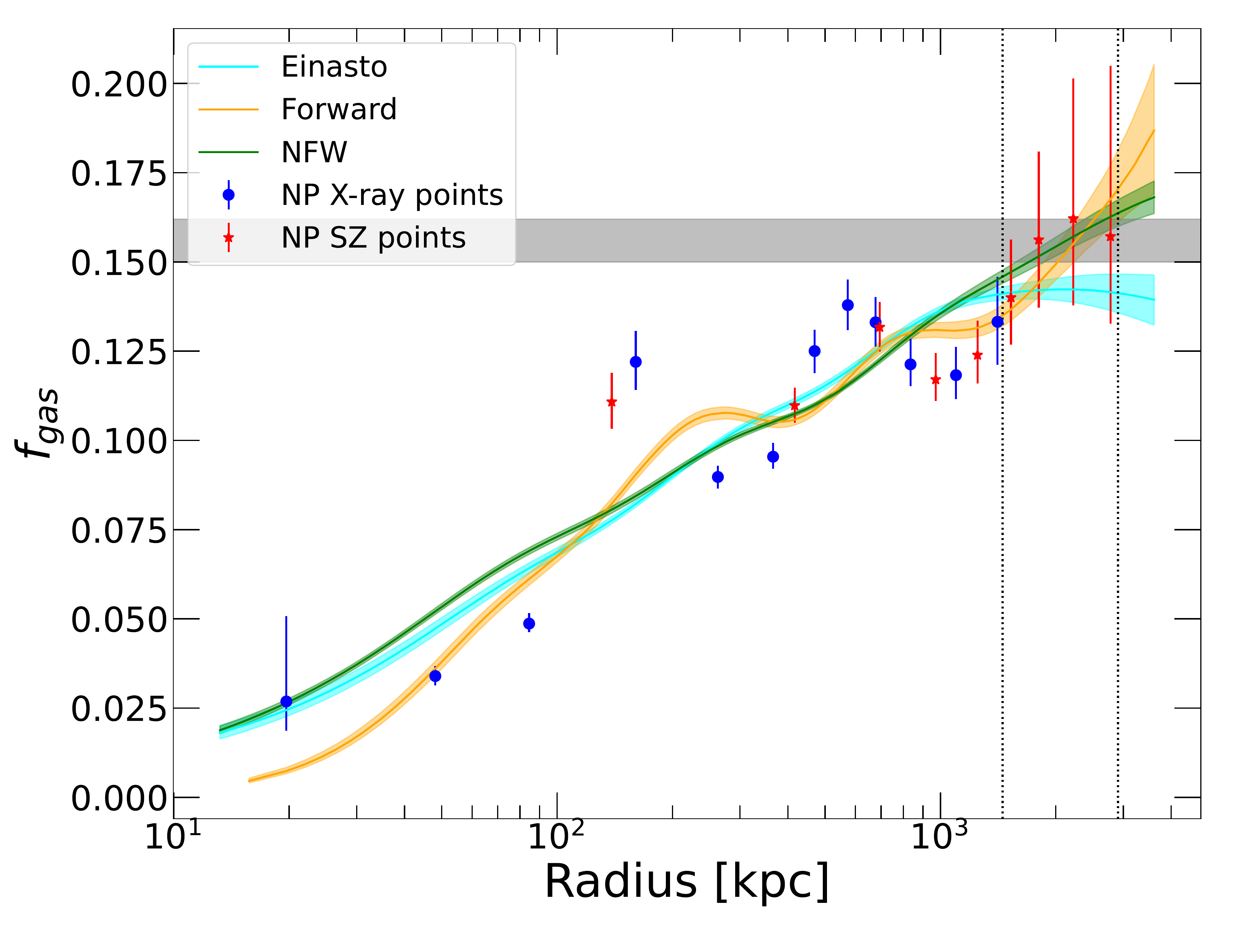}
                                \hspace{0.5cm}
                                \includegraphics[width=0.37\textwidth]{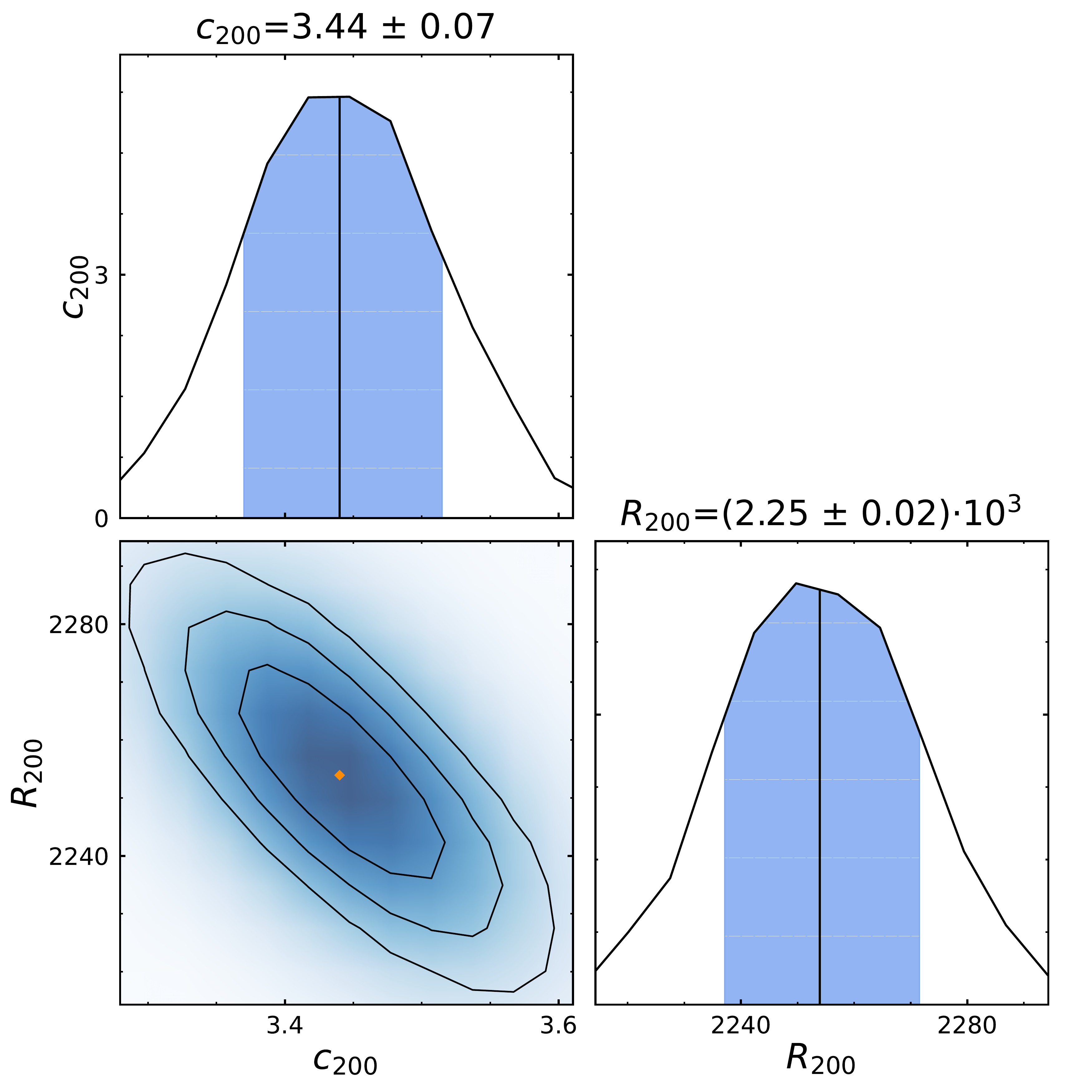}
        }}}
        \caption{Same as Fig. \ref{fig:a1795} but for A2142. } 
\end{figure*}

\begin{figure*}
        \centerline{\resizebox{\hsize}{!}{\vbox{
                                \includegraphics[width=0.45\textwidth]{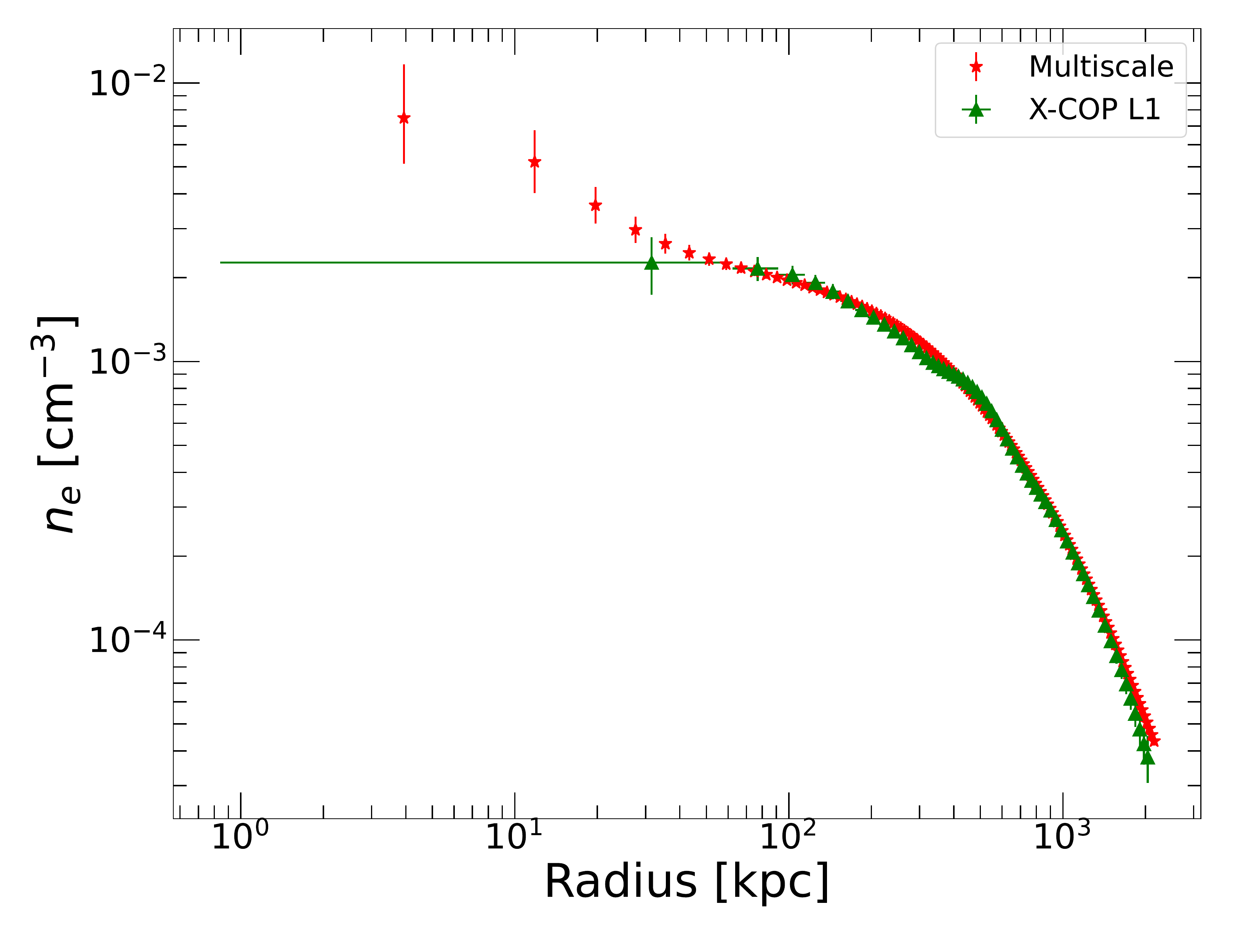}
                                \includegraphics[width=0.45\textwidth]{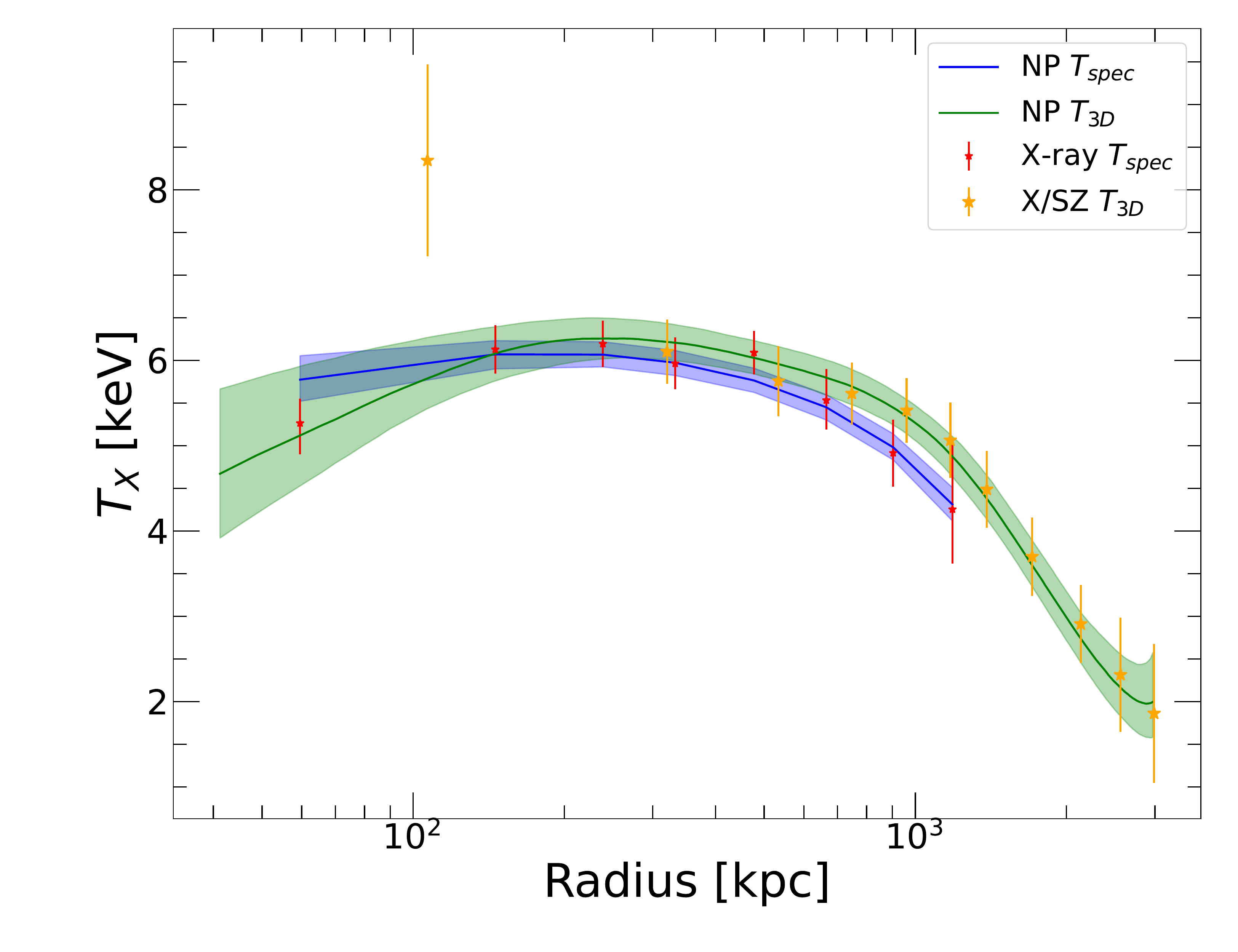}\\
                                
                                \includegraphics[width=0.45\textwidth]{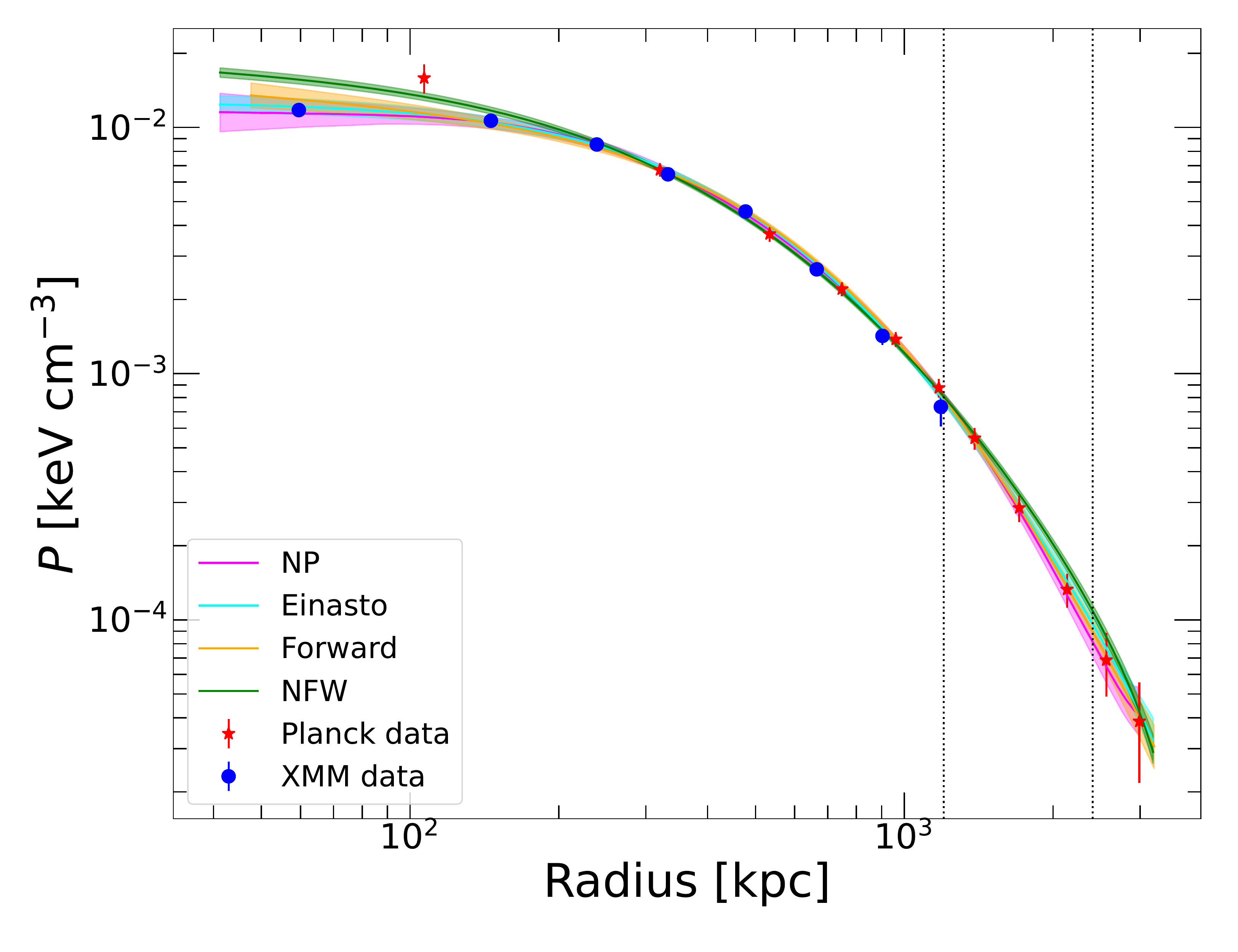}
                                \includegraphics[width=0.45\textwidth]{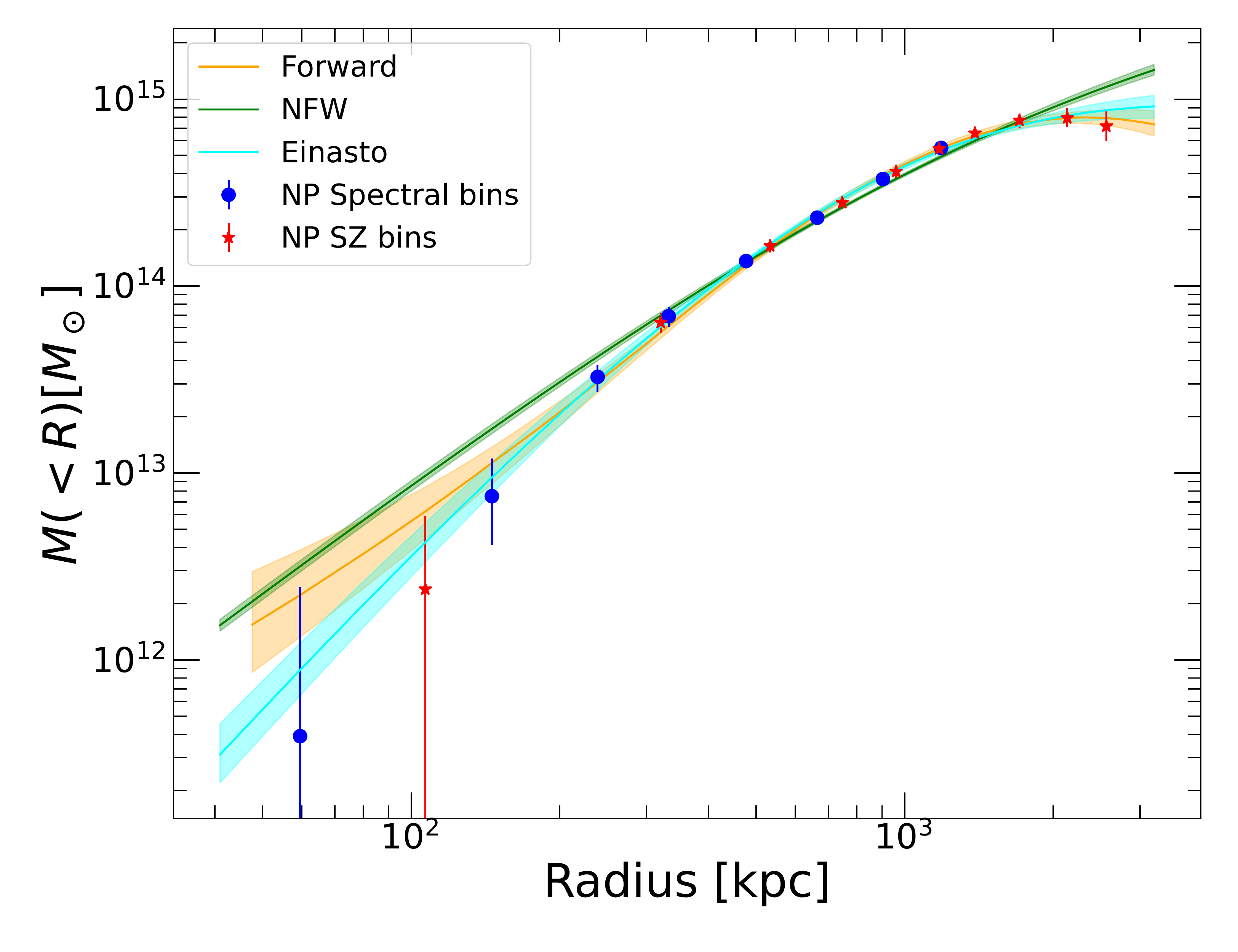}\\
                                
                                \includegraphics[width=0.45\textwidth]{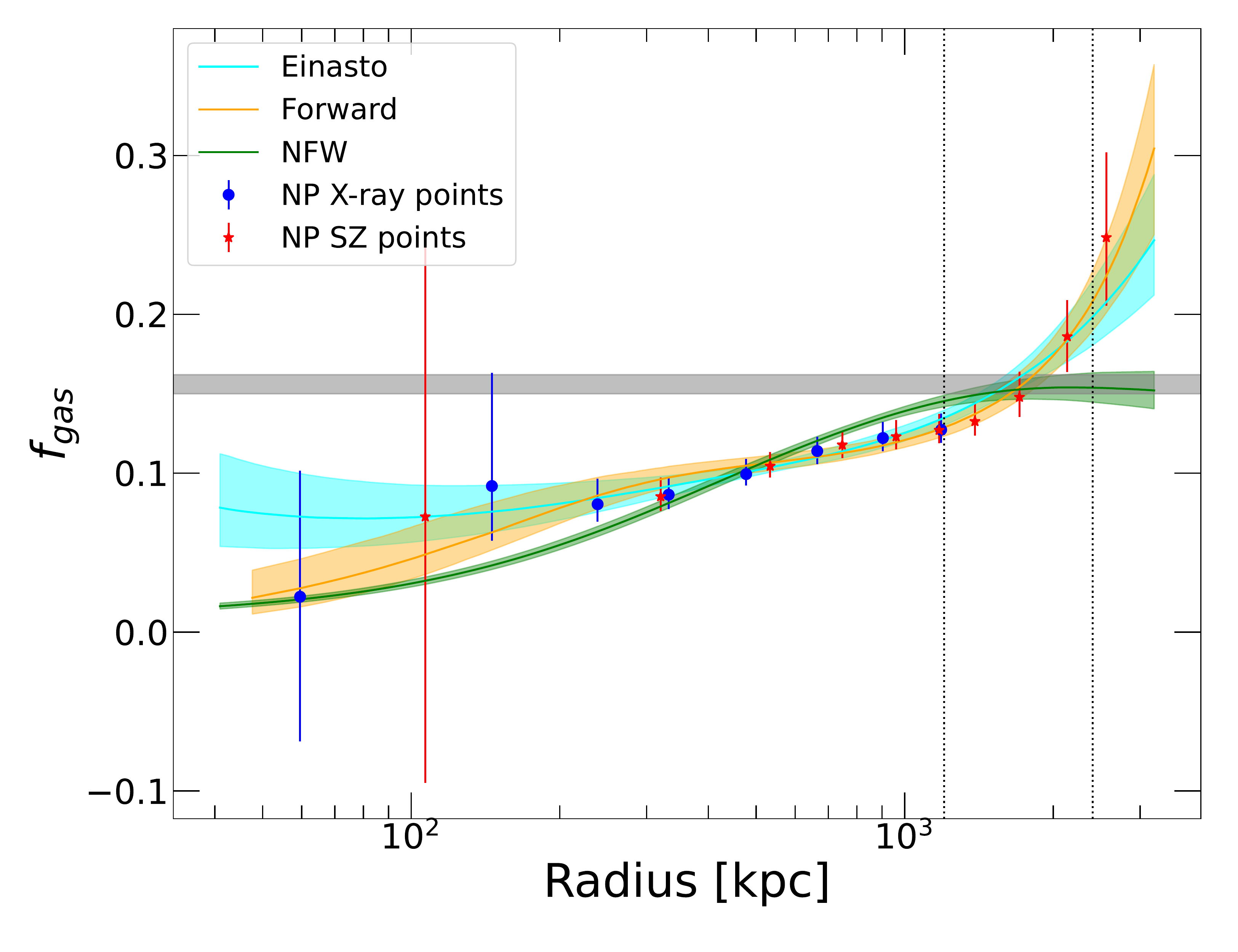}
                                \hspace{0.5cm}
                                \includegraphics[width=0.37\textwidth]{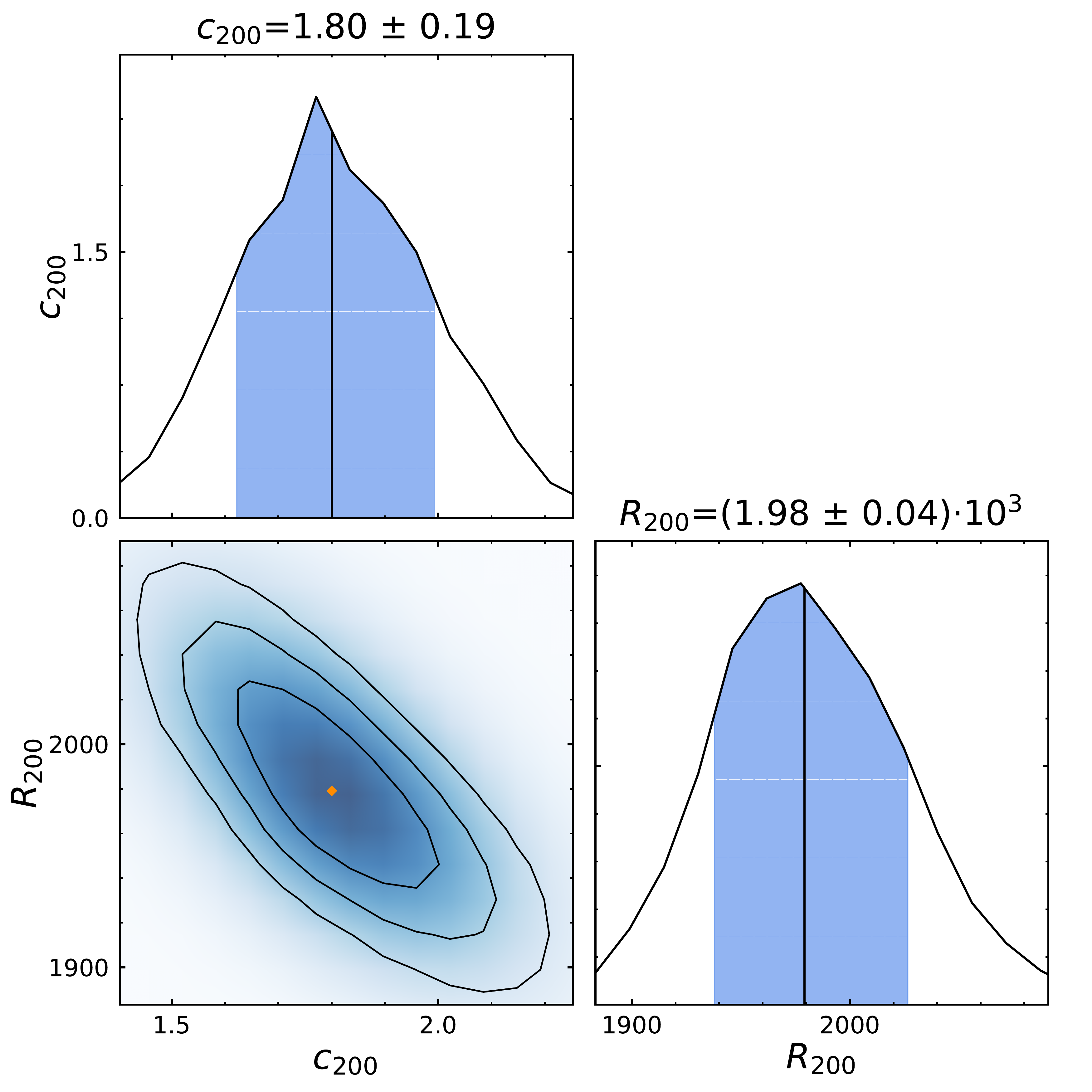}
        }}}
        \caption{Same as Fig. \ref{fig:a1795} but for A2255. } 
\end{figure*}

\begin{figure*}
        \centerline{\resizebox{\hsize}{!}{\vbox{
                                \includegraphics[width=0.45\textwidth]{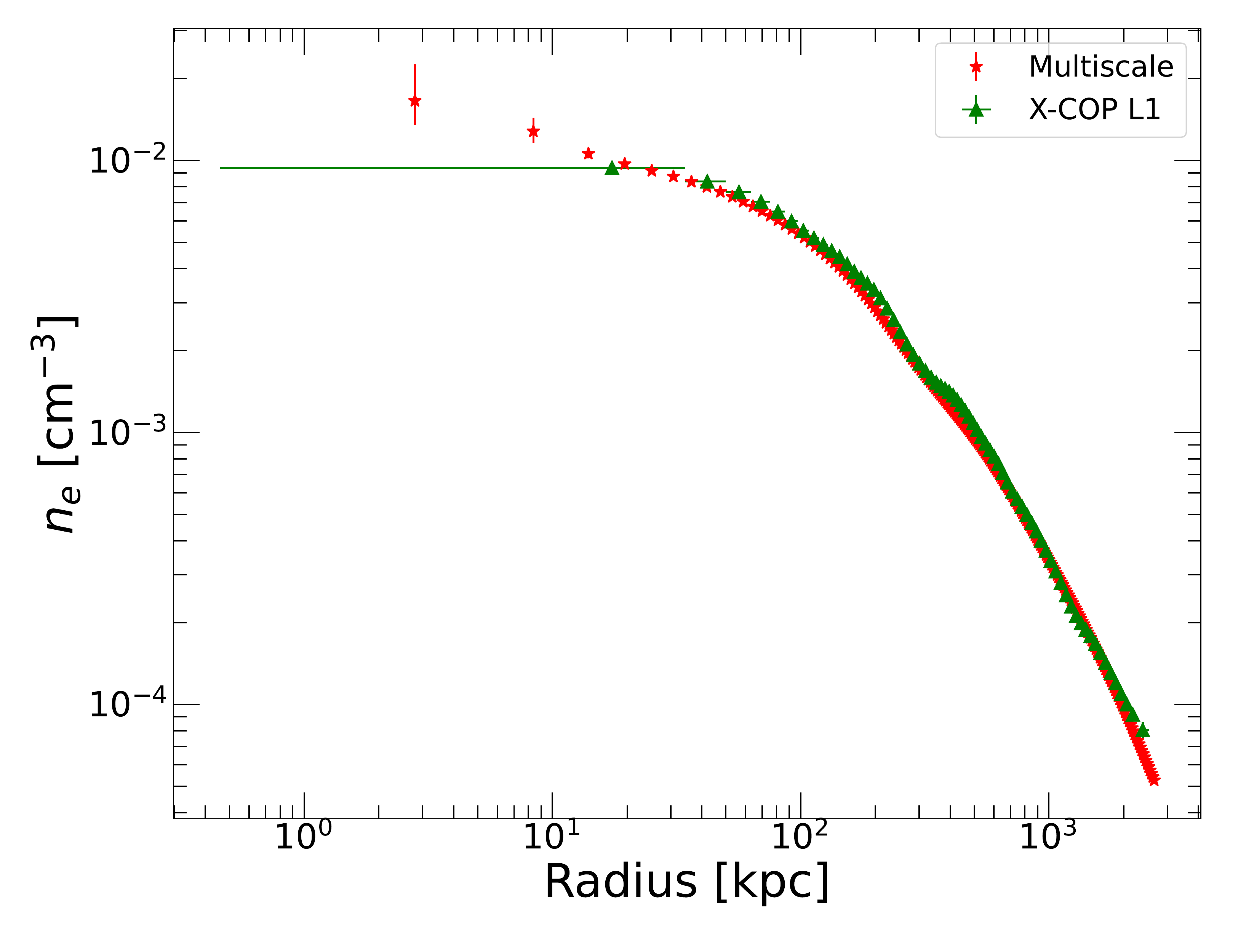}
                                \includegraphics[width=0.45\textwidth]{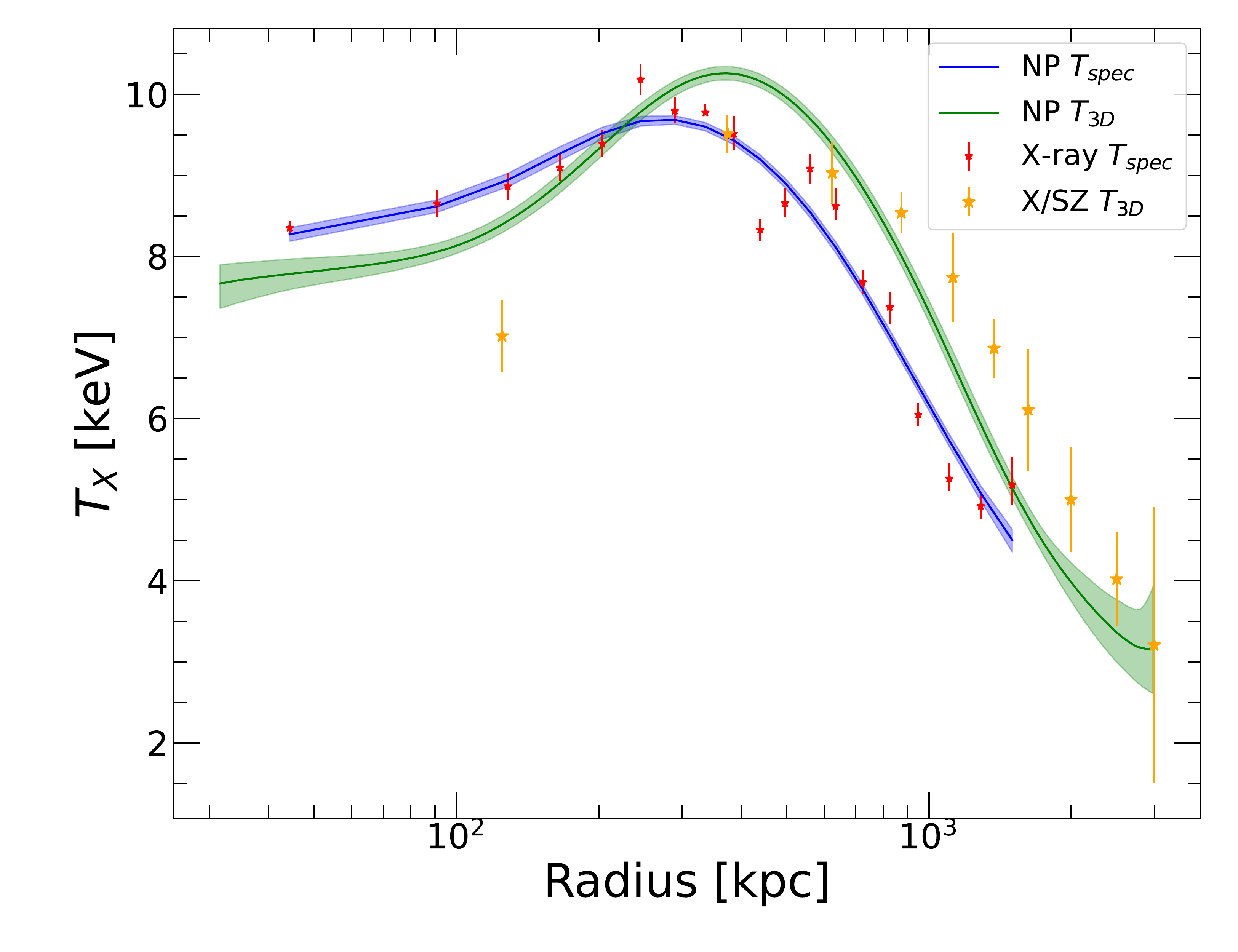}\\
                                
                                \includegraphics[width=0.45\textwidth]{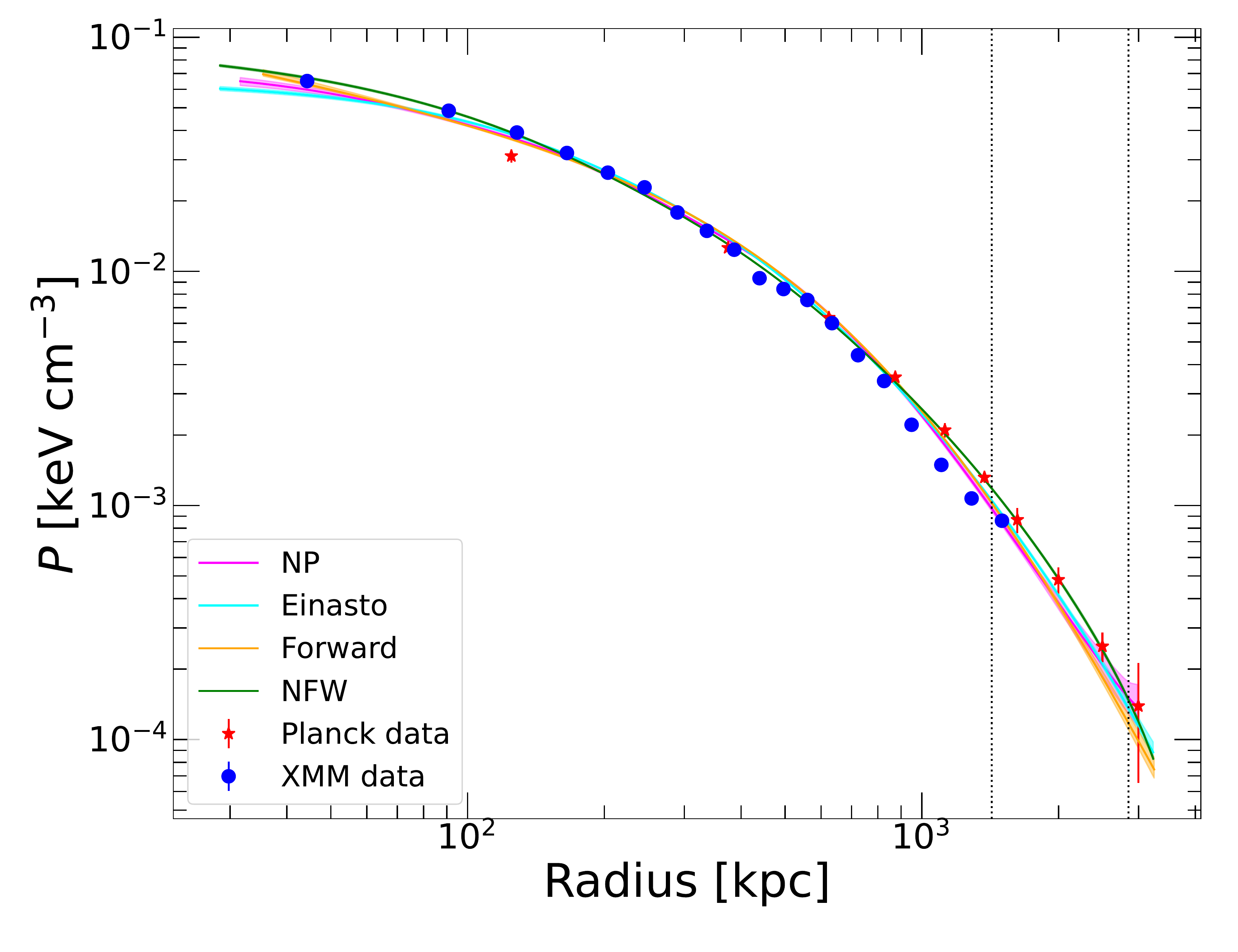}
                                \includegraphics[width=0.45\textwidth]{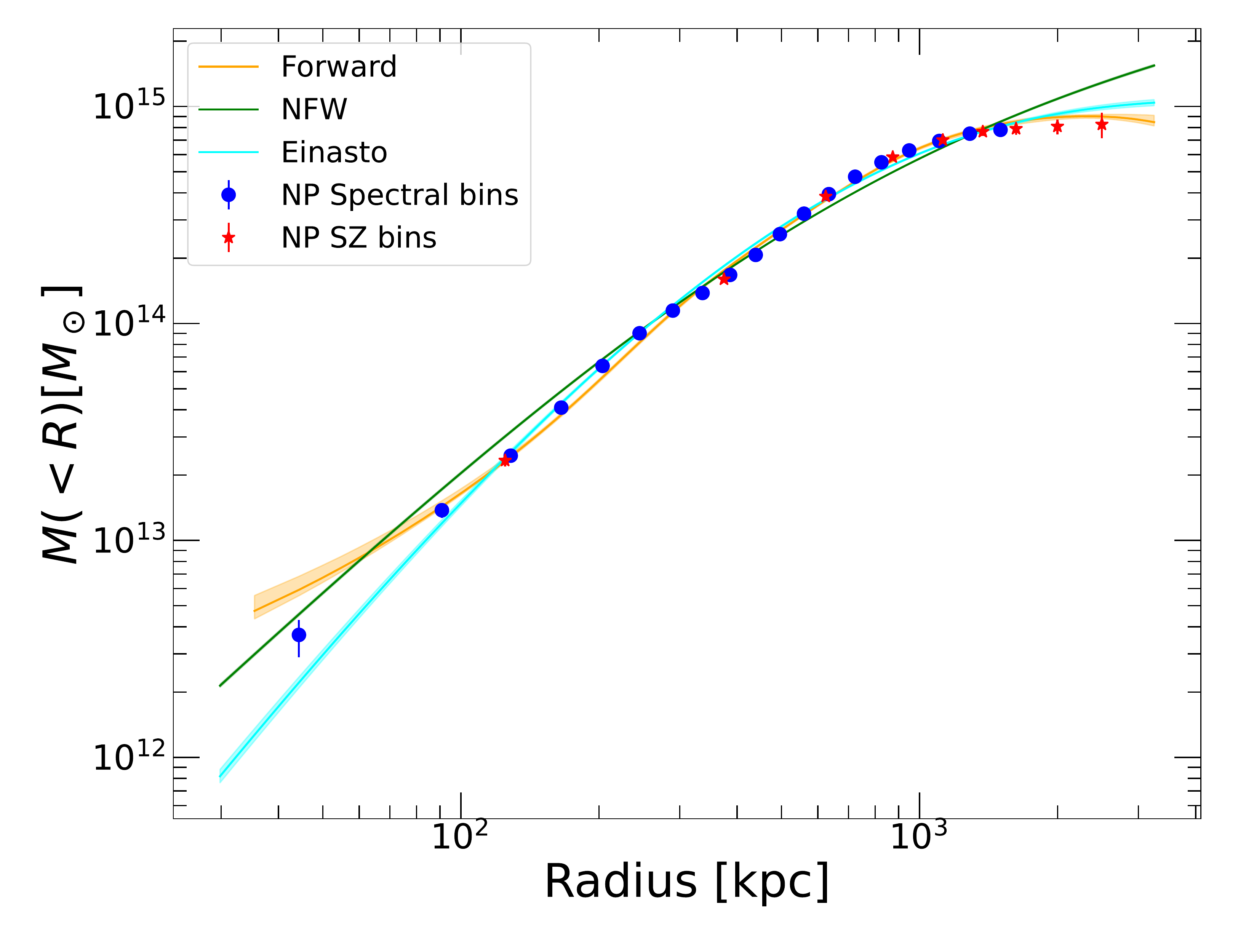}\\
                                
                                \includegraphics[width=0.45\textwidth]{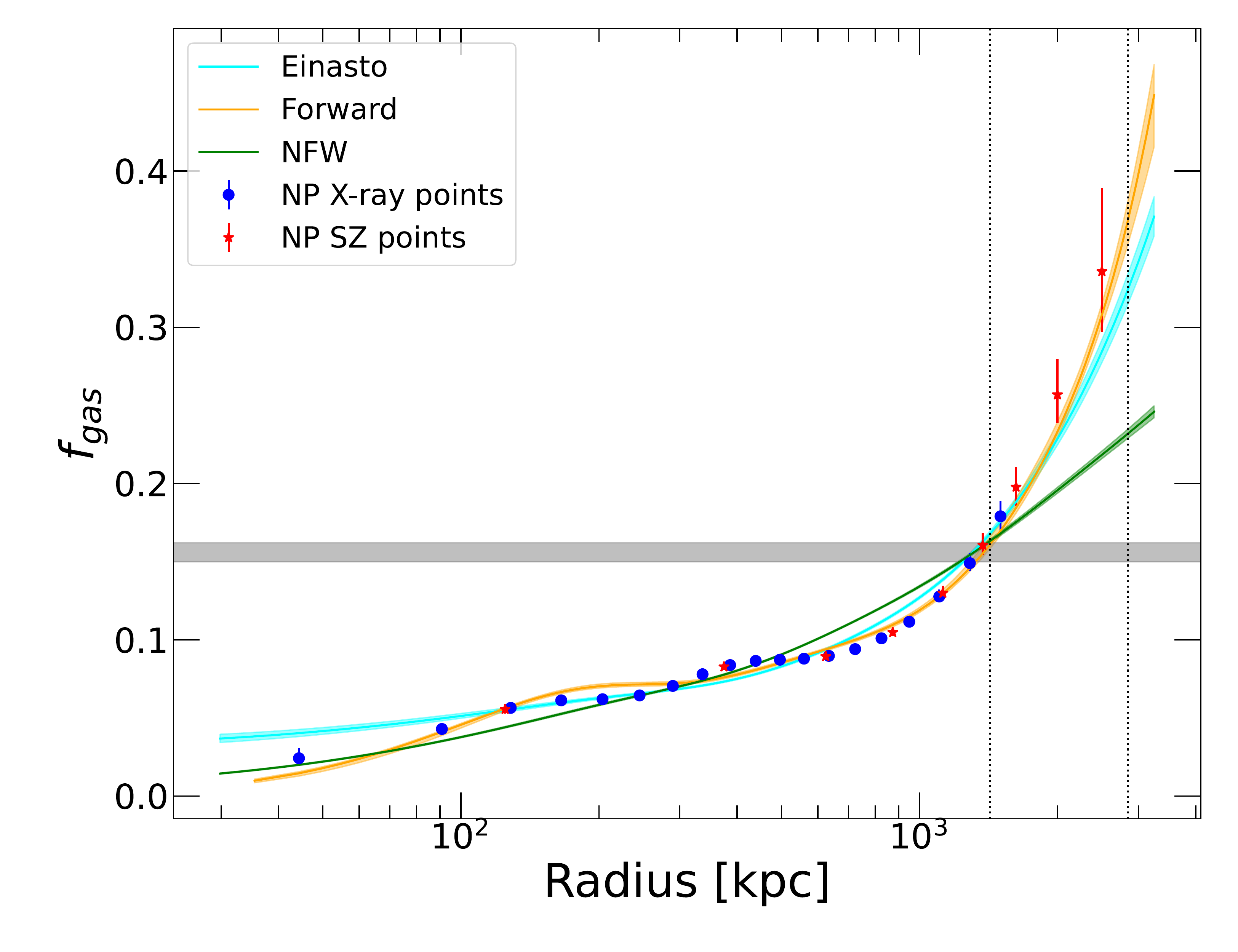}
                                \hspace{0.5cm}
                                \includegraphics[width=0.37\textwidth]{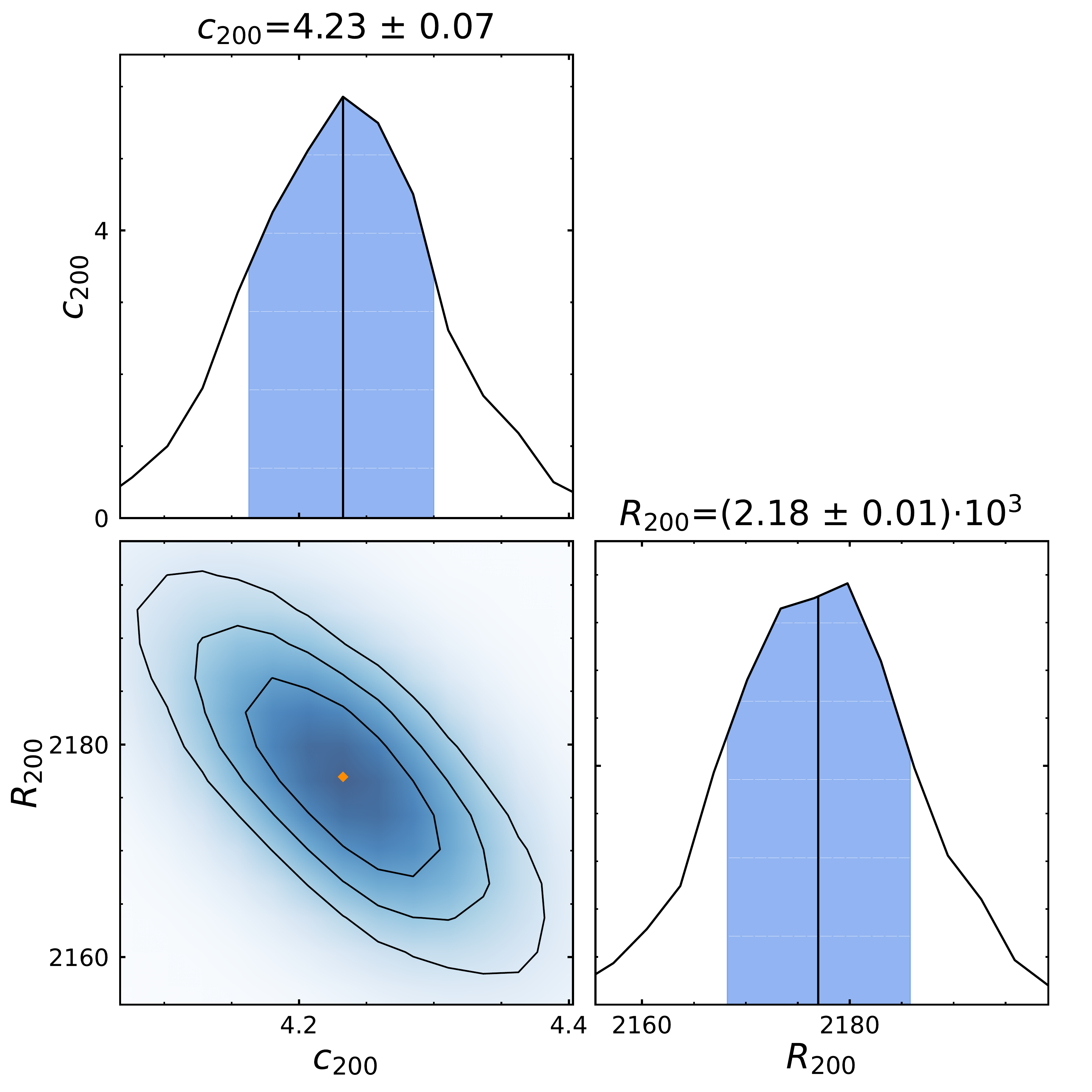}
        }}}
        \caption{Same as Fig. \ref{fig:a1795} but for A2319. } 
\end{figure*}

\begin{figure*}
        \centerline{\resizebox{\hsize}{!}{\vbox{
                                \includegraphics[width=0.45\textwidth]{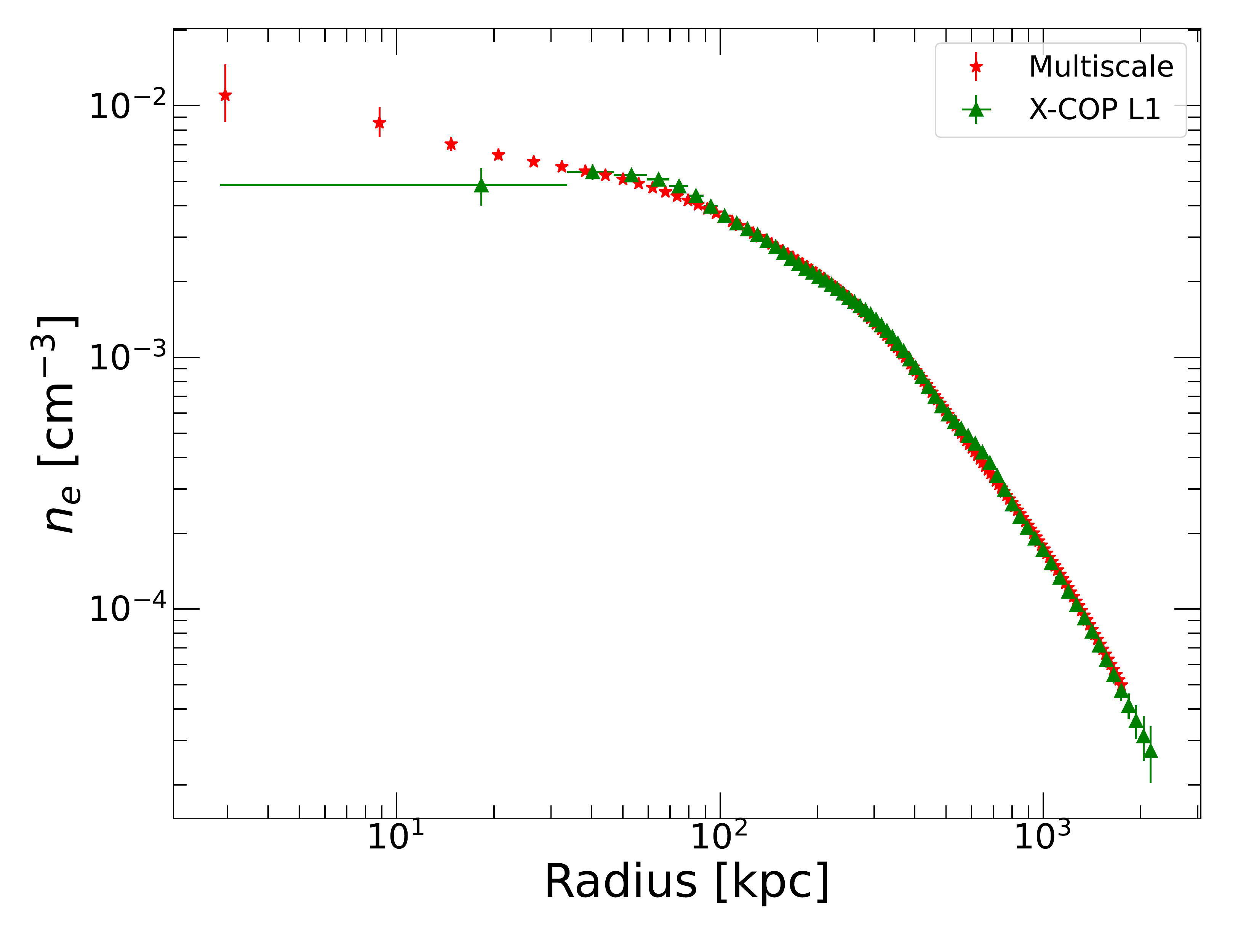}
                                \includegraphics[width=0.45\textwidth]{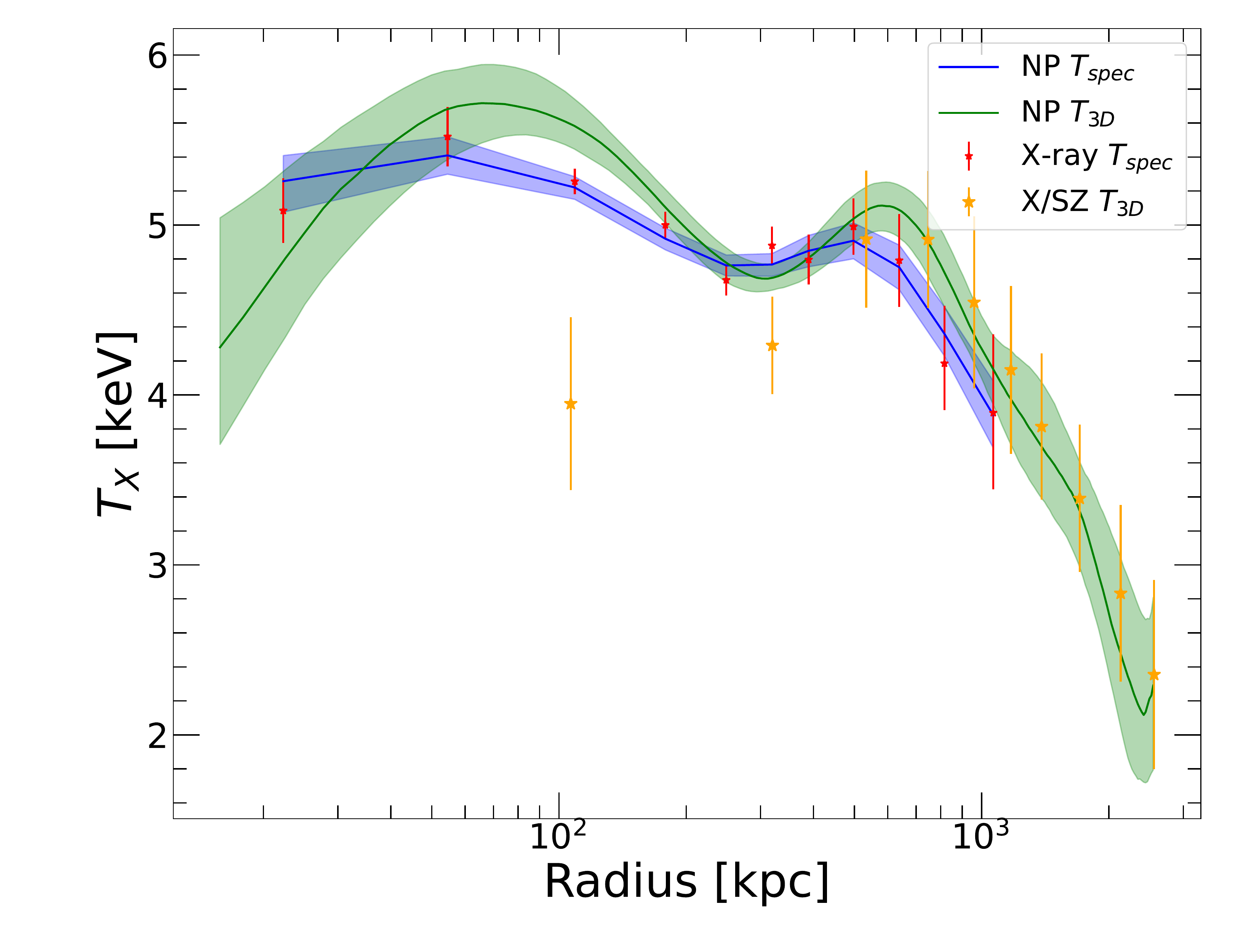}\\
                                
                                \includegraphics[width=0.45\textwidth]{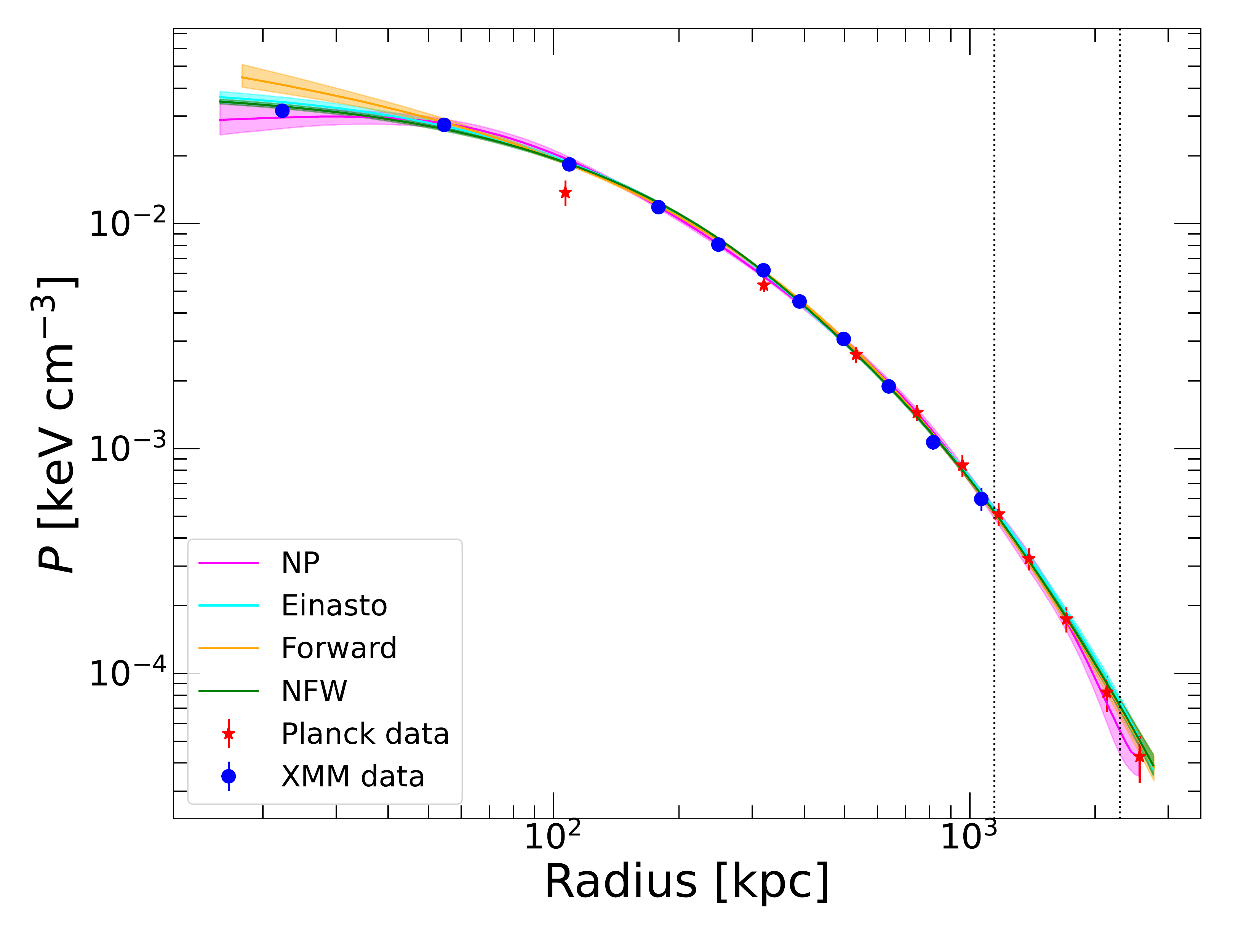}
                                \includegraphics[width=0.45\textwidth]{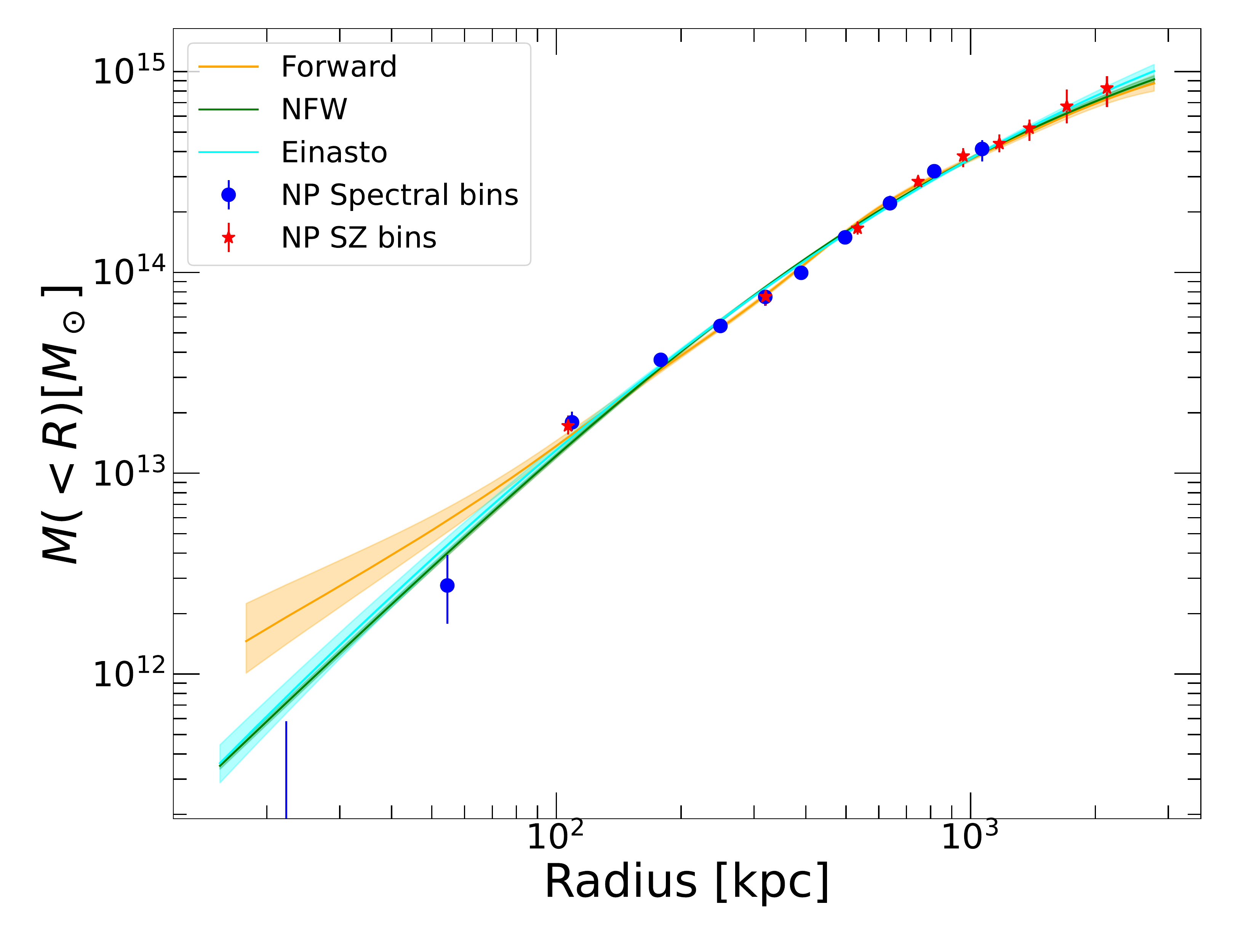}\\
                                
                                \includegraphics[width=0.45\textwidth]{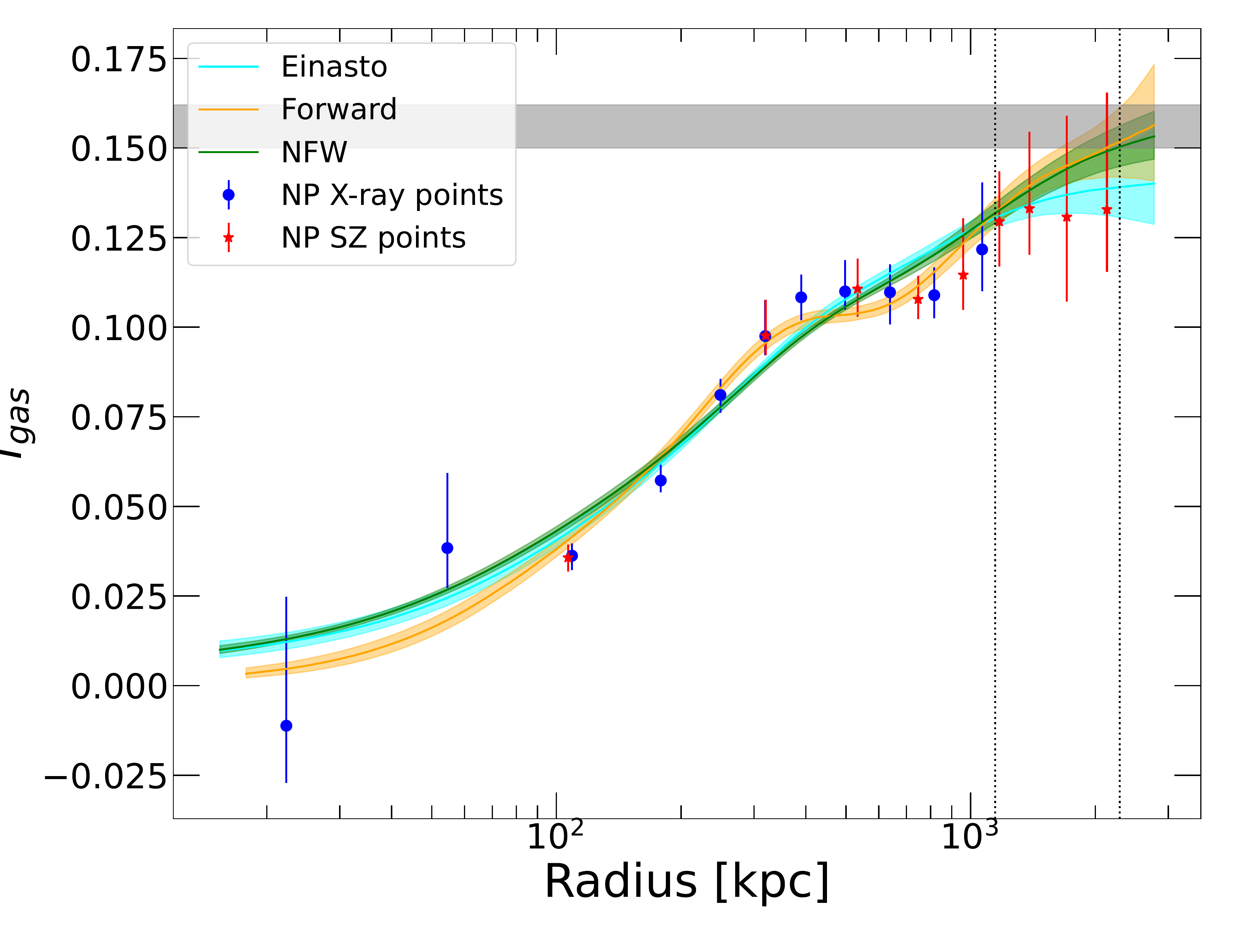}
                                \hspace{0.5cm}
                                \includegraphics[width=0.37\textwidth]{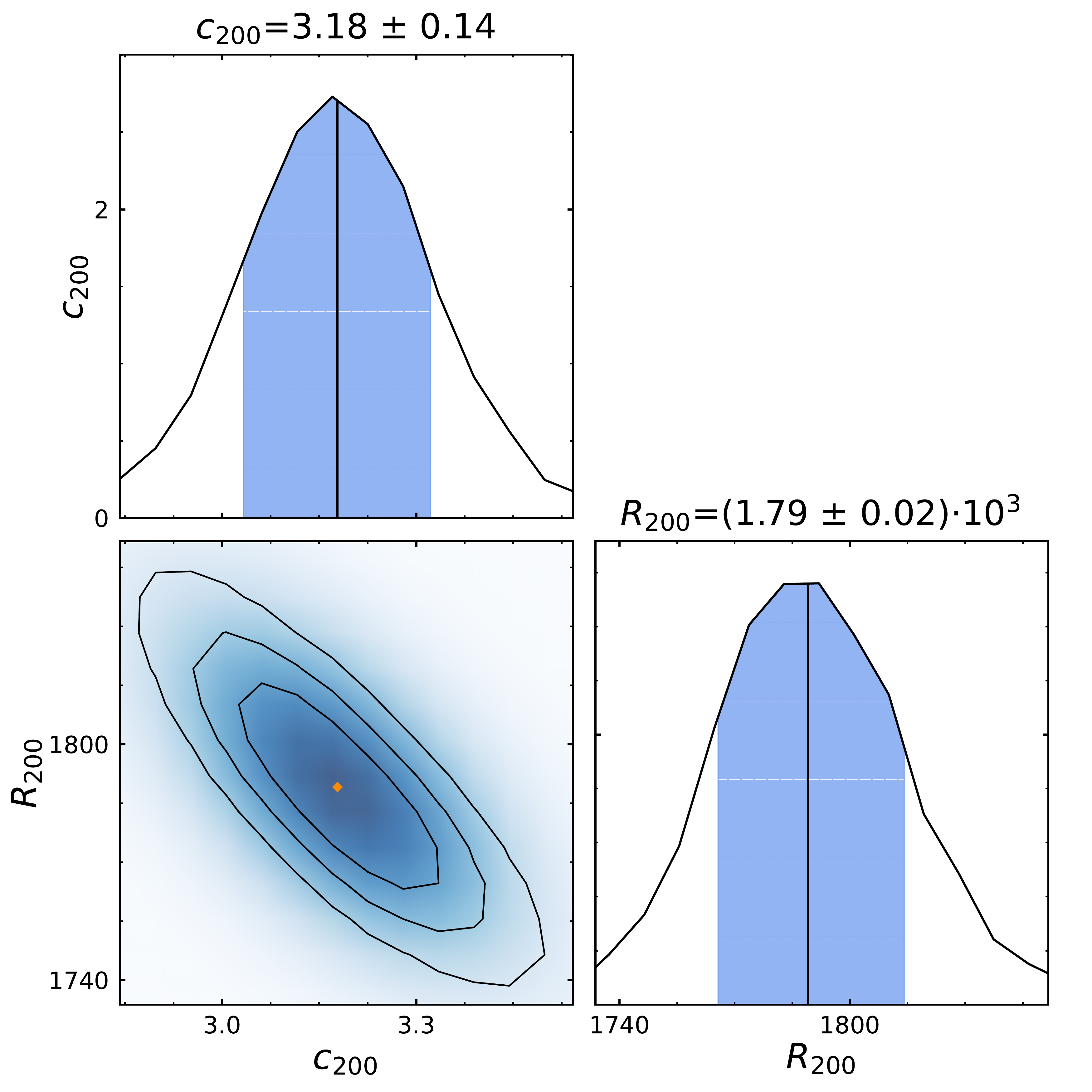}
        }}}
        \caption{Same as Fig. \ref{fig:a1795} but for A3158. } 
\end{figure*}

\begin{figure*}
        \centerline{\resizebox{\hsize}{!}{\vbox{
                                \includegraphics[width=0.45\textwidth]{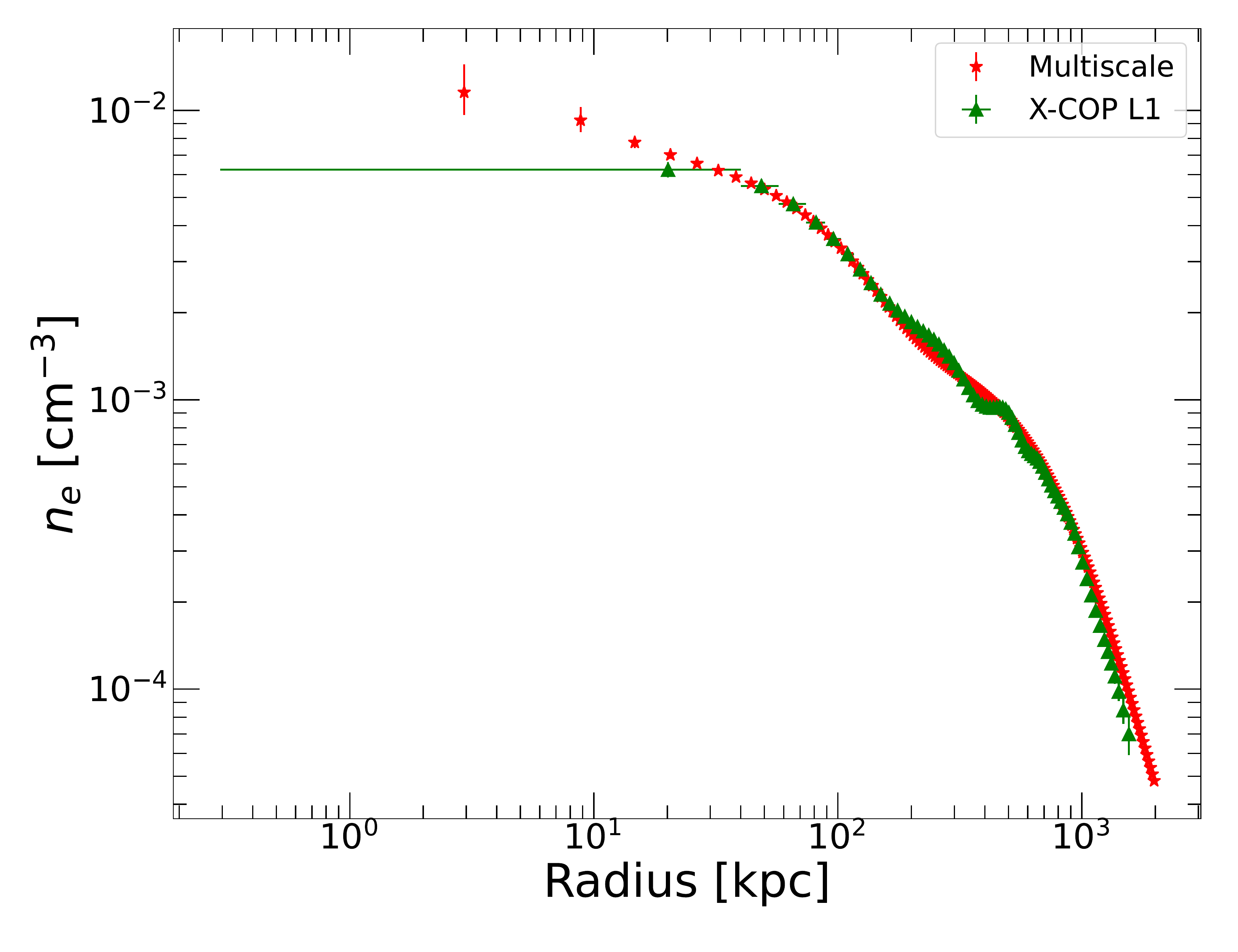}
                                \includegraphics[width=0.45\textwidth]{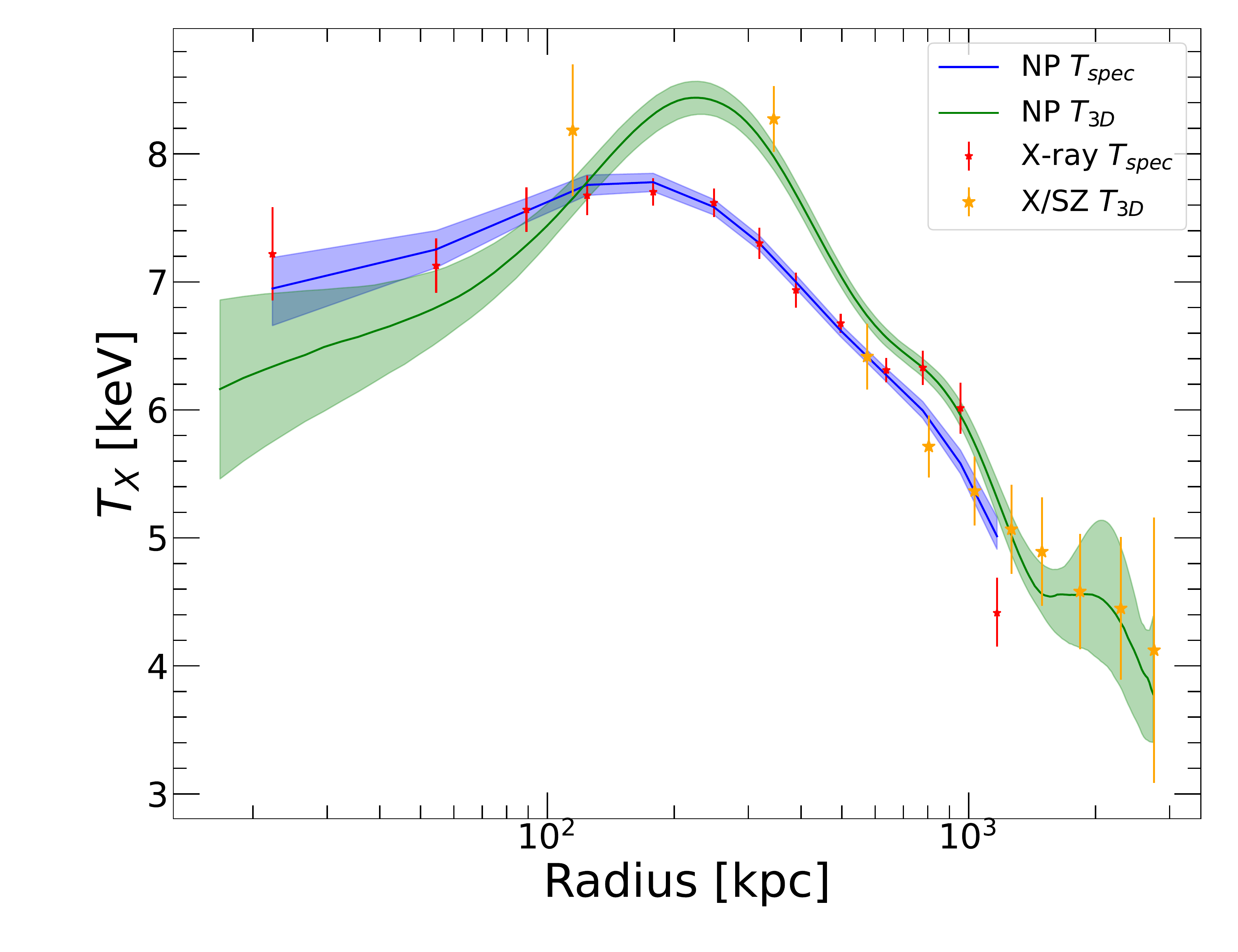}\\
                                
                                \includegraphics[width=0.45\textwidth]{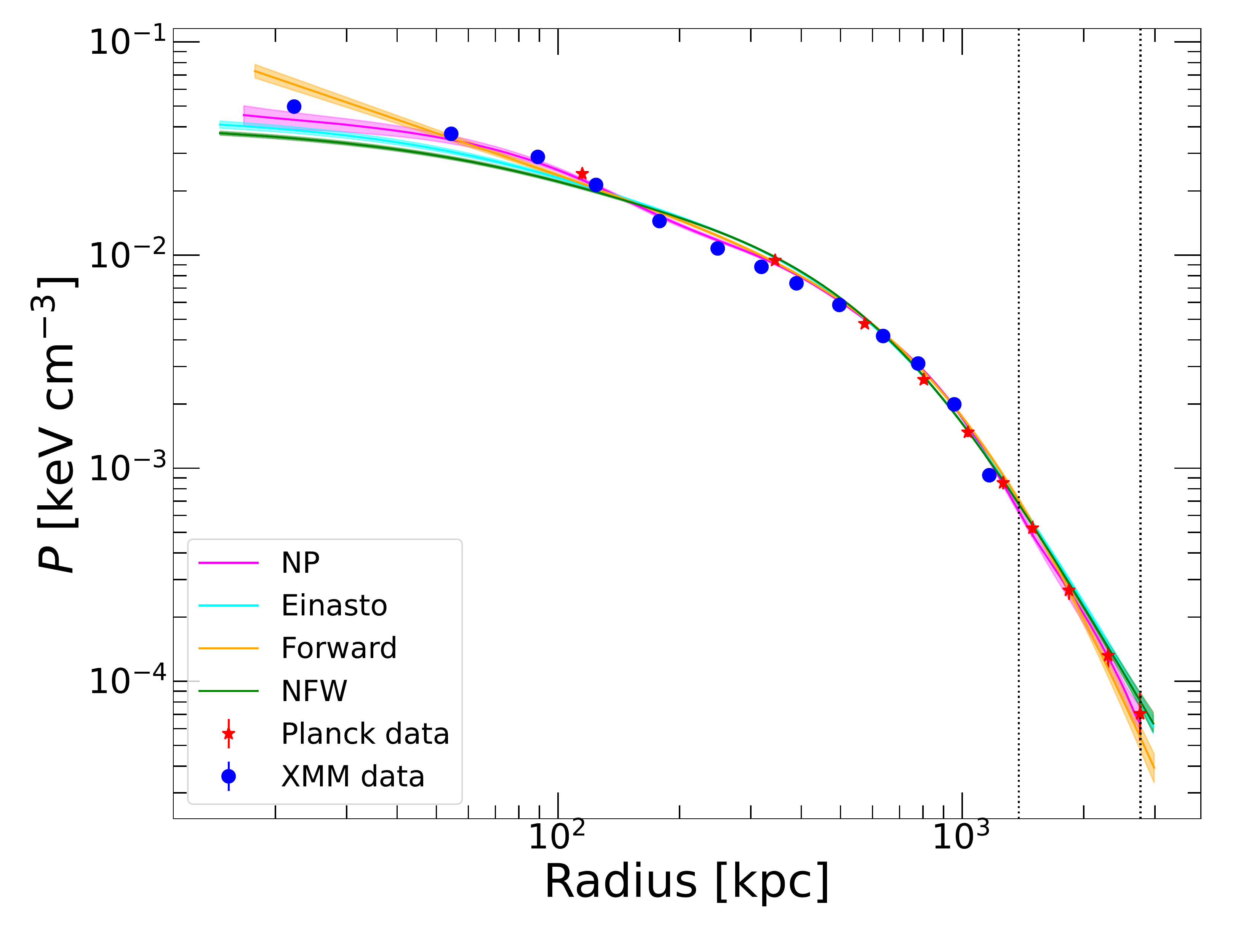}
                                \includegraphics[width=0.45\textwidth]{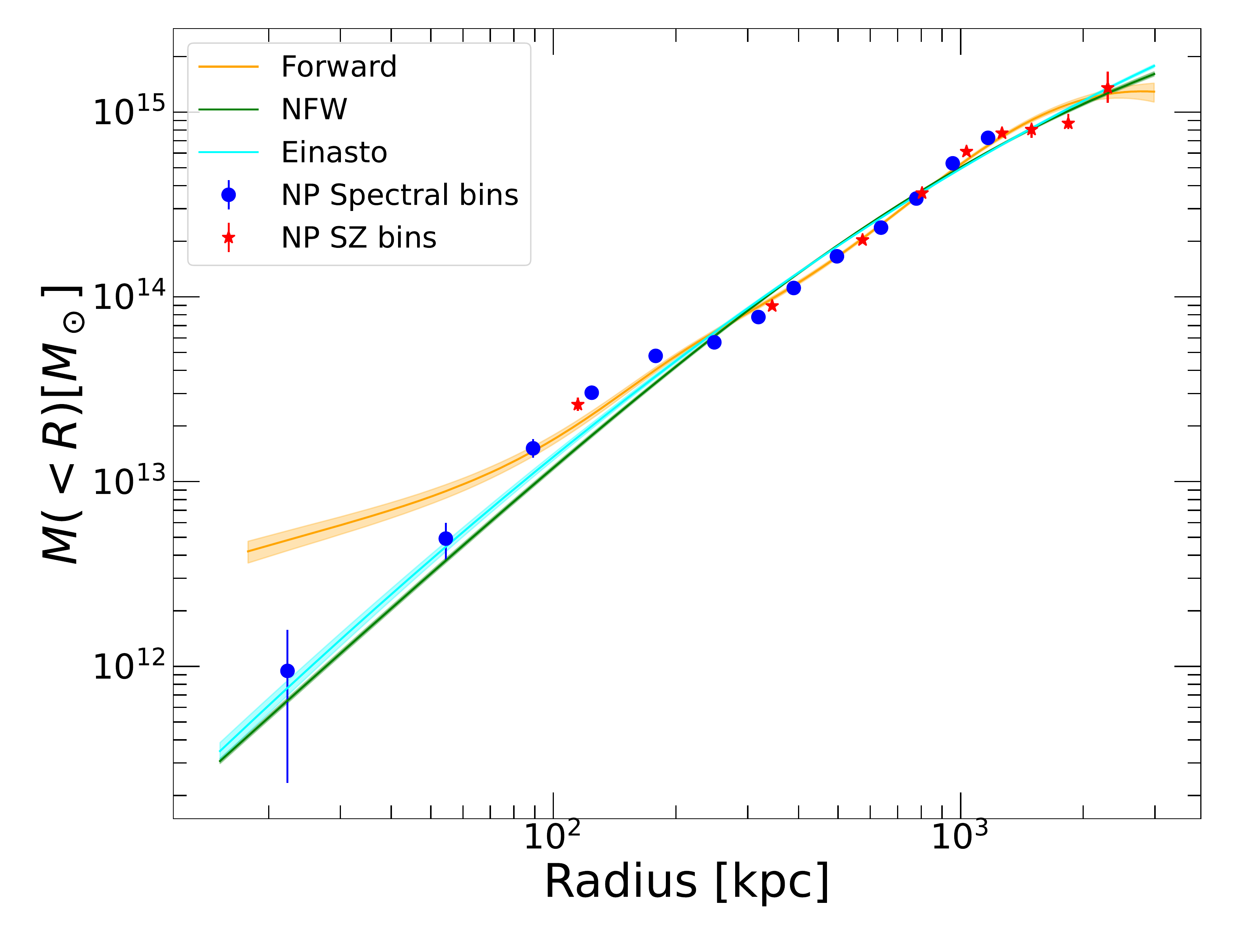}\\
                                
                                \includegraphics[width=0.45\textwidth]{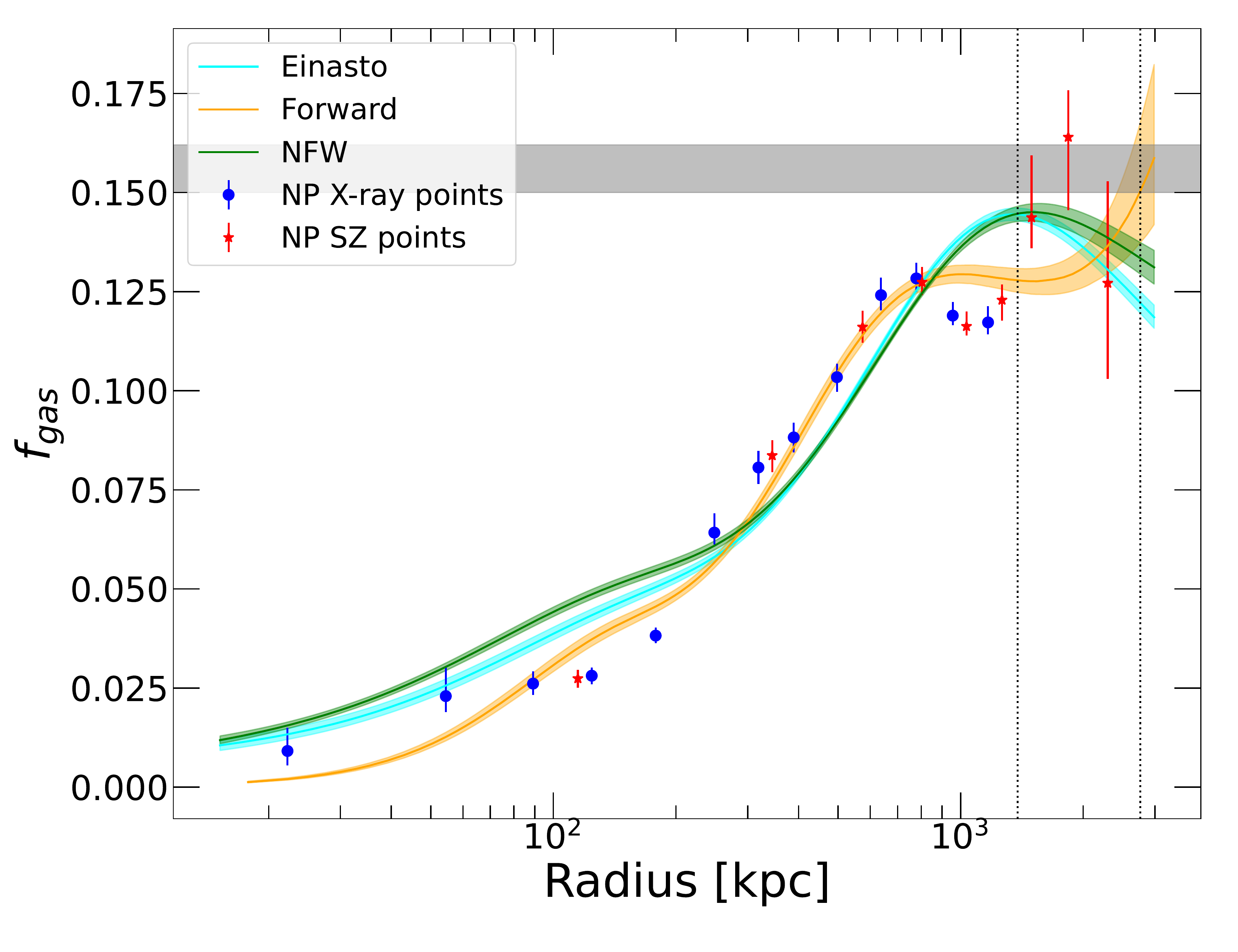}
                                \hspace{0.5cm}
                                \includegraphics[width=0.37\textwidth]{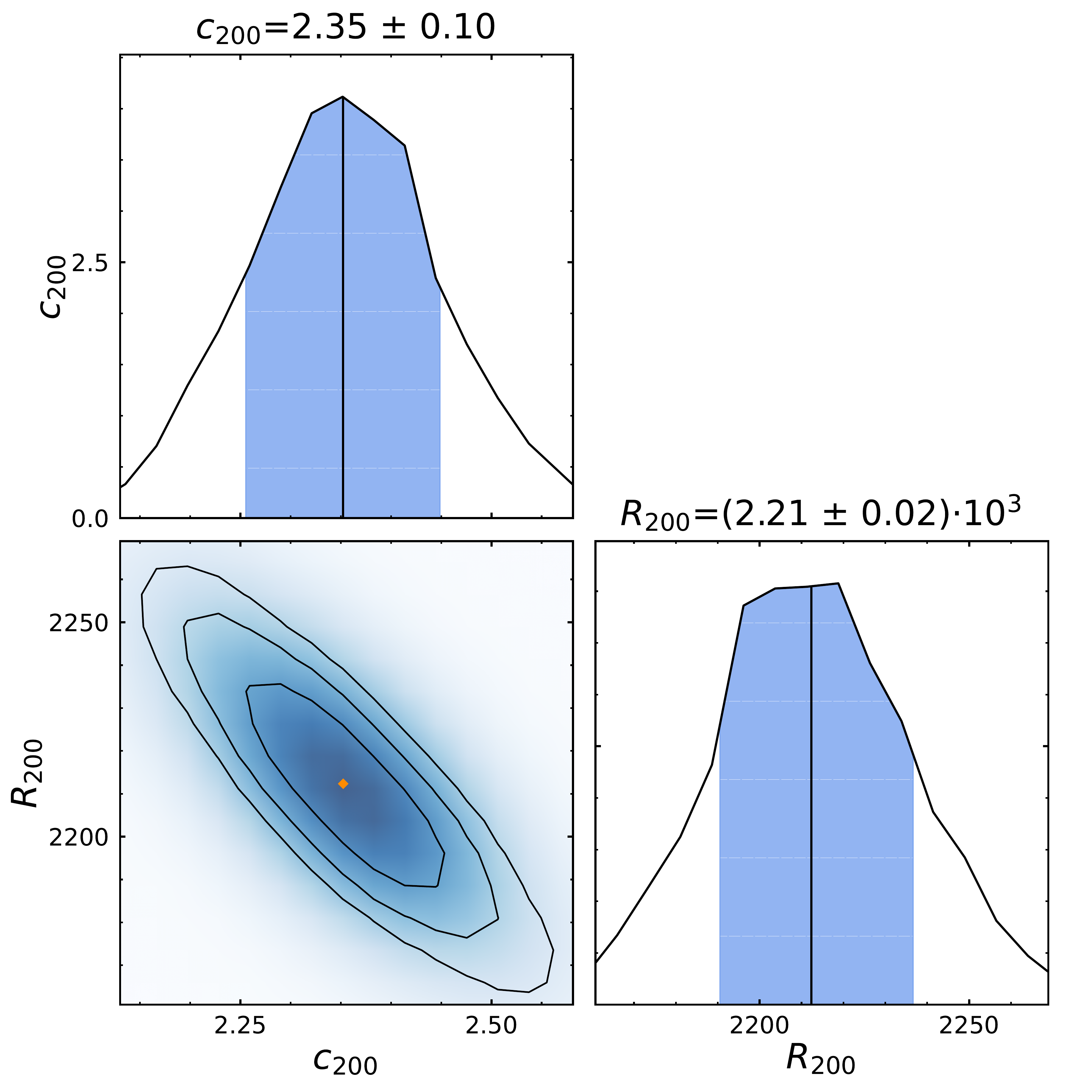}
        }}}
        \caption{Same as Fig. \ref{fig:a1795} but for A3266. } 
\end{figure*}

\begin{figure*}
        \centerline{\resizebox{\hsize}{!}{\vbox{
                                \includegraphics[width=0.45\textwidth]{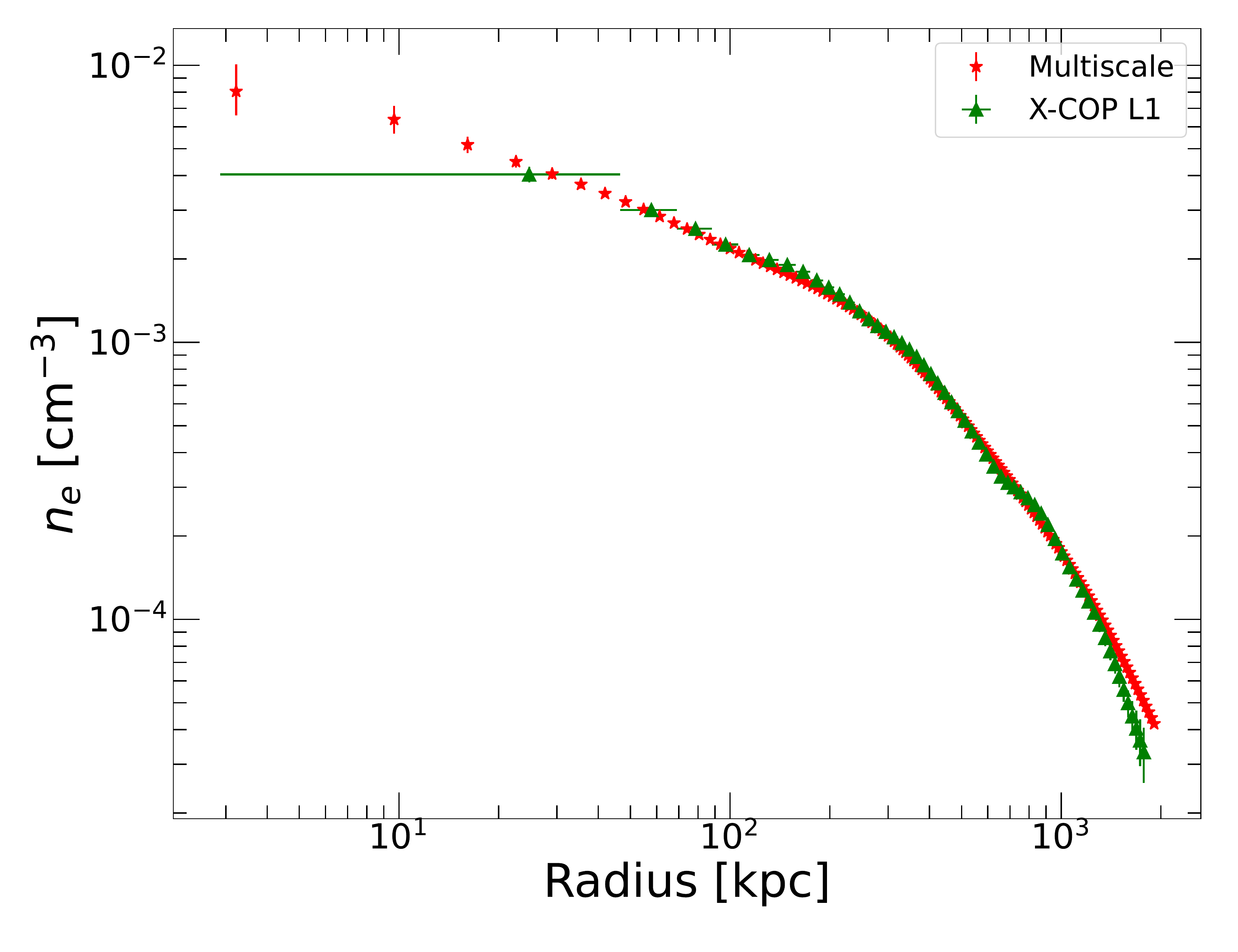}
                                \includegraphics[width=0.45\textwidth]{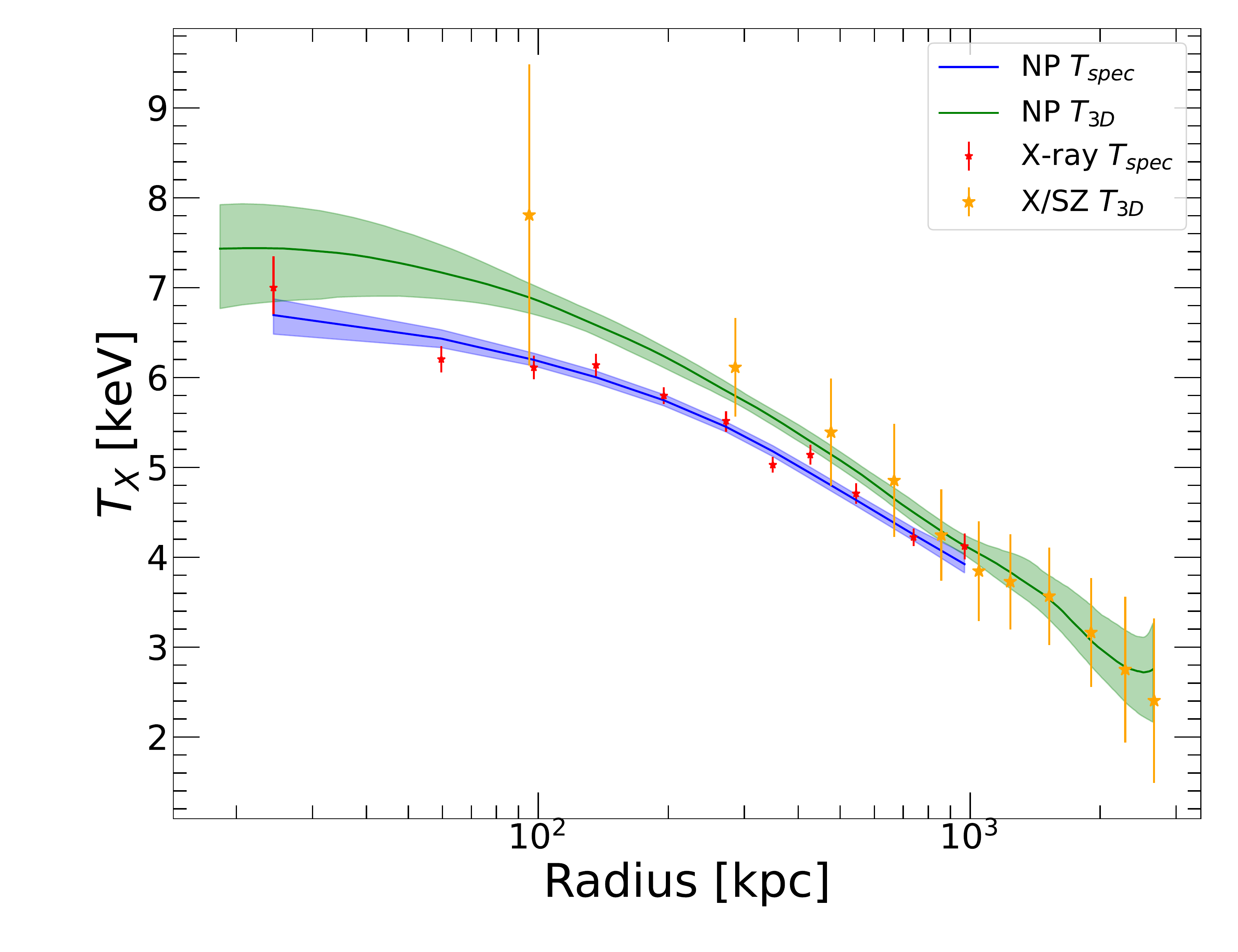}\\
                                
                                \includegraphics[width=0.45\textwidth]{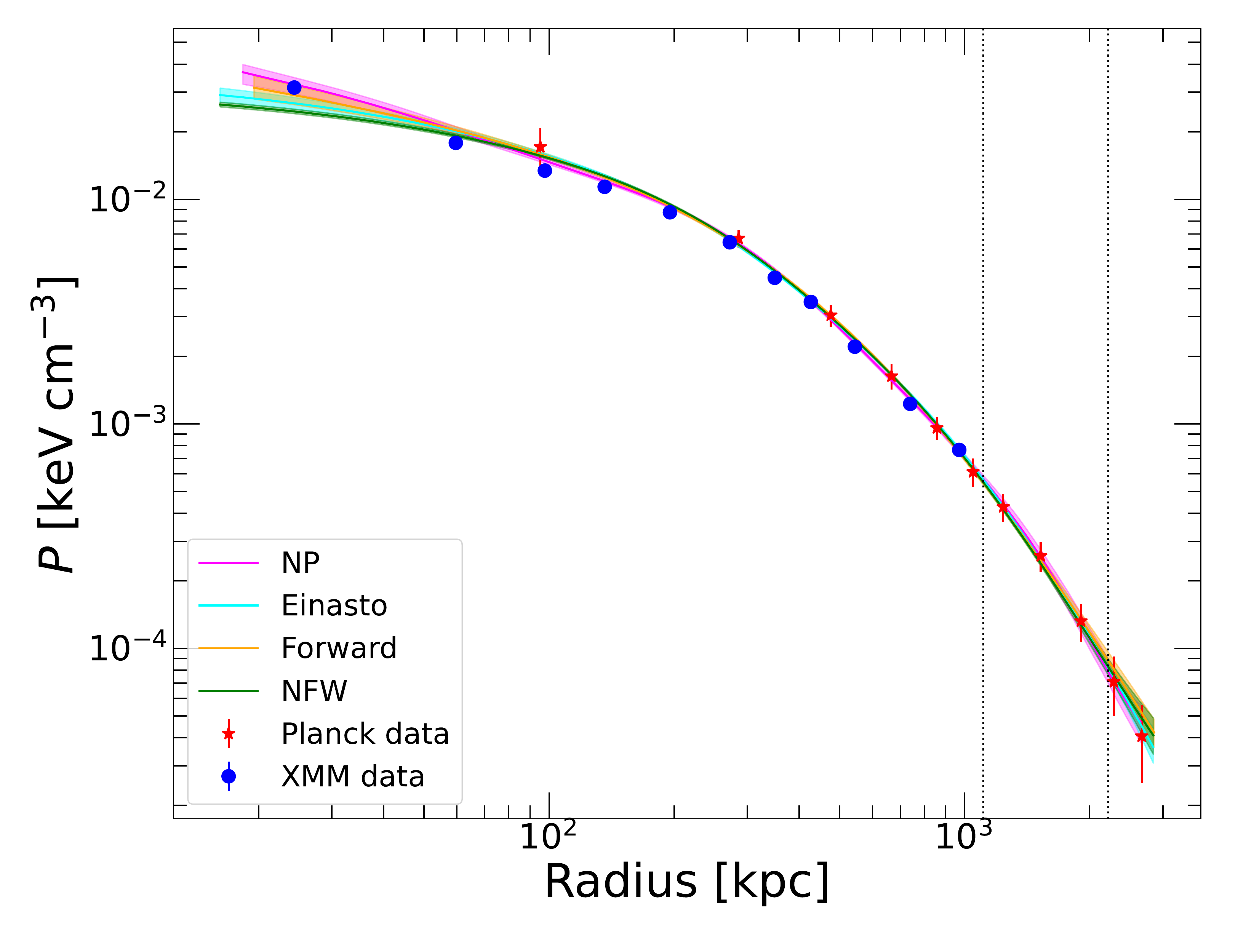}
                                \includegraphics[width=0.45\textwidth]{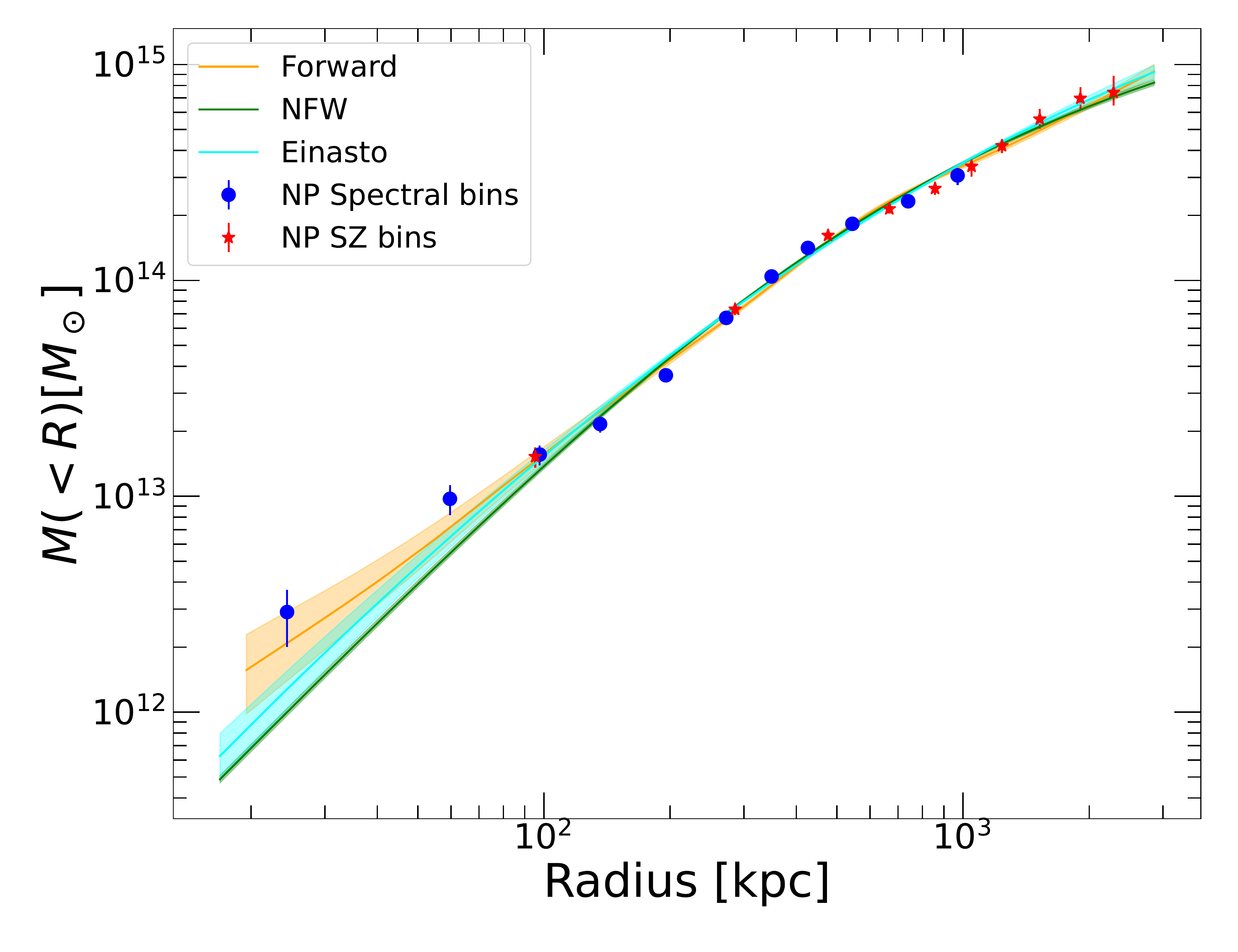}\\
                                
                                \includegraphics[width=0.45\textwidth]{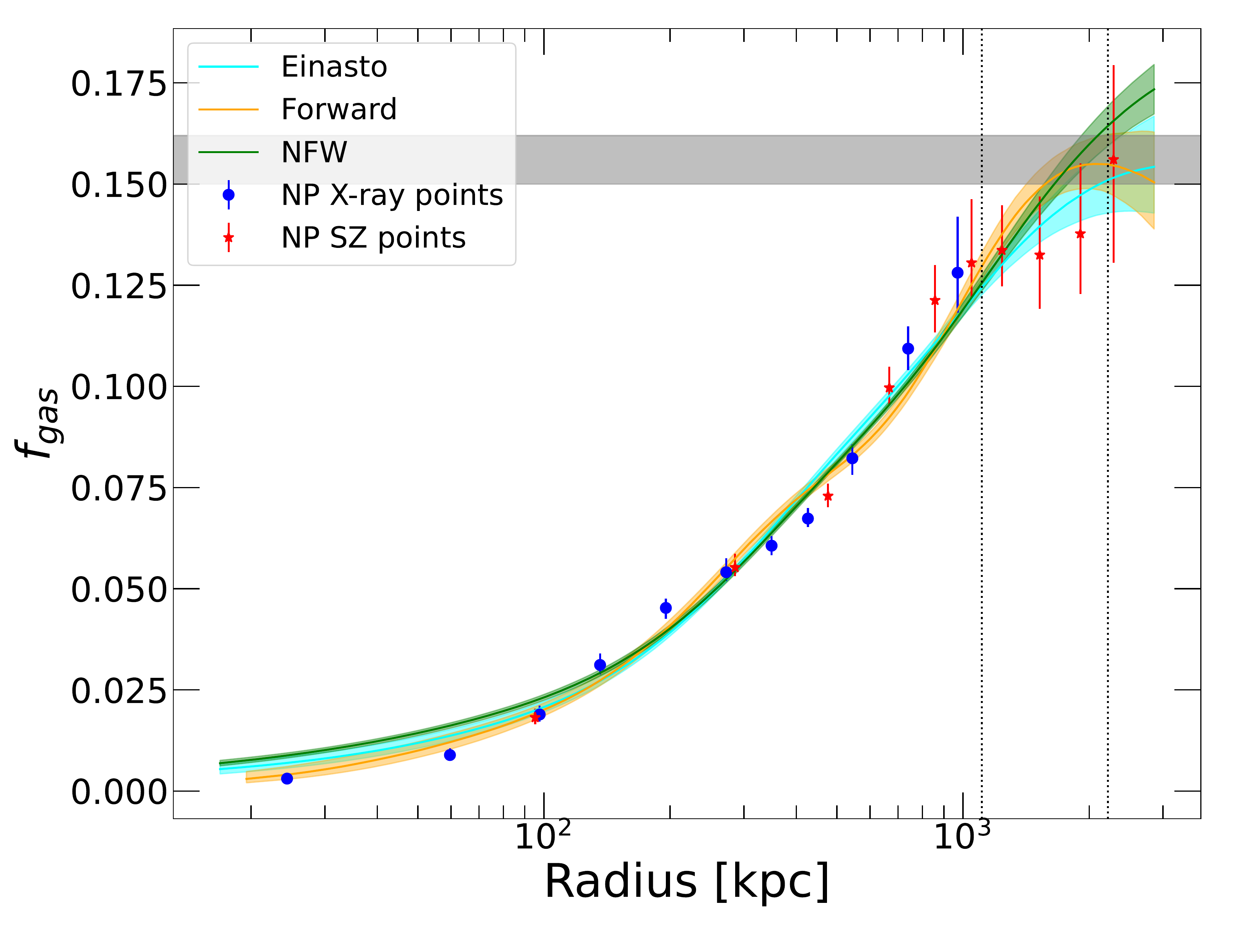}
                                \hspace{0.5cm}
                                \includegraphics[width=0.37\textwidth]{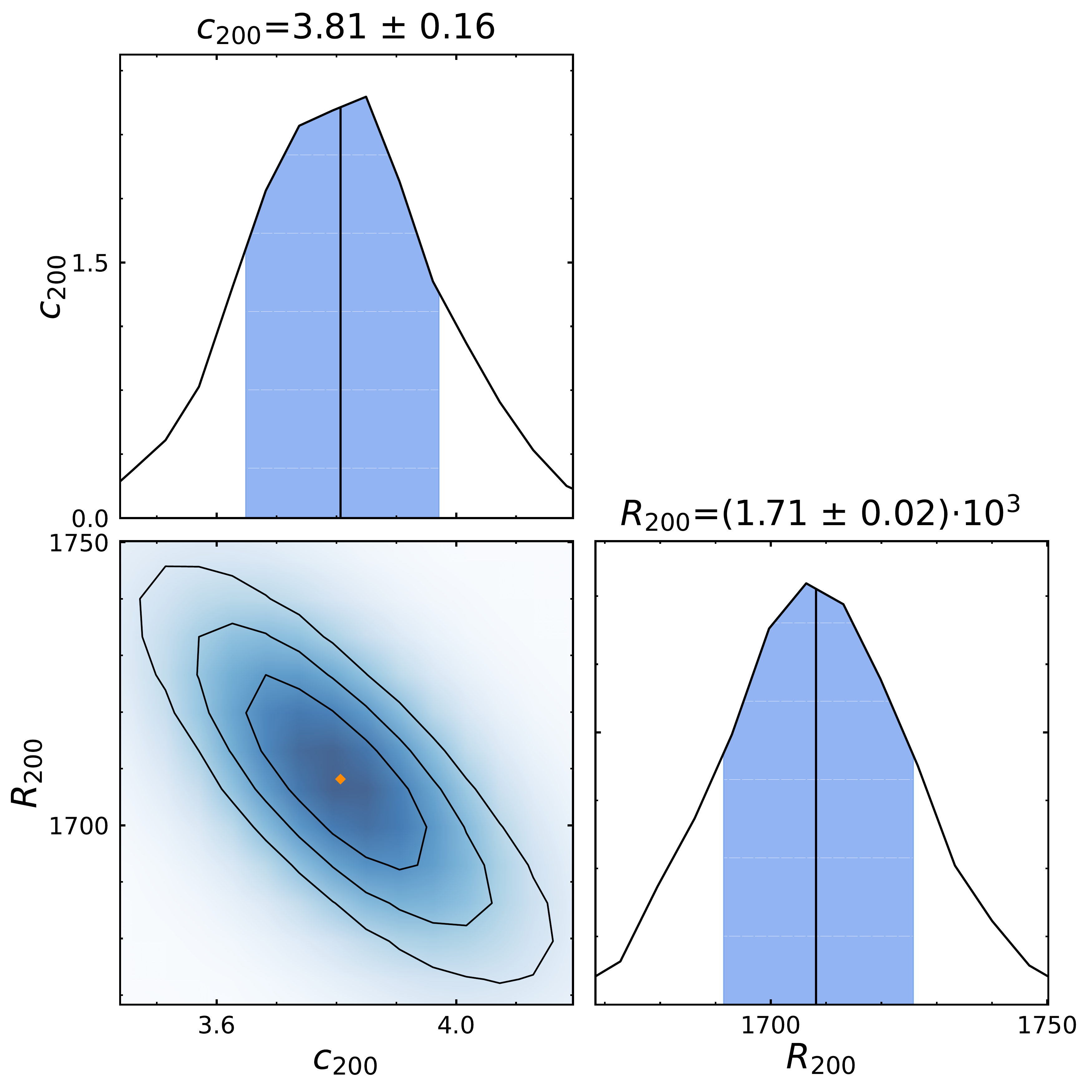}
        }}}
        \caption{Same as Fig. \ref{fig:a1795} but for RXC1825. } 
\end{figure*}

\begin{figure*}
        \centerline{\resizebox{\hsize}{!}{\vbox{
                                \includegraphics[width=0.45\textwidth]{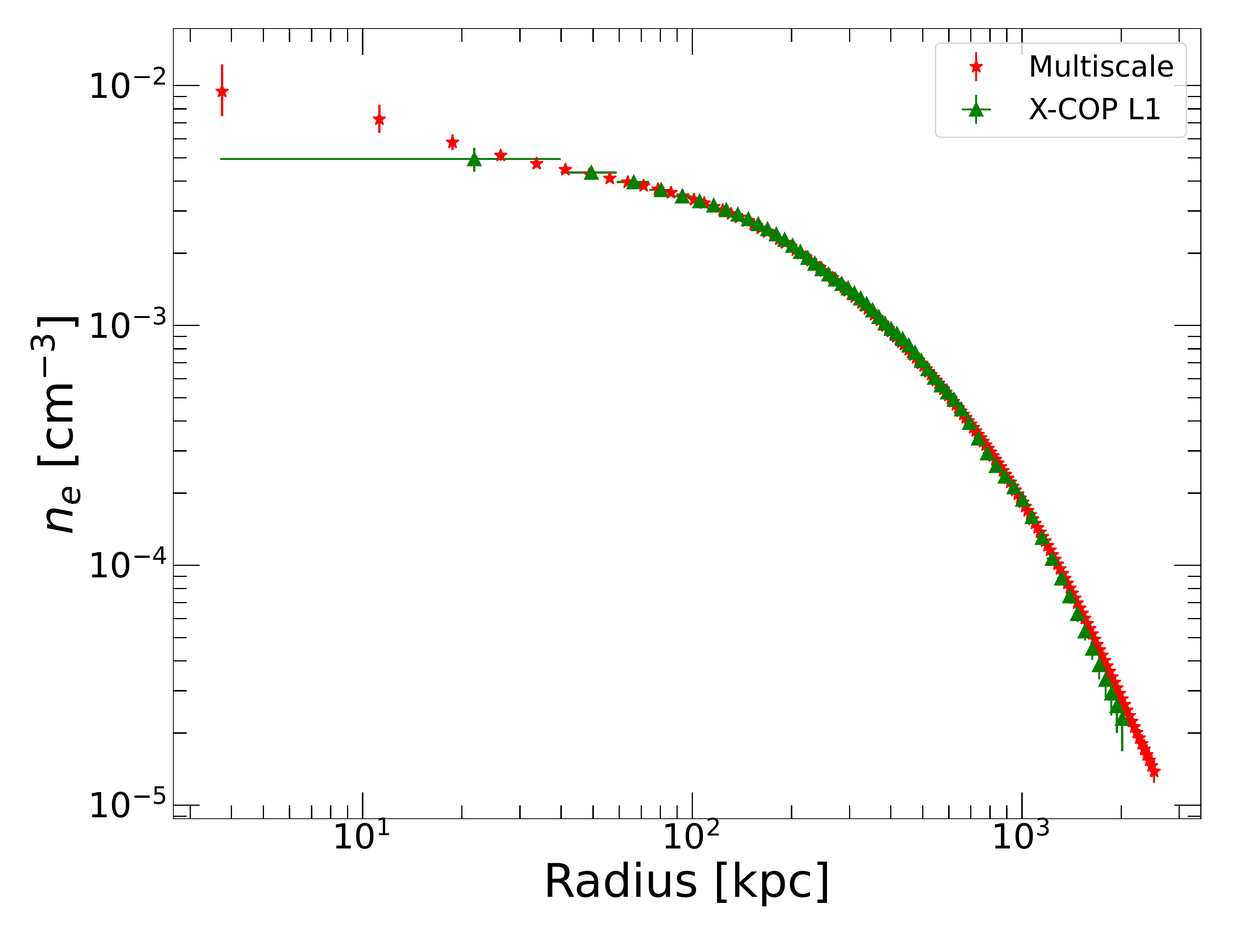}
                                \includegraphics[width=0.45\textwidth]{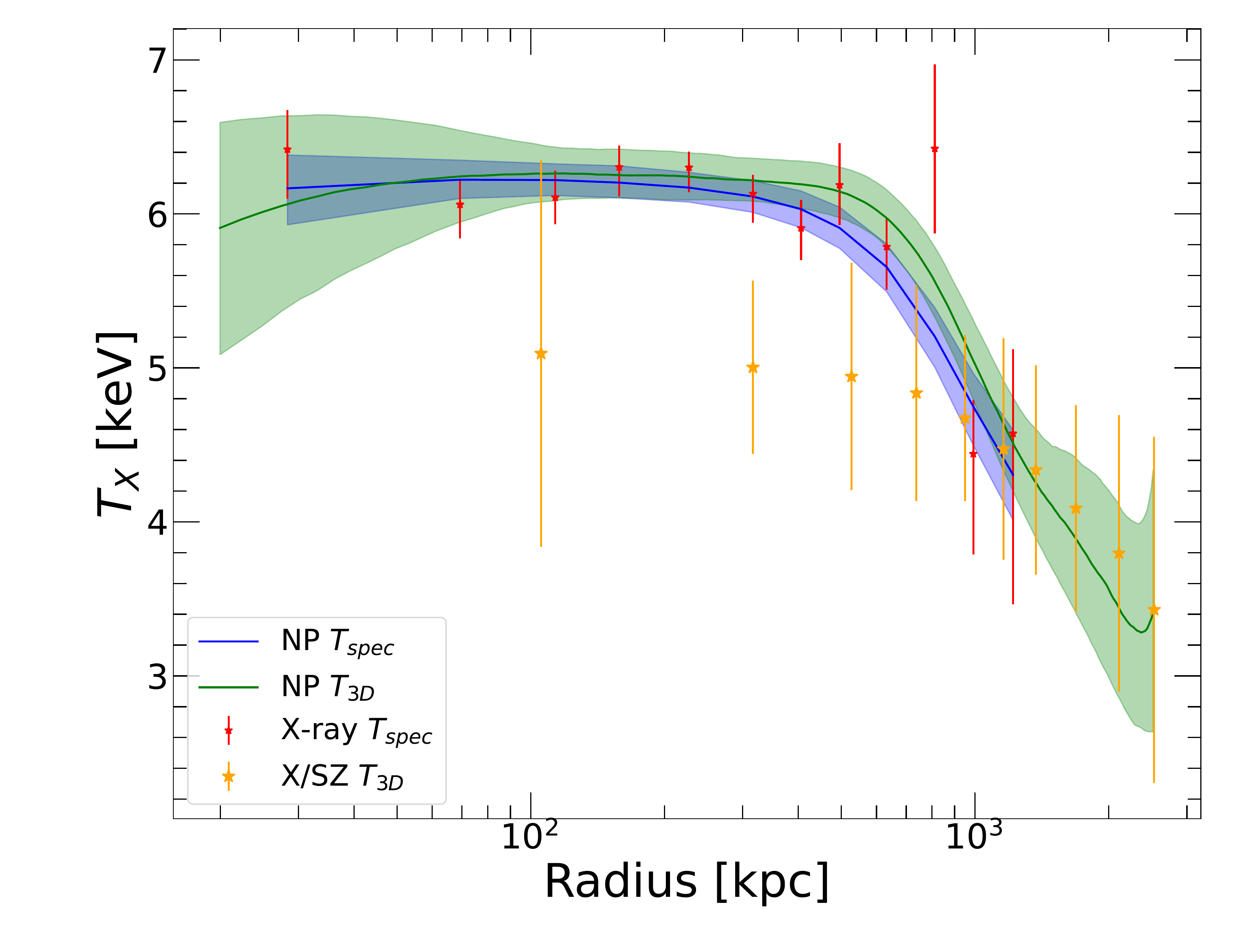}\\
                                
                                \includegraphics[width=0.45\textwidth]{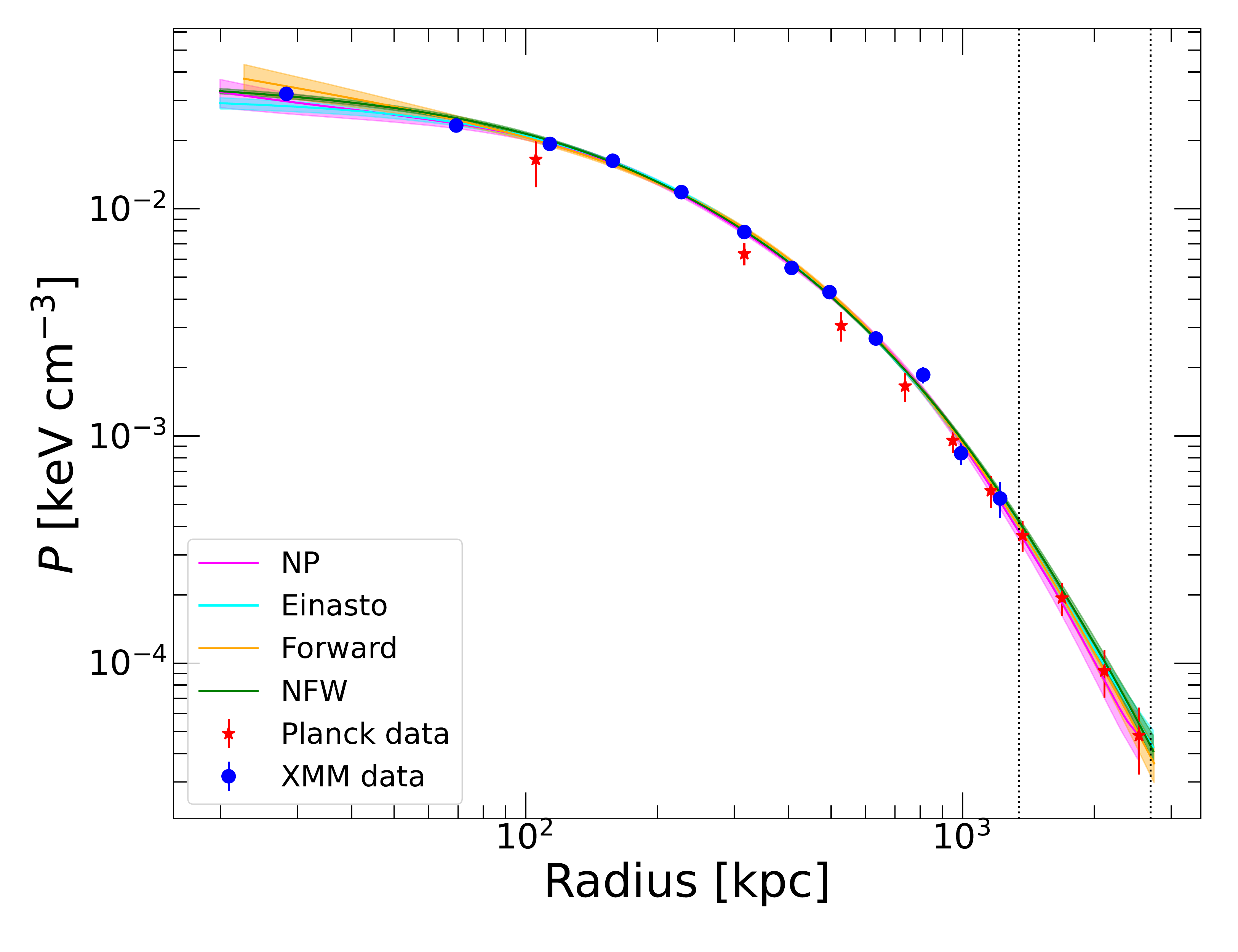}
                                \includegraphics[width=0.45\textwidth]{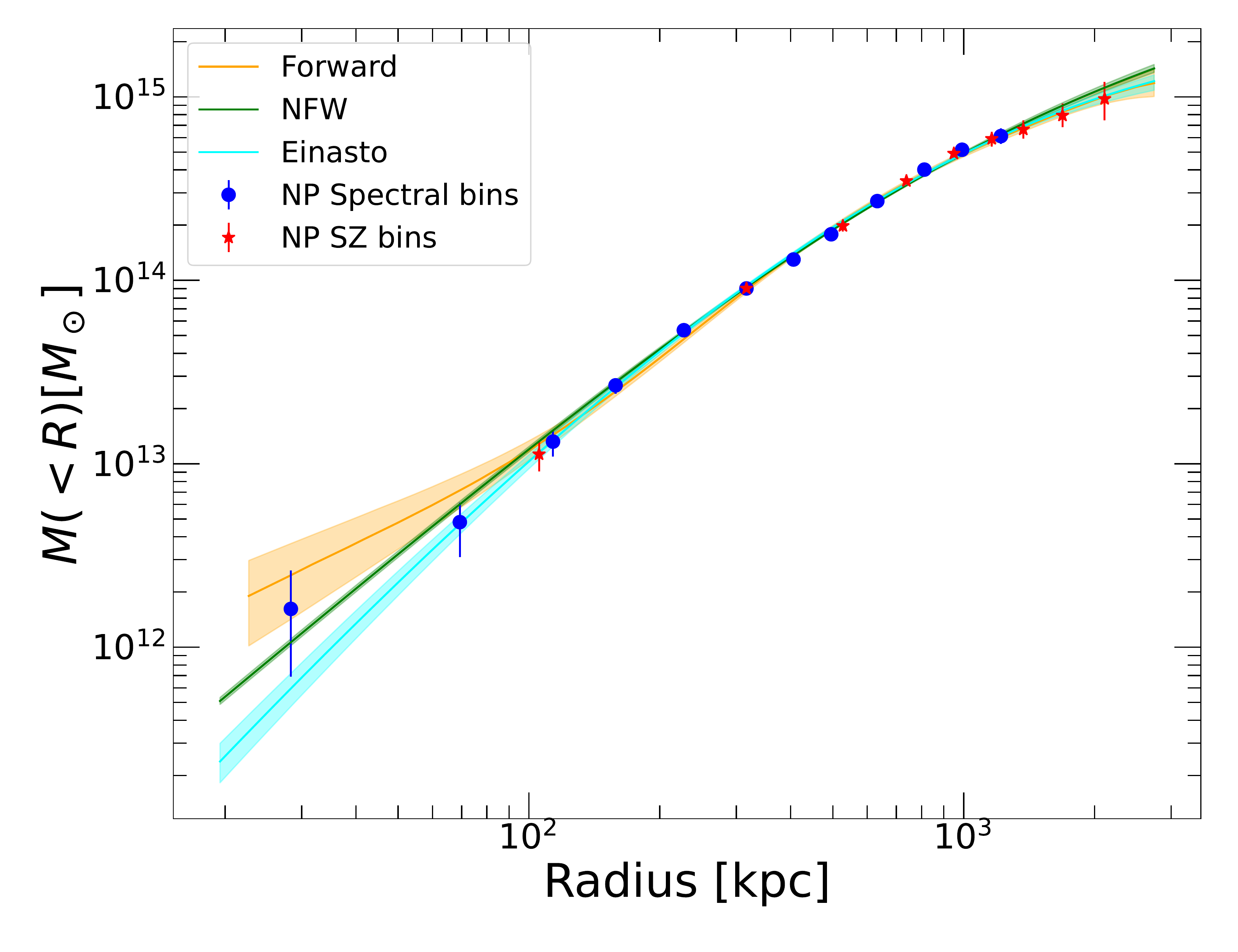}\\
                                
                                \includegraphics[width=0.45\textwidth]{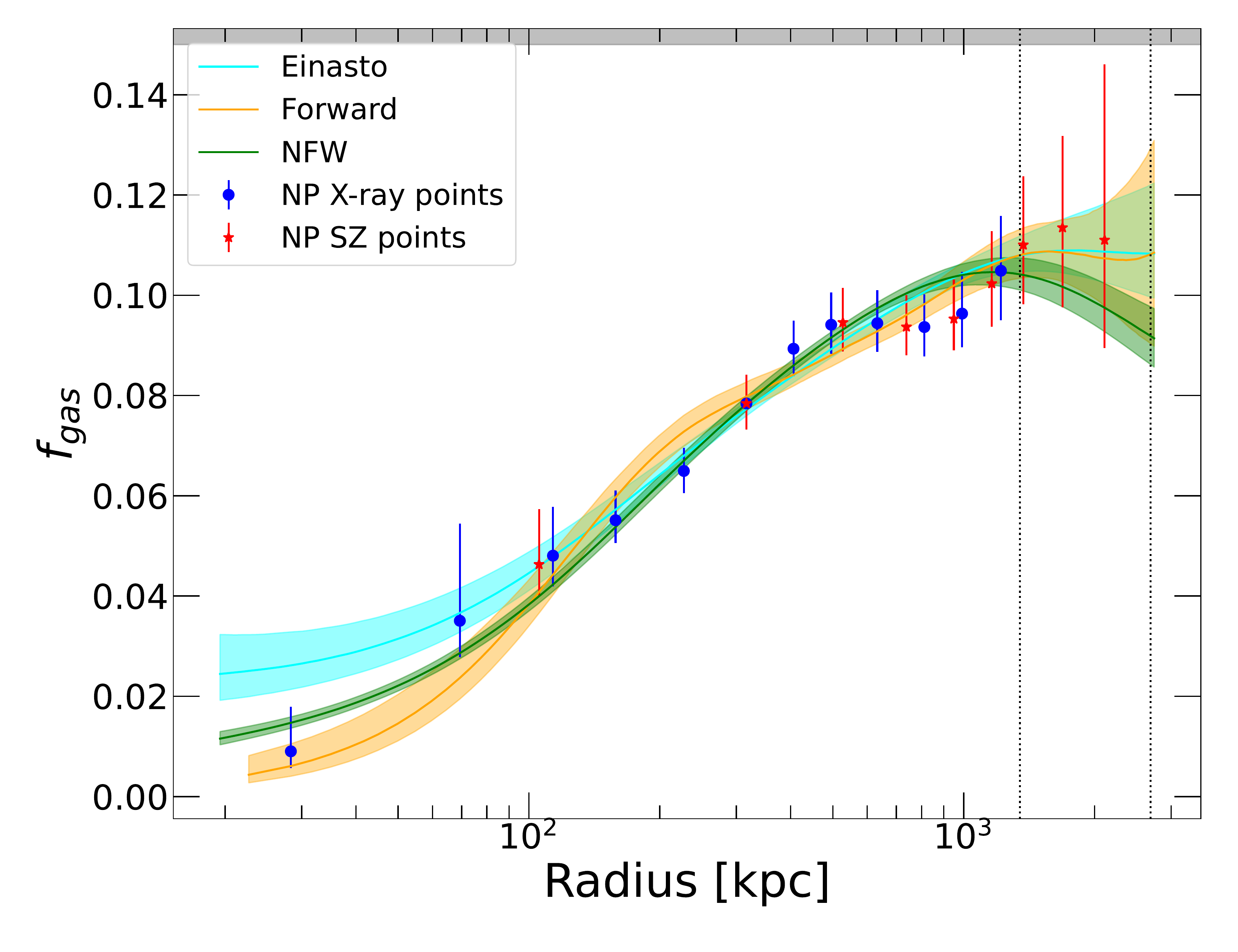}
                                \hspace{0.5cm}
                                \includegraphics[width=0.37\textwidth]{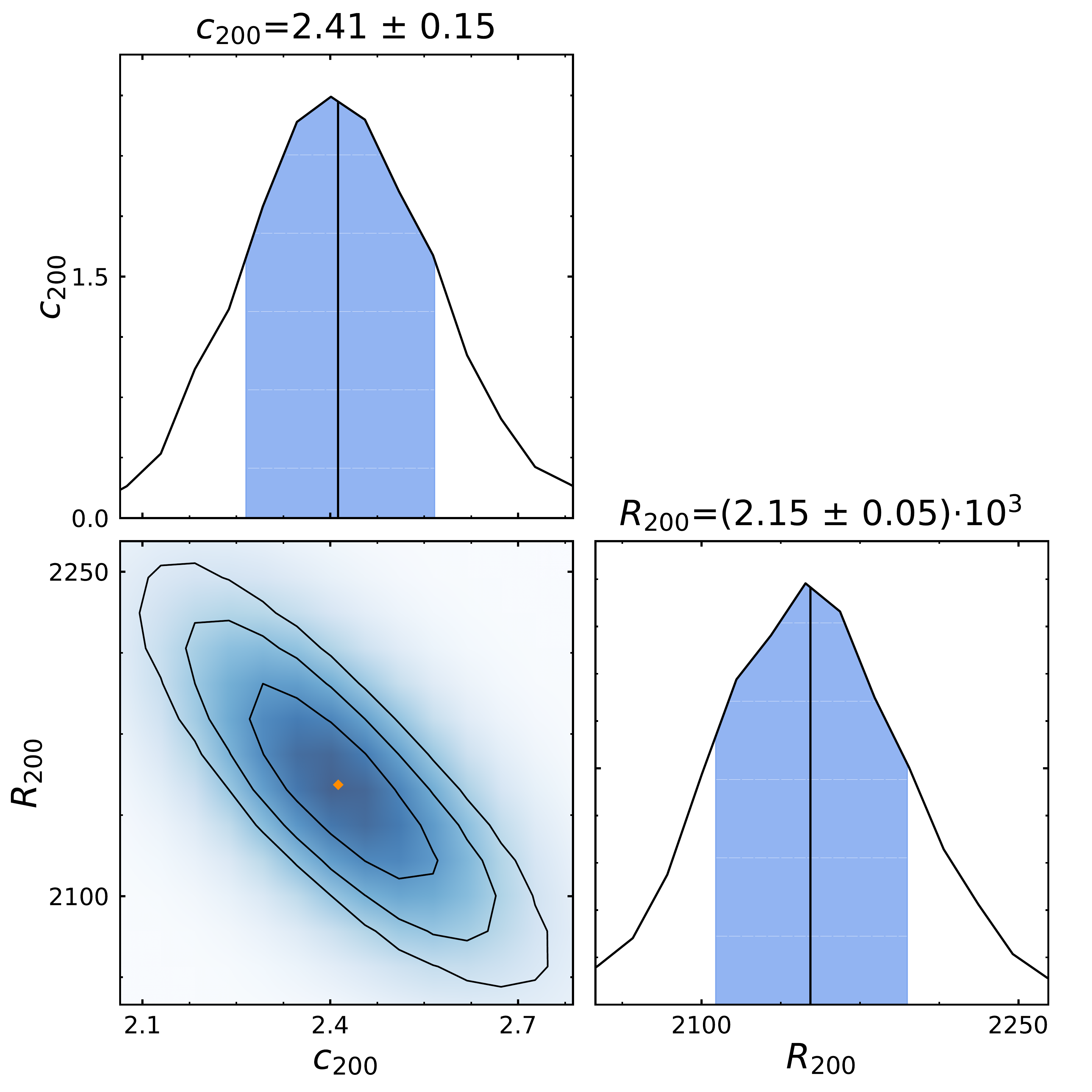}
        }}}
        \caption{Same as Fig. \ref{fig:a1795} but for Zw1215. } 
\end{figure*}

\end{document}